\documentclass[12pt]{st}
\usepackage{epsfig}\parskip 5pt plus 1pt
\newcommand{\nn}{\nonumber}

\newcommand{\be}{\begin{equation}}
\newcommand{\ee}{\end{equation}}
\newcommand{\bea}{\begin{eqnarray}}
\newcommand{\eea}{\end{eqnarray}}
\newcommand{\tc}{ \theta_{13} }
\newcommand{\tatm}{ \theta_{23} }

\newcommand{\He}{\ensuremath{^6{\mathrm{He}\,}}}
\newcommand{\Ne}{\ensuremath{^{18}{\mathrm{Ne}\,}}}

\def\nue{\ensuremath{\nu_{e}}}
\def\nubare{\ensuremath{\overline{\nu}_{e}}}

\def\simge{\mathrel{%
   \rlap{\raise 0.511ex \hbox{$>$}}{\lower 0.511ex \hbox{$\sim$}}}}
\def\simle{\mathrel{
   \rlap{\raise 0.511ex \hbox{$<$}}{\lower 0.511ex \hbox{$\sim$}}}}

\begin{document}
\begin{frontmatter}
\thispagestyle{empty}
\begin{flushright}
{IFT-UAM/CSIC-05-27}\\
{ROMA-1406-05}
\end{flushright}
\title{The impact of solar and atmospheric parameter uncertainties on the measurement 
of $\theta_{13}$ and $\delta$}
\author[Madrid]{A. Donini},
\author[Roma]{D. Meloni} and
\author[Madrid]{S. Rigolin}
\address[Madrid]{I.F.T. and Dep. F\'{\i}sica Te\'{o}rica, 
Univ. Autonoma Madrid., E-28049, Madrid, Spain}
\address[Roma]{INFN, Sezione di Roma e Dip. di Fisica, 
Univ. di Roma ``La Sapienza'', P.le A. Moro 2, I-00185 Roma, Italy} 
\vspace{.3cm}
\begin{abstract}
  We present in this paper the analysis of the measurement of the unknown PMNS parameters 
  $\theta_{13}$ and $\delta$ at future LBL facilities performing complete three parameters fits, 
  each time fully including in the fit one of the atmospheric and solar oscillation parameters 
  within its present (future) error. We show that, due 
  to the presence of degeneracies, present uncertainties on $\theta_{23}$ and $\Delta m^2_{23}$  
  worsen significantly the precision on ($\theta_{13}$,$\delta$) at future LBL experiments. 
  Only if a precision on the atmospheric parameters at least similar to what expected at T2K-I 
  is reached, then the sensitivities to $\theta_{13}$ and $\delta$ that have been presented 
  in the literature for many facilities (where $\theta_{23}$ and $\Delta m^2_{23}$ are generally 
  considered as fixed external inputs) can indeed be almost recovered.
  On the other hand, the impact on this measurement of the uncertainties on the solar parameters, 
  $\theta_{12}$ and $\Delta m^2_{12}$ is already negligible. 
  Our analysis has been performed using three reference setups: the SPL Super-Beam and the 
  standard low-$\gamma$ $\beta$-Beam, both aiming toward a Mton Water \v Cerenkov detector located 
  at $L=130$ km; the 50 GeV Neutrino Factory with a 40 kton Magnetized Iron Detector to look for 
  the ``golden channel'' $\nu_e \to \nu_\mu$ with baseline $L=3000$ km and a 4 kton Emulsion Cloud 
  Chamber to look for the ``silver channel'' $\nu_e \to \nu_\tau$ with baseline $L=732$ km.
\end{abstract}

\vspace*{\stretch{2}}
\begin{flushleft}
\end{flushleft}
\end{frontmatter}
\newpage

\section{Introduction}
\label{sec:intro}
The results of atmospheric, solar, accelerator and reactor \cite{exp} neutrino experiments show 
that flavour mixing occurs not only in the hadronic sector, as it has been known for long, but 
in the leptonic sector as well. The experimental results point to two very distinct mass 
differences\footnote{A third mass difference, $\Delta m^2_{LSND} \sim 1$ eV$^2$, suggested by 
the LSND experiment \cite{lsnd}, has not being confirmed yet \cite{boone} and will not be 
considered in this paper.}, $\Delta m^2_{sol} \approx 8.2 \times 10^{-5}$ eV$^2$ and 
$|\Delta m^2_{atm}| \approx 2.5 \times 10^{-3}$ eV$^2$.
Only two out of the four parameters of the three-family leptonic mixing matrix $U_{PMNS}$ 
\cite{neutrino_osc} are known: $\theta_{12} \approx 32^\circ$ and $\theta_{23}\approx 45^\circ$. 
The other two parameters, $\theta_{13}$ and $\delta$, are still unknown: for the mixing angle 
$\theta_{13}$ direct searches at reactors \cite{chooz} and three-family global analysis of the 
experimental data \cite{globalfit,Gonzalez-Garcia:2004jd} give the upper bound $\theta_{13} 
\leq 11.5^\circ$, whereas for the leptonic CP-violating phase $\delta$ we have no informations 
whatsoever. Two additional discrete unknowns are the sign of the atmospheric mass difference and 
the $\theta_{23}$-octant (if $\theta_{23} \neq 45^\circ$). 

The full understanding of the leptonic mixing matrix constitutes, together with the discrimination 
of the Dirac/Majorana character and the measure of its absolute mass scale, the main 
neutrino-physics goal for the next decade. However, strong correlations between $\theta_{13}$ and 
$\delta$ \cite{Cervera:2000kp} and the presence of parametric degeneracies in the 
($\theta_{13},\delta$) parameter space, \cite{Burguet-Castell:2001ez}-\cite{Barger:2001yr}, make 
the simultaneous measurement of the two variables extremely difficult. Several setups have been 
proposed to face these problems and perform this task, the first option being Super-Beam's
(of which T2K \cite{Itow:2001ee} is the first approved one). New machines have been also proposed, 
such as the $\beta$-Beam \cite{Zucchelli:sa} or the Neutrino Factory \cite{Geer:1997iz}.

In the literature, however, the simultaneous measurement of $\theta_{13}$ and $\delta$ has been 
normally studied considering the solar and atmospheric mixing parameters as external quantities 
fixed to their best fit values (see for example Ref.~\cite{Apollonio:2002en} and refs. therein; 
see also \cite{varinuovi} for some recent papers). 
This is clearly an approximation that has been adopted to get a first insight on the problems 
related to the ($\theta_{13},\delta$) measurement. However, the experimental uncertainties on 
these parameters can in principle affect the measurement of the unknowns, and it seems important 
to perform an analysis that goes beyond the two-parameters fits presented in the literature.

In this paper we therefore study, in a systematic way, the impact that ``solar'' (i.e. $\theta_{12}$ 
and $\Delta m^2_{12}$) and ``atmospheric'' (i.e. $\theta_{23}$ and $\Delta m^2_{23}$) parameters 
uncertainties have on the measurement of $\theta_{13}$ and $\delta$ at three of the many proposed 
setups. By doing this we want to catch the characteristic features of the inclusion of external 
parameters uncertainties in a ($\theta_{13},\delta$) measurement. 
A complete six-dimensional fit\footnote{To which one could add in principle other variables such
as the matter parameter or systematic errors, \cite{Huber:2002mx,Burguet-Castell:2005pa}.} 
requires a really hard computing effort. The authors of Refs.~\cite{Huber:2002mx,Huber:2004ug} 
obviate this problem marginalizing over all the external parameters and reducing the fit to 
a two-dimentional one. Our approach, conversely, 
consists of a series of three-parameters fits (taking $\theta_{12},\Delta m^2_{12},\theta_{23}$ 
and $\Delta m^2_{23}$ in turn as the third fitting variable) to be compared with standard 
two-parameters fits in $\theta_{13}$ and $\delta$. In this way, we realized that the atmospheric 
parameters are the external inputs whose uncertainties are more important in the reconstruction 
of ($\theta_{13},\delta$), and that must be better measured in future experiments. 
We have also tried to compare our results with other methods that have been proposed to deal 
with external parameter uncertainties in the measurement of $\theta_{13}$ and $\delta$ such as 
the inclusion of a covariance matrix in two-parameters $\chi^2$'s \cite{Burguet-Castell:2001ez} 
or the so-called {\it CP-coverage} \cite{Huber:2002mx}.

We consider here, as exemplificative setups, three CERN-based facilities: 
\begin{itemize}
\item the 4 MWatt SPL Super-Beam~\cite{Gomez-Cadenas:2001eu} and a $\gamma \sim 100$ 
$\beta$-Beam~\cite{Bouchez:2003fy} both aimed at the Fr\'ejus tunnel where a 440 kT fiducial 
volume UNO-like Water \v Cerenkov detector \cite{Jung:1999jq} could be located with a $L=130$ 
km baseline. 
\item the CERN-based 50 GeV Neutrino Factory (see Ref.~\cite{Apollonio:2002en} and refs. therein), 
with two detectors of different characteristics to take advantage of both the ``golden'' \cite{Cervera:2000kp} 
and ``silver'' \cite{Donini:2002rm} channels $\nu_e \to \nu_\mu,\nu_\tau$. The two detectors considered are 
a 40 kT magnetized iron detector \cite{Cervera:2000vy} located at $L=3000$ km and a 4 kT emulsion cloud 
chamber \cite{Autiero:2003fu} located at $L=732$ km in the Gran Sasso tunnel.
\end{itemize}

By comparing the results at these three, very different, facilities, we deduce that the impact of the 
atmospheric parameters uncertainties is a common problem that future experiments looking for $\theta_{13}$ 
and $\delta$ will have to face. Of course, this analysis can be done for any of the different setups 
proposed in the literature and not analyzed here. Our intention is mainly to address, in this paper, 
the problem of how uncertainties in the atmospheric and solar parameters affect the measurement of 
($\theta_{13},\delta$) at setups that have been thoroughly discussed than to present a comprehensive 
comparison between two- and three-parameters fits at all of the facilities proposed in the literature.

The paper is organized as follows: in Sect.~\ref{sec:setup} we shortly introduce the three facilities and
the neutrino-nucleon cross-section; in Sect.~\ref{sec:setup:par} we remind the central values and the uncertainties
of solar and atmospheric parameters; in Sect.~\ref{sec:parcor} we review the parametric degeneracies in the measurement of $\theta_{13}$ and $\delta$ 
in appearance and disappearance channels; in Sect.~\ref{sec:fit} we introduce the statistical approach used in the paper;
in Sect.~\ref{sec:uncert} we present our results for the measurement of $\theta_{13}$ and $\delta$ taking into account
the uncertainties on solar and atmospheric parameters; in Sect.~\ref{sec:sensitivity} we show the CP-violation discovery
potential of the considered facilities taking into account the uncertainties on atmospheric parameters; 
in Sect.~\ref{sec:concl} we eventually draw our conclusions. In App.~A we compare our statistical approach with other methods; 
in App.~B we present three-parameters fits for the three considered setups for different choices of the input pair ($\bar \theta_{13},\bar \delta$).

\section{The experimental setup} 
\label{sec:setup}

In this section we describe, briefly, the three facilities that we will use in 
the following and we remind the neutrino-nucleon cross-section used throughout the paper.

\subsection{The $\beta$-Beam}
\label{sec:setup:bb}

\begin{figure}[t!]
\vspace{-0.5cm}
\begin{center}
\begin{tabular}{cc}
\hspace{-0.3cm} \epsfxsize7.5cm\epsffile{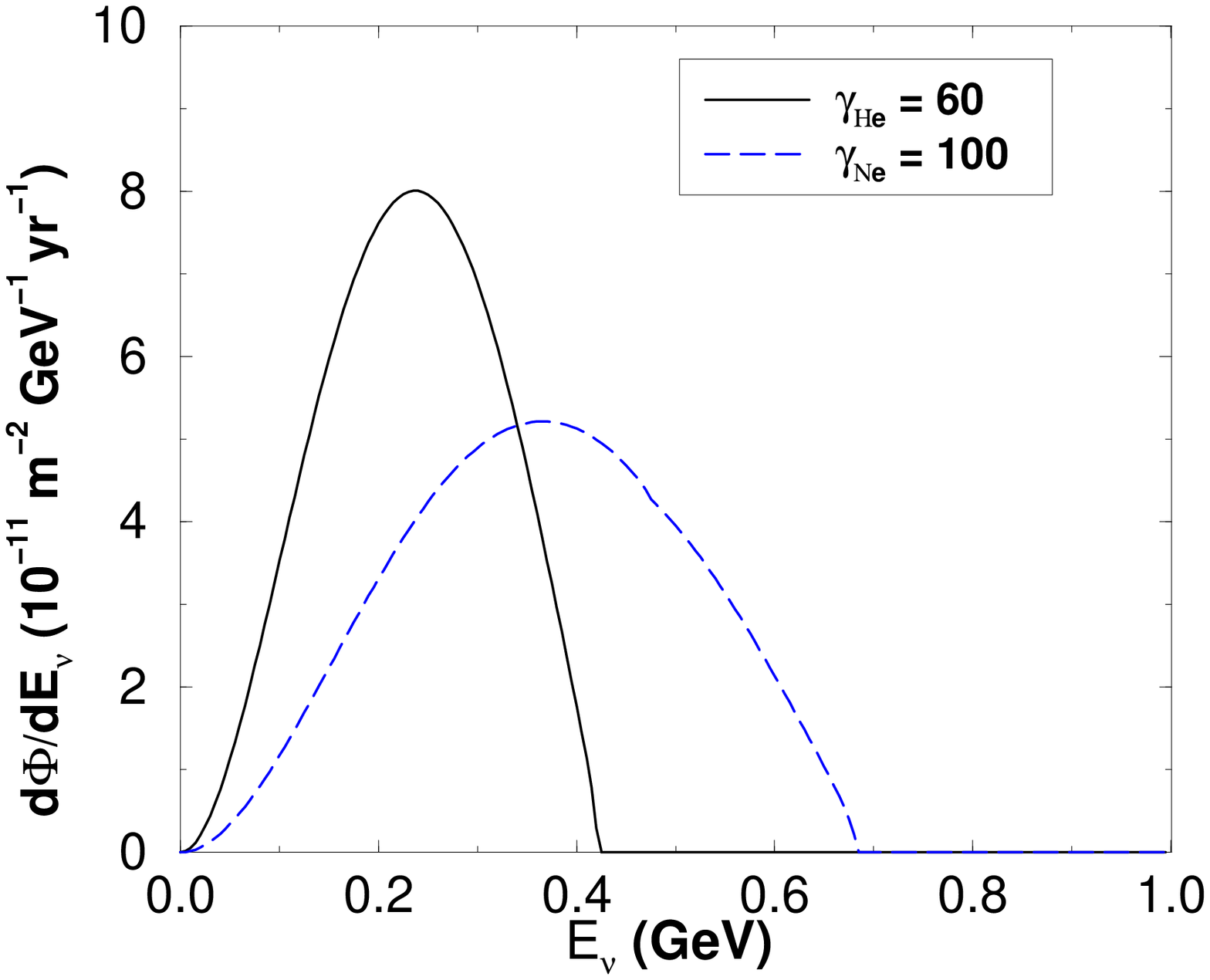}&
\hspace{-0.3cm} \epsfxsize7.5cm\epsffile{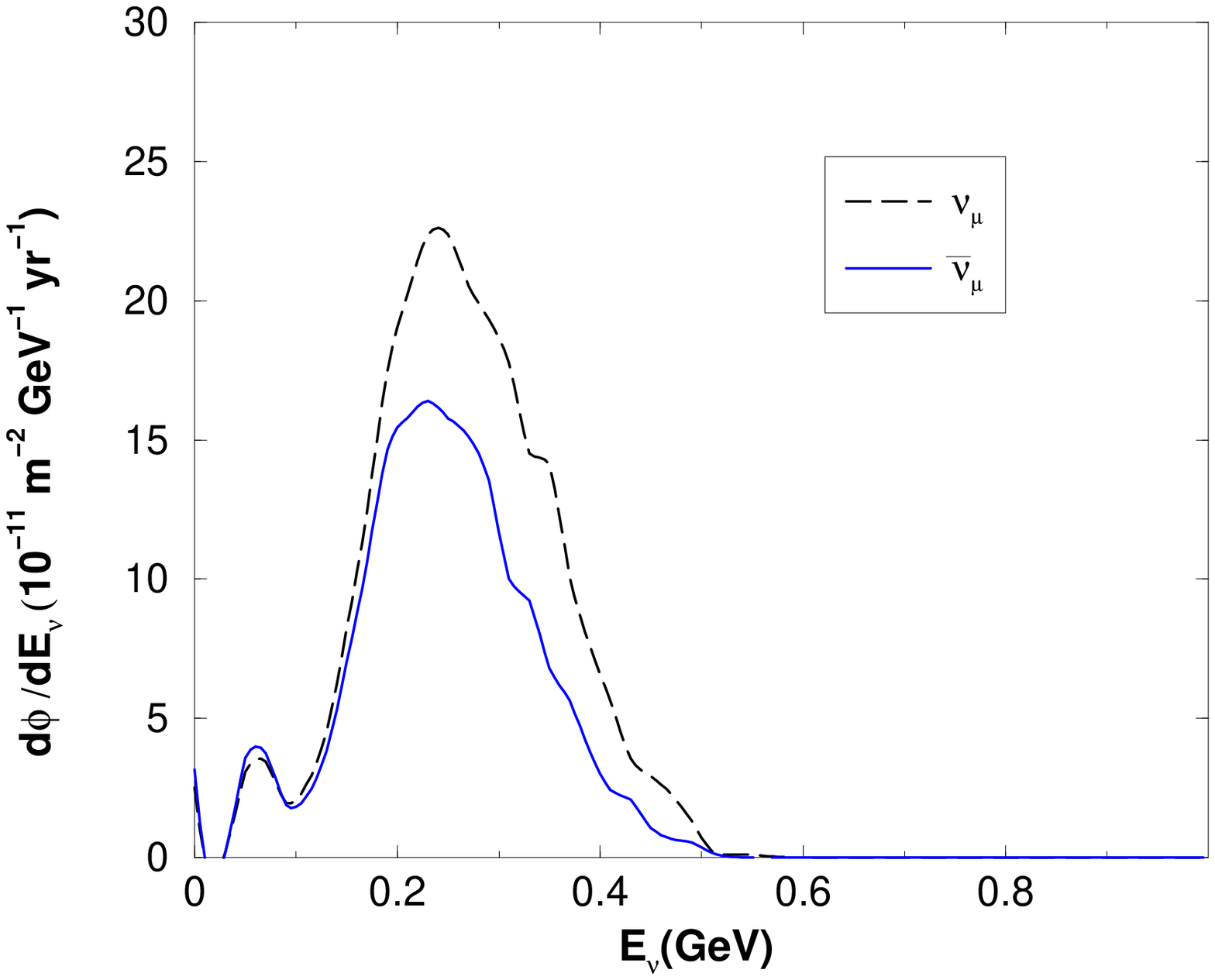} \\
\end{tabular}
\caption{\it \label{fig:fluxes}
Left: $\beta$-Beam fluxes at the Fr{\'{e}}jus location ($L = 130$ km) \cite{Donini:2004hu};
Right: SPL Super-Beam fluxes at the Fr{\'{e}}jus location (130 km baseline) \cite{gilardoni}. }
\end{center}
\end{figure}

The $\beta$-Beam concept was first introduced in Ref.~\cite{Zucchelli:sa}.  
It involves producing a beam of $\beta$-unstable heavy ions, accelerating them to some reference
energy, and allowing them to decay in the straight section of a storage ring, resulting in a 
very intense neutrino beam. The chosen ions are $^6$He, to produce a pure $\bar{\nu}_e$ beam, 
and $^{18}$Ne, to produce a $\nu_e$ beam. We follow the setup proposed in Ref.~\cite{Bouchez:2003fy}: 
the $\gamma$ ratio for the two ions has been fixed to
$\gamma(^6{\rm He})/\gamma(^{18}{\rm Ne})=3/5$, in order to have both ions 
circulating in the storage ring at the same time; the $\gamma$ value has been 
fixed to $\gamma_{^{18}{\rm Ne}} = 100$ (i.e., $\gamma_{^6 {\rm He}} = 60$) to tune the 
neutrino/antineutrino mean energy at the maximum of the $\nu_e \to \nu_\mu$ oscillation probability
for the CERN to Fr\'ejus baseline. 
A flux of $2.9 {\times} 10^{18}$ \He\ decays/year and $1.1{\times}10^{18}$ \Ne\ decays/year is assumed. 
Fig.~\ref{fig:fluxes}(left) shows the $\beta$-Beam neutrino fluxes computed at $L = 130$ km, 
keeping $m_e \neq 0$ \cite{Burguet-Castell:2003vv} and taking into account the three different decay modes
of $^{18}$Ne \cite{Donini:2004hu}.
The mean energy of the \nubare, \nue\ beams for this setup is 0.23~GeV and 0.37~GeV, respectively. 
Clearly, energy resolution is very poor at such low energy, given the influence of Fermi motion and 
other nuclear effects. Therefore, in the following all the sensitivities are computed for a counting 
experiment with no energy cuts \cite{Casper:2002sd}.
The $\beta$-Beam is a clean environment to produce electron-type neutrinos: the main sources of 
systematic error are the overall flux normalization (that can be controlled with a near detector), 
the definition of the fiducial volume of the detector and the neutrino-nucleon cross-sections. 
Alternative $\beta$-Beam proposals can be found in Refs.~\cite{Burguet-Castell:2003vv,Serreau:2004kx,Terranova:2004hu,Burguet-Castell:2005pa}.

\subsection{The Super-Beam}
\label{sec:setup:sb}

A Super-Beam is a conventional neutrino beam with a proton intensity higher 
than that of existing (or under construction) beams such as K2K \cite{Nakaya:2005gz}, NuMI \cite{Kopp:2004zs} and the CNGS \cite{Kodama:2004db}. 
With respect to the $\beta$-Beam and the Neutrino Factory, neutrino beams of a new design, 
it has the advantage of a well known technology. On the other hand, the flux 
composition (with $\nu_\mu$ as the main component for a $\pi^+$ focusing, plus a
small but unavoidable admixture of $\bar{\nu}_\mu$, $\nu_e$ and $\bar{\nu}_e$) 
limits its sensitivity to $\nu_\mu \to \nu_e$ oscillations.

We follow the setup proposed in Ref.~\cite{Gomez-Cadenas:2001eu} as a reference:
a 2.2 GeV proton beam of 4~MWatt power (the SPL), with neutrino fluxes computed in a full simulation of 
the beamline in Ref.~\cite{gilardoni}, assuming a decay
tunnel length of 60~m. The corresponding fluxes are shown in Fig.~\ref{fig:fluxes}(right).
Notice that this beam was designed, originally, as the first stage of a would-be Neutrino Factory, 
and it has not been optimized as a facility to look for $\nu_\mu \to \nu_e$ on its own.
Such an optimization has been presented in Ref.~\cite{Campagne:2004wt}. 
Also in this case, as it was for the $\beta$-beam, the main source of systematic error are
the poorly known neutrino-nucleon cross-sections, the definition of the fiducial volume in the far detector
and the overall normalization of the flux (with the additional problem of new background coming from 
neutrino species not present in the $\beta$-Beam flux). For this setup, also, we consider two tentative values 
of systematic error: an ``optimistic'' 2\% and a ``pessimistic'' 5\%.

\subsection{The Neutrino Factory}
\label{sec:setup:nf}

\begin{figure}[t!]
\vspace{-0.5cm}
\begin{center}
\begin{tabular}{cc}
\hspace{-0.3cm} \epsfxsize7.5cm\epsffile{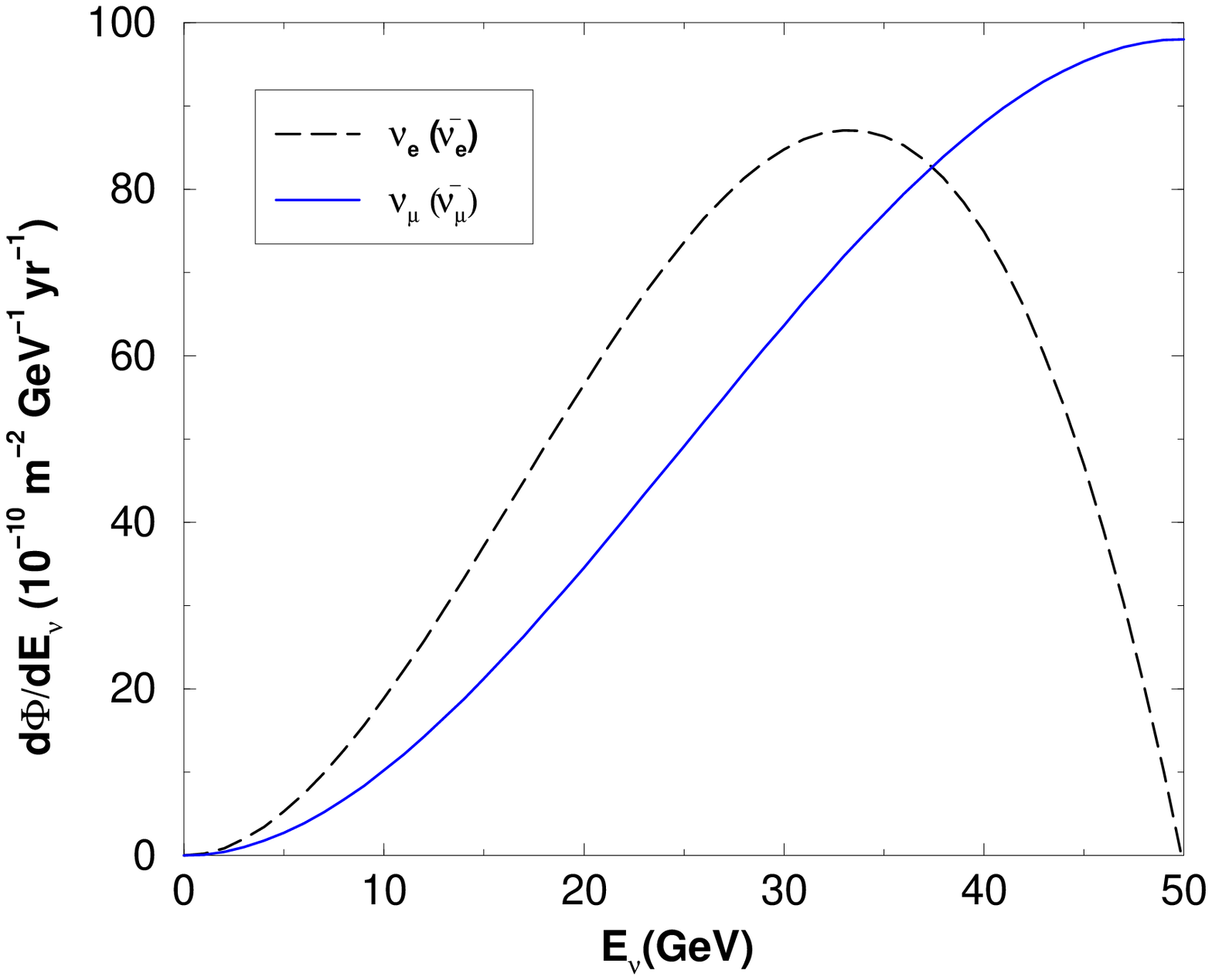} &
\hspace{-0.3cm} \epsfxsize7.5cm\epsffile{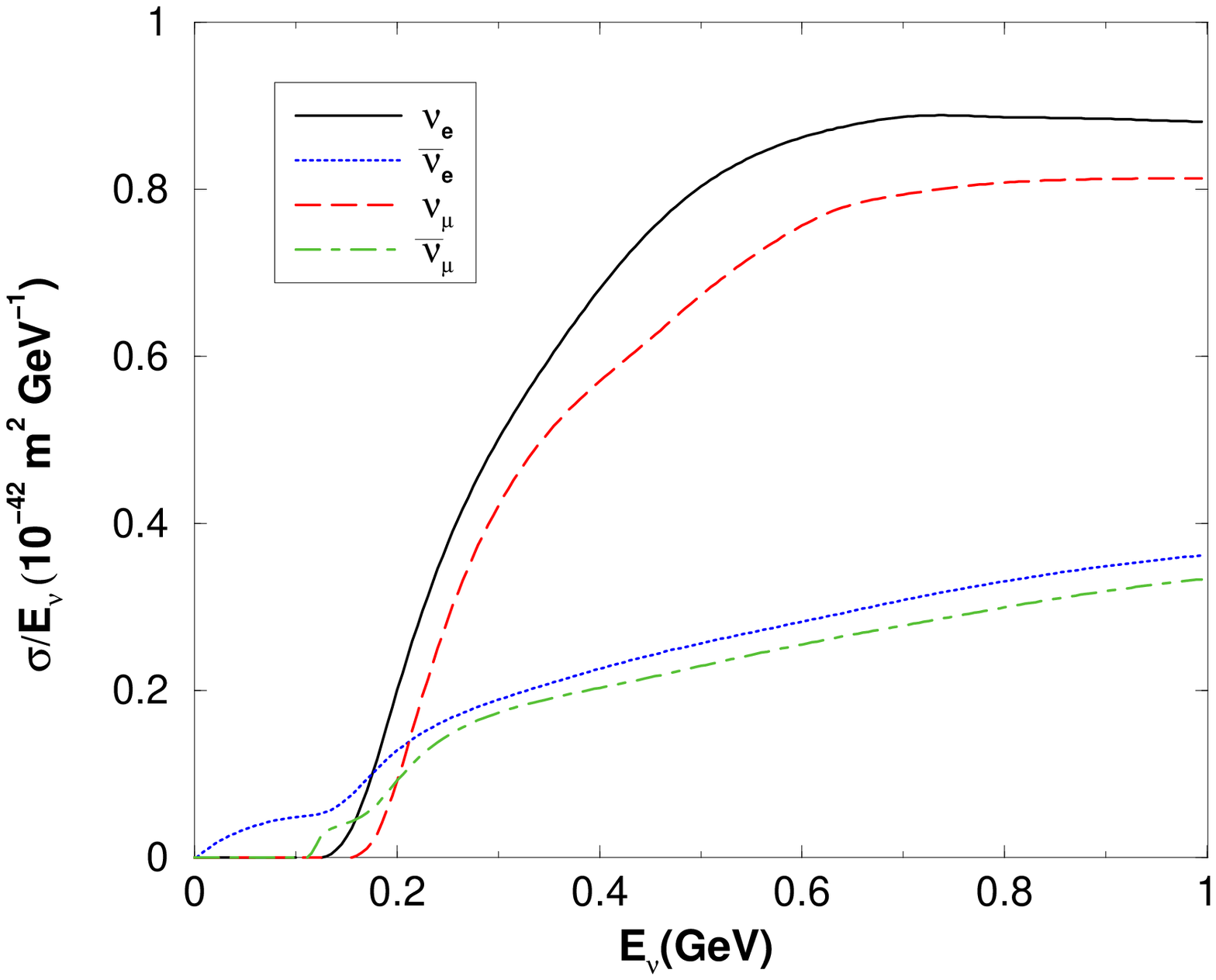}\\
\end{tabular}
 \caption{\it
Left: 50 GeV Neutrino Factory fluxes at the Gran Sasso location ($L = 732$ km) \cite{Cervera:2000kp};
Right: $ \nu N$ and $\bar \nu N$ cross-sections on water \cite{lipari}. }
\label{fig:fluxes2}
\end{center}
\end{figure}

The Neutrino Factory that we consider consists of a SPL-like Super-Beam
and a 50 GeV muon storage ring \cite{Blondel:2000gj}, with $2\times 10^{20}$ muons decaying in the straight
section of the storage ring per year. Five years of data taking for each muon polarity is envisaged.
Two detectors of different technology are considered: a 40 kT Magnetized Iron Detector (MID) at $L = 2810$ km;
and a 4 kT Emulsion Cloud Chamber (ECC) at $L= 732$ or $2810$ km.
This proposal corresponds to the design of a possible CERN-based Neutrino Factory Complex, 
with detectors located at the Gran Sasso Laboratory (the ECC) and at a second site to be defined (the MID and possibly the ECC).
Each one of these detectors is especially optimized to look for a particular signal: 
the ``golden'' channel $\nu_e \to \nu_\mu$ for the 40 kT MID, and the ``silver'' channel $\nu_e \to \nu_\tau$ for the 4 kT ECC. 
The corresponding neutrino fluxes are shown in Fig.~\ref{fig:fluxes2}(left).

The detectors background and systematics for this specific facility have been studied in details in Ref.~\cite{Cervera:2000vy} (the
Magnetized Iron Detector) and in Ref.~\cite{Autiero:2003fu} (the Emulsion Cloud Chamber).

\subsection{The neutrino cross-section}
\label{sec:setup:xs}

An important source of systematic error is our present poor knowledge of the $\nu N$ and $\bar \nu N$ cross-sections
for energies below 1~GeV~\cite{Zeller:2003ey}: either there are very few data (the case of neutrinos) or there are no 
data at all (the case of antineutrinos). On top of that, the few available data have generally not been taken on 
the target used in the experiments (either water, iron or lead), and the extrapolation from different nuclei is complicated by nuclear effects 
that at the considered energies play an important role. 
For definiteness we show in Fig.~\ref{fig:fluxes2}(right) the cross-sections on water used for the Water \v Cerenkov detector throughout 
the paper \cite{lipari}. Notice that we also used cross-sections on iron and lead for the MID and the ECC, respectively.

\section{The leptonic mixing parameters}
\label{sec:setup:par}

In Tab.~\ref{tab:input} we remind the values of solar and atmospheric sector parameters used in the paper, their present
uncertainties and the errors expected after a round of new experiments. 

In particular, in the second column of Tab.~\ref{tab:input} we report the input values for 
$\theta_{12},\theta_{23}, \Delta m^2_{12}$ and $\Delta m^2_{23}$ used in the paper. They correspond to the present 
best fit values for solar and atmospheric parameters \cite{Gonzalez-Garcia:2003qf} with the only exception of 
$\theta_{23}$, for which we do not use the present best fit value, $\theta_{23}=45^\circ$, but $\theta_{23}=40^\circ$ 
to make manifest the impact of possible octant degeneracies on the results \cite{Fogli:1996pv}. 
Notice that throughout the paper the experimentally measured atmospheric mass difference 
(whose present best fit value will be labelled as  $\Delta m^2_{atm}$) will be fitted with the three-family 
parameter $|\Delta m^2_{23}| = |m^2_3 - m^2_2|$ (see \cite{Fogli:2001wi} for a different convention). 
For the solar mass difference, on the other hand, we can unambiguously identify the three-family
parameter $\Delta m^2_{12} = m^2_2 - m^2_1$ with the experimentally measured quantity, $\Delta m^2_{sol}$.
In the third column of Tab.~\ref{tab:input} we report the present uncertainties on each of the parameters.
Finally, in the fourth column, we present the uncertainties on solar and atmospheric parameters that are expected 
to be achieved with ongoing or planned experiments. For an estimate of the reduction of solar parameter uncertainties 
we refer to Ref.~\cite{Choubey:2004tu}. For an estimate of the reduction of atmospheric parameter uncertainties 
we refer to the Letter of Intent of the T2K-phase I experiment, \cite{Itow:2001ee}.
The expected error on $|\Delta m^2_{23}|$ for the central value $|\Delta m^2_{23}|=\Delta m^2_{atm}=2.5\times 10^{-3}$ 
eV$^2$ is a function of the sign of the atmospheric mass difference, something that will not be measured at T2K-I. 
For this reason we present both spreads specifying the chosen hierarchy, using the results of an analysis yet to 
appear, \cite{Enrique}. 

The T2K-I improved bounds are used to analyse the impact of the expected atmospheric 
uncertainties in the measurement of ($\theta_{13},\delta$) at the $\beta$-Beam and the Neutrino Factory\footnote{An updated detailed computation 
of the expected errors on atmospheric parameters that can be obtained at this facility is lacking, \cite{Barger:1999fs,Bueno:2000fg}.}. 
On the other hand, in Sect.~\ref{sec:disSB} it will be shown that the $\nu_\mu$ 
disappearance channel at the SPL Super-Beam can improve significantly the present uncertainties on the atmospheric parameters. 
This measure will therefore be combined with the appearance channel when analysing the impact of the atmospheric 
uncertainties in the measure of ($\theta_{13},\delta$) at the Super-Beam.

\begin{table}[hbtp]
\hspace{-1cm}
\begin{center}
\begin{tabular}{|c|c|c|c|} \hline
    Solar Sector           & Central values       &  Present \cite{Gonzalez-Garcia:2003qf} & Expected (KamLand) \cite{Choubey:2004tu} \\ \hline \hline
$\tan^2 \theta_{12}$       &     $0.39$           & $0.30-0.54$  &    $0.30-0.51$     \\ \hline
$\Delta m^2_{12}$ (eV$^2$) & $8.2 \times 10^{-5}$ & $(7.5-9.1) \times 10^{-5}$ &  
                                                    $(7.7-8.7) \times 10^{-5}$        \\ \hline \hline
 Atmospheric Sector  & Central values             &  Present  \cite{Gonzalez-Garcia:2003qf}   & Expected (T2K-phaseI) \cite{Enrique}\\ \hline \hline
$\tan^2 \theta_{23}$ &     $0.7$                  & $0.53-2.04$  &    $0.62-0.85$ / $1.21-1.66$     \\ \hline
$|\Delta m^2_{23}|$ (eV$^2$) & $2.5 \times 10^{-3}$ & $(1.7-3.5) \times 10^{-3}$ & 
                                                    $s_{atm} = +, (2.42-2.61) \times 10^{-3}$        \\
                             &                      &                            & 
                                                    $s_{atm} = -, (2.46-2.64) \times 10^{-3}$        \\ \hline \hline
                     &                            &              & Expected (SPL, this paper) \\ \hline
$\tan^2 \theta_{23}$ &     $0.7$                  &     $0.53-2.04$         &    $0.53-2.04$     \\ \hline
$|\Delta m^2_{23}|$ (eV$^2$) & $2.5 \times 10^{-3}$ &$(1.7-3.5) \times 10^{-3}$ & 
                                                    $s_{atm} = +, (2.30-2.75) \times 10^{-3}$ \\
                             &                      & & 
                                                    $s_{atm} = -, (2.40-2.90) \times 10^{-3}$ \\ \hline \hline
\end{tabular}
\end{center}
\caption{\label{tab:input} \it Central values and allowed ranges for solar and
atmospheric parameters.
}
\end{table}

\section{Statistical approach} 
\label{sec:fit}

In this section we describe the statistical approach used in the paper to estimate the impact 
of uncertainties in the atmospheric and solar parameters in the measurement of ($\theta_{13},\delta$). 

The obvious approach would be to fit the data in $(N_\alpha + 2)$ parameters, 
with $N_\alpha$ the number of external parameters that are allowed to vary in a given range (e.g., the solar and atmospheric
parameters plus the matter density). This procedure, however, is increasingly time-consuming as the number of parameters to be 
fitted goes up. For this reason, in order to understand how any single parameter affect the measurement, we perform three-parameters fits
in $\theta_{13},\delta$ and one of the following parameters in turn: $\theta_{12}$, $\Delta m^2_{12}$, 
$\theta_{23}$ and $\Delta m^2_{23}$, each of them allowed to vary uniformly in the ranges of Tab.~\ref{tab:input}
(no gaussian priors are introduced). The matter density has been considered as a fixed quantity throughout the paper \cite{Dziewonski:1981xy}. 

To perform the three-parameters fits we have constructed grids for the expected number of 
charged-current events for each facility, each grid in the two unknowns, $\theta_{13}$ and $\delta$, plus 
the two measured parameters in the solar sector ($\theta_{12},\Delta m^2_{12}$) or in the atmospheric sector
($\theta_{23},\Delta m^2_{23}$). When studying the impact of solar parameter uncertainties we have fixed
the atmospheric parameters to $\theta_{23} = 40^\circ$ and $\Delta m^2_{23} = 2.5 \times 10^{-3}$ eV$^2$
and computed four different grids, for $s_{atm}= \pm 1;s_{oct} = \pm 1$. 
When studying the impact of atmospheric parameter uncertainties we have fixed the solar parameters
to $\theta_{12} = 32^\circ$ and $\Delta m^2_{12} = 8.2 \times 10^{-5}$ eV$^2$. In this 
case, only two grids must be computed, one for each value of $s_{atm}$: the octant-degeneracy need not to
be considered as an external (discrete) input, since $\theta_{23}$ is one of the free parameters in the grid. 

When a three-parameters fit is performed, the other parameters in the grid are fixed to the corresponding 
present best fit value for $\theta_{12},\Delta m^2_{12}$ and $\Delta m^2_{23}$ or to $\theta_{23}=40^\circ$ 
for the atmospheric angle (to take into account possible octant and mixed degeneracies, 
that would disappear for maximal mixing). This procedure is used to study the effect of one parameter at a time
on the ($\theta_{13},\delta$) measure. The $\chi^2$ function is:
\bea
\left [ \chi^2(\theta_{13}, \delta, x) \right ]_{\alpha\beta} =
\sum_\pm \left[\frac{
N^\pm_{\alpha\beta} (\theta_{13}, \delta, x; s_{atm}, s_{oct} )-
N^\pm_{\alpha\beta} (\bar \theta_{13}, \bar \delta, \bar x;  \bar s_{atm}, \bar s_{oct} )
}{\delta N^\pm_{\alpha\beta}}\right]^2 \, ,
\label{chi2}
\eea
with $x$ any of the parameters to be fitted in addition to $\theta_{13}$ and $\delta$, $\pm$ refers to neutrinos 
or antineutrinos and $N^\pm_{\alpha\beta}$ is the number of charged leptons $l^\pm_\beta$ observed in the detector 
for a $\nu_\alpha (\bar \nu_\alpha)$ beam. 
The error on the sample $N^\pm_{\alpha\beta}$ is: 
\be
(\delta N^\pm_{\alpha\beta})^2= \sigma^2_{N^\pm_{\alpha\beta}} + (\epsilon^\pm_\beta \,N^\pm_{\alpha\beta})^2 
                                                               + (\epsilon^\pm_\beta \, B^\pm_{\alpha\beta})^2 \, ,
\nn
\ee
where $\sigma_{N^\pm_{\alpha\beta}}$ is the statistical error on $N^\pm_{\alpha\beta}$ (Gaussian or Poissonian, depending 
on the corresponding statistics), $B^\pm_{\alpha\beta}$ is the sum of beam and detector backgrounds for the considered 
channel, computed as in Refs.~\cite{Gomez-Cadenas:2001eu,Bouchez:2003fy}, and $\epsilon^\pm_\beta$ is the total systematic 
error for the considered channel at a given facility. No covariance matrix for the non-fitted parameters has been considered. 
The three-parameters $\chi^2$ function defines a three-dimensional 90\% CL contour that is 
eventually projected onto the ($\theta_{13},\delta$) plane to perform a direct comparison with 
the standard two-parameters 90\% CL contours for the considered
setups\footnote{A preliminary result obtained by means of this procedure has been presented in 
\cite{Meloni:2004ee}.} \cite{Bouchez:2003fy,Burguet-Castell:2003vv,Donini:2004hu}. 

A discussion on the statistical approach chosen and its difference with existing approaches \cite{Burguet-Castell:2001ez,Huber:2002mx} is mandatory and can be found in App.~A.

\section{Parameter correlations and degeneracies}
\label{sec:parcor}

Parameter correlations and degeneracies arise in the determination of $\theta_{13}$ and $\delta$ at future neutrino 
experiments, as it has been studied in many papers \cite{Burguet-Castell:2001ez}-\cite{Barger:2001yr}.
The problem is due to the strong correlation between these two parameters
in the appearance transition probabilities ($\nu_e \to \nu_\mu,\nu_\tau$ and $\nu_\mu \to \nu_e$) 
and in the present (and near future) ignorance of two discrete unknowns, the sign of the atmospheric mass difference 
$\Delta m^2_{23}$ and the $\theta_{23}$-octant, that can be parametrized by the sign variables 
$s_{atm}=~\mbox{sign}[\Delta m^2_{23}]$ and $s_{oct}=~\mbox{sign} [\tan(2\theta_{23})]$ 
that take the values $\pm 1$ for $\Delta m^2_{23} > 0 (< 0)$ and $\theta_{23} < 45^\circ (> 45^\circ)$, respectively.
Solving the systems of equations corresponding to the four distinct choices of $s_{atm}$ and $s_{oct}$:
\bea
\label{eq:ene0int}
N^\pm_{\alpha\beta} (\bar \theta_{13},\bar \delta; \bar s_{atm},\bar s_{oct}) &=&
N^\pm_{\alpha\beta} (\theta_{13},\delta; s_{atm}=\bar s_{atm}; s_{oct}=\bar
s_{oct}) \, ,\\
\label{eq:ene0sign}
N^\pm_{\alpha\beta}(\bar \theta_{13}, \bar \delta; \bar s_{atm}, \bar s_{oct} )&=&
N^\pm_{\alpha\beta} ( \theta_{13}, \delta; s_{atm} = -\bar s_{atm}, s_{oct} = \bar
s_{oct}) \, , \\
\label{eq:ene0t23}
N^\pm_{\alpha\beta}(\bar \theta_{13}, \bar \delta; \bar s_{atm}, \bar s_{oct}) &=&
N^\pm_{\alpha\beta} ( \theta_{13},  \delta; s_{atm} = \bar s_{atm}, s_{oct} = -\bar
s_{oct}) \, , \\
\label{eq:ene0t23sign}
N^\pm_{\alpha\beta}(\bar \theta_{13}, \bar \delta; \bar s_{atm}, \bar s_{oct} )&=&
N^\pm_{\alpha\beta} ( \theta_{13},  \delta; s_{atm} = -\bar s_{atm}, s_{oct} =
-\bar s_{oct}) \, ,
\eea
(with $N^\pm_{\alpha \beta}$ defined in the previous section) will result, in general, in the input pair 
($\bar \theta_{13},\bar \delta$) plus seven additional solutions
(the {\it clones}) to form an eightfold degeneracy: the {\em intrinsic clone} (Eq.~\ref{eq:ene0int}), 
the {\em sign clones} (Eq.~\ref{eq:ene0sign}), the {\em octant clones} (Eq.~\ref{eq:ene0t23}) and the 
{\em mixed clones} (Eq.~\ref{eq:ene0t23sign}). A complete theoretical analysis of the clones location 
has been presented in Ref.~\cite{Donini:2003vz}, where an algorithm to numerically find each clone location 
in the ($\theta_{13},\delta$) plane as a function of the considered experimental setup and of the 
input parameters has been given. A similar approach can be applied to study the presence of degeneracies in
the disappearance channels $\nu_e \to \nu_e$ and $\nu_\mu \to \nu_\mu$. 

\subsection{Correlation and degeneracies in $\nu_e$ disappearance}
\label{sec:disBB}

The $\nu_e$ disappearance probability 
does not depend on the CP violating phase $\delta$ and on the atmospheric $\theta_{23}$ mixing angle. 
The $\theta_{13}$ measurement is, therefore, not affected by $(\theta_{13}-\delta)$ correlations 
nor by the $s_{oct}$ ambiguity. 
The $\nu_e \to \nu_e$ matter oscillation probability, expanded at second order
in the small parameters $\theta_{13}$ and $(\Delta m^2_{12}L/E)$ reads:
\bea
P^\pm_{ee}&=&1-\left(\frac{\Delta_{23}}{B_\mp}\right)^2 \sin^2(2\theta_{13}) \, \sin^2\left(\frac{B_\mp\,L}{2}\right)
- \left(\frac{\Delta_{12}}{A}\right)^2 \sin^2(2\theta_{12}) \,\sin^2\left(\frac{A\,L}{2}\right) \, , \nn\\ 
\label{eq:disnue}
\eea
where $\Delta_{23}=\Delta m^2_{23}/2 E$, $\Delta_{12}=\Delta m^2_{12}/2 E$, $A = \sqrt{2} G_F N_e$
and $B_\mp=|A\mp\Delta_{23}|$ with $\pm$ for neutrinos (antineutrinos),
respectively. This formula describes reasonably well the behaviour of the transition probability in the energy range 
covered by the considered $\beta$-Beam setup ($L\sim 100$ km and $E_\nu\sim$ 100 MeV) and it illustrates clearly that
two sources of ambiguities are still present in $\nu_e$ disappearance, 
$s_{atm}$ (for large values of $\theta_{13}$, i.e. in the ``atmospheric'' region)
and the $(\theta_{13}-\theta_{12})$ correlation (for small values of $\theta_{13}$, i.e. in the ``solar'' region).
A $\beta$-Beam could in principle improve our present errors on the solar parameters through $\nu_e$ disappearance. 
We have checked that this is not the case for the considered setup: at large $\theta_{13}$ the second term in 
eq.~(\ref{eq:disnue}) dominates over the last term, that is more sensitive to solar parameters. On the other hand, 
for small $\theta_{13}$ the statistics is too low to improve present uncertainties on $\theta_{12}$ and $\Delta m^2_{12}$ 
(remind that energy and baseline of the low-$\gamma$ $\beta$-Beam has not been chosen to fulfill this task, and therefore 
our results are not surprising at all). Eventually, in Ref.~\cite{Donini:2004iv} it has been shown that if systematic 
errors cannot be controlled better than at 5\%, the $\beta$-Beam disappearance channel does not improve the CHOOZ 
bound on $\theta_{13}$.

Eq.~(\ref{eq:disnue}) can be also applied to reactor experiments aiming to a precise measurement of $\theta_{13}$ in 
a ``degeneracy-free'' environment. For the typical baseline and energy of a reactor experiment (e.g., $L = 1.05$ km 
and $\langle E_\nu \rangle = 4$ MeV for the Double-Chooz proposal, \cite{Ardellier:2004ui}) we can safely consider 
antineutrino propagation in vacuum. As a consequence, no sensitivity to $s_{atm}$ is expected at these experiments, 
since $B_\mp \to \Delta_{23}$ for $\Delta_{23} \gg A$. It is very difficult that reactor experiments could test small 
values of $\theta_{13}$, and thus the $\theta_{13}-\theta_{12}$ correlation (significant only in the ``solar'' region) 
can also be neglected.

\subsection{Correlation and degeneracies in $\nu_\mu$ disappearance}
\label{sec:disSB}

A Super-Beam facility can perform an independent measurement of the atmospheric parameters via the $\nu_\mu$ disappearance channel: 
these kind of facilities should in principle reduce the error on the atmospheric mass difference to less than 10 \%
and on the atmospheric angle to $\sim 10$ \% \cite{Minakata:2004pg}. It is thus interesting to study, as for the $\nu_e$ disappearance
channel, the presence of parameter correlations and degeneracies. 
The vacuum oscillation probability expanded to the second order in the small
parameters $\tc$ and $(\Delta_{12}L/E)$ \cite{Akhmedov:2004ny} is:
\bea
P(\nu_\mu \to \nu_\mu) & = & 1-  \left [ \sin^2 2\tatm-s^2_{23} \sin^2 2\tc \cos
    2\tatm \right ]\, \sin^2\left(\frac{\Delta_{23}L}{2}\right) \cr
& - & \left(\frac{\Delta_{12} L}{2}\right) [s^2_{12} \sin^2 2\tatm + \tilde{J} 
s^2_{23} \cos \delta] \, \sin(\Delta_{23} L) \cr
& - & \left(\frac{\Delta_{12} L}{2}\right)^2 [c^4_{23} \sin^2 2\theta_{12}+
s^2_{12} \sin^2 2\tatm \cos(\Delta_{23} L)] \, ,
\label{probdismu}
\eea
where $\tilde{J}=\cos \tc \sin 2\theta_{12}\sin 2\theta_{13}\sin 2\theta_{23}$.
The first term in the first parenthesis is the dominant one and is symmetric under $\tatm \to \pi/2-\tatm$.
This is indeed the source of our present ignorance on $s_{oct}$. This symmetry is lifted by the other terms, 
that introduce a mild CP-conserving $\delta$-dependence also, albeit through subleading effects very difficult 
to isolate. We present our results for the $\nu_\mu$ disappearance channel in the ($\theta_{23},\Delta m^2_{23}$) 
plane: as a consequence, we do not need to specify the $\theta_{23}$-octant, since the interval $\theta_{23} \in 
[36^\circ,55^\circ]$ is spanned explicitly. 

\begin{figure}[h!]
\begin{center}
\begin{tabular}{c c}
\hspace{-1.5cm}\epsfxsize8.5cm\epsffile{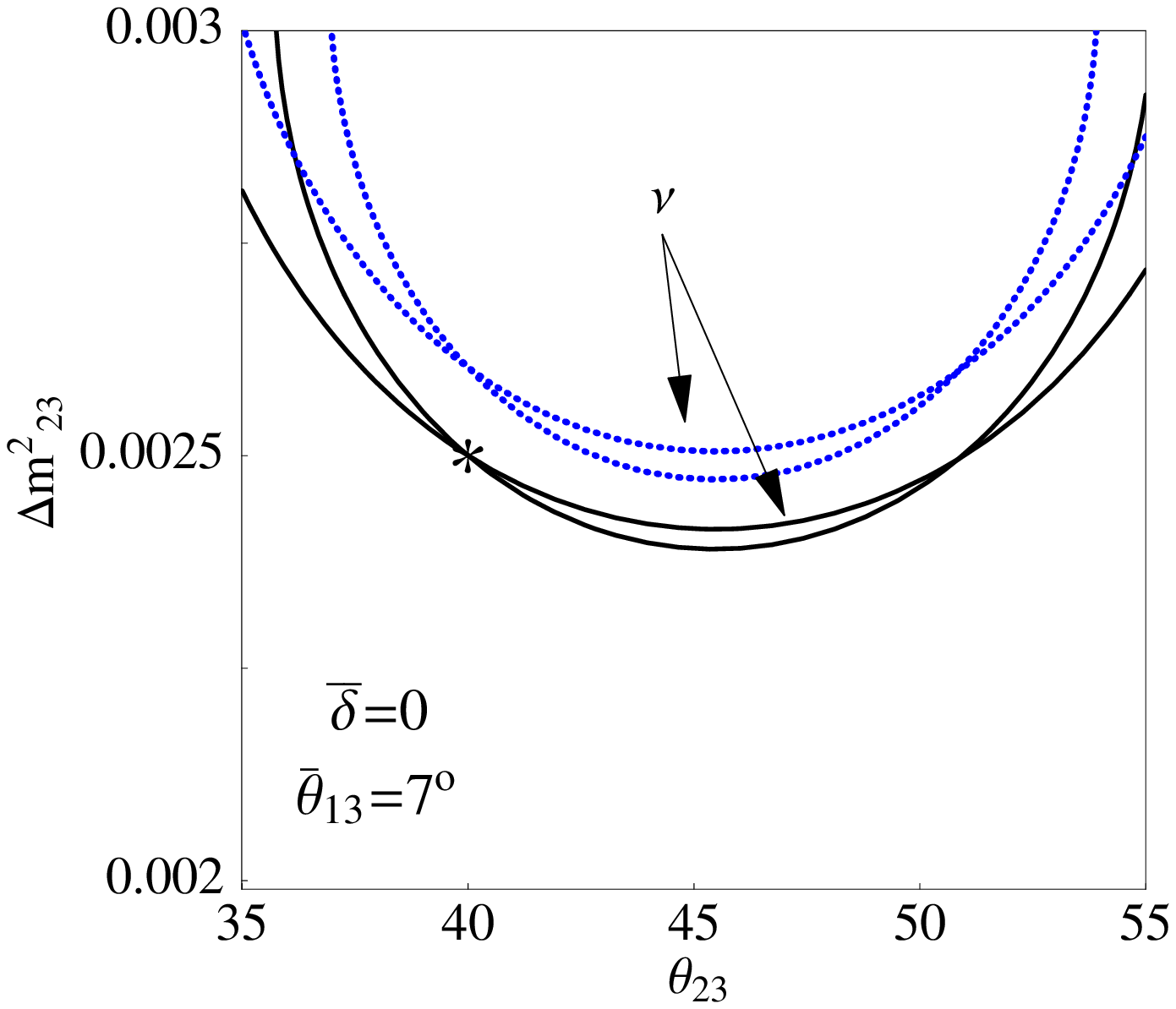} & 
\epsfxsize8.5cm\epsffile{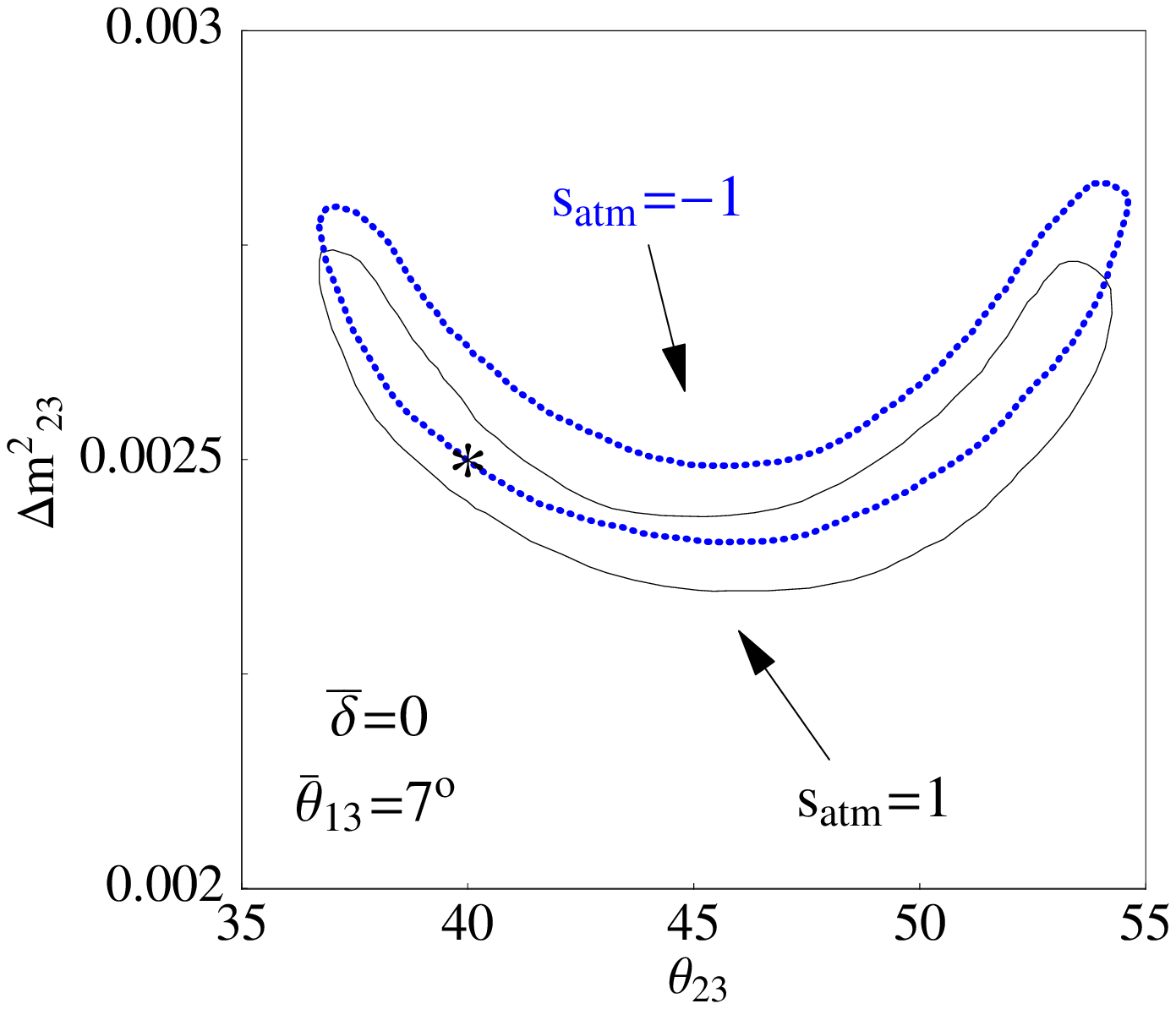}
\end{tabular}
\caption{\it 
Left: ENE curves in the ($\tatm,|\Delta m^2_{23}|$) plane at the SPL Super-Beam facility, 
for $s_{atm} = \bar s_{atm}$ (solid) and $s_{atm} = - \bar s_{atm}$ (dotted) and 
$\bar \theta_{23} = 40^\circ$, $\Delta m^2_{atm} = 2.5 \times 10^{-3}$ eV$^2$.
Right: Projection of the three-parameters 90 \% CL contour in ($\tatm,|\Delta m^2_{23}|,\theta_{13}$)
at the Super-Beam facility for the input point $\bar \theta_{23} = 40^\circ$, $\Delta m^2_{atm} = 2.5 \times 10^{-3}$ eV$^2$ 
and $\bar \theta_{13} = 7^\circ$ onto the ($\tatm,|\Delta m^2_{23}|$) plane. Again, 
the solid (dotted) line stands for $s_{atm} = \bar s_{atm}$ ($s_{atm} = - \bar s_{atm}$).
}
\label{fig:sbdis} 
\end{center}
\end{figure}

Solving the two systems of equations: 
\be
\label{eq:enedismu}
N^{\pm}_{\mu\mu}( \bar \theta_{23}, \Delta m^2_{atm}; \bar s_{atm})= 
N^{\pm}_{\mu\mu} ( \theta_{23}, |\Delta m^2_{23}|; \bar s_{atm}) \, , 
\ee
\be
\label{eq:enedismusign}
N^{\pm}_{\mu\mu}( \bar \theta_{23}, \Delta m^2_{atm}; \bar s_{atm})=
N^{\pm}_{\mu\mu} ( \theta_{23}, |\Delta m^2_{23}|; -\bar s_{atm}) \, , 
\ee
four different solutions are found for $\bar \theta_{23} \neq 45^\circ$: two solutions from eq.~(\ref{eq:enedismu}), 
the input value $\theta_{23} = \bar \theta_{23}$ and $\theta_{23} \simeq \pi/2 - \bar \theta_{23}$, being 
the second solution not exactly at $\theta_{23} = \pi/2 - \bar \theta_{23}$ due to the small 
$\theta_{23}$-octant asymmetry; and two more solutions from eq.~(\ref{eq:enedismusign}) 
at a different value of $|\Delta m^2_{23}|$ \cite{Donini:2004iv}. 
In eq.~(\ref{probdismu}) we can see that changing sign to $\Delta m^2_{23}$ the second term becomes positive: 
a change that must be compensated with an increase in $|\Delta m^2_{23}|$ to give 
$P^\pm_{\mu\mu}(\Delta m^2_{atm}; \bar s_{atm}) = P^\pm_{\mu\mu}(|\Delta m^2_{23}|; - \bar s_{atm})$.
The two solutions of eq.~(\ref{eq:enedismusign}) corresponding to the wrong choice of $s_{atm}$ can be observed 
in Fig.~\ref{fig:sbdis}(left), where equal-number-of-events (ENE) curves are computed for both 
$s_{atm} = \bar s_{atm}$ (solid) and $s_{atm} = - \bar s_{atm}$ (dotted) at the considered Super-Beam facility. 
The two intersections are notably off the input pair $\bar \tatm = 40^\circ$, $\Delta m^2_{23} = \Delta m^2_{atm}$. 
As expected, the two {\it sign clones} are located at $|\Delta m^2_{23}| \geq \Delta m^2_{atm}$
and are almost symmetric with respect to $\tatm=45^\circ$. The shift in the vertical axis is a function of $\theta_{13}$
and $\delta$. 
If $\bar \theta_{23} = 45^\circ$ only two solutions (corresponding to different choices of $s_{atm}$) are expected. 

We must also stress that such an uncertainty can be enhanced once we take into account 
that $\theta_{13}$ and $\delta$ are completely unknown \cite{Minakata:2004pg} (although the impact of this last parameter
in $\nu_\mu$ disappearance is expected to be rather small). To gain some feeling on the precision 
that can be expected in a $\nu_\mu$ disappearance measurement at the SPL Super-Beam facility, 
we performed a full three-parameters analysis in $\theta_{23},|\Delta m^2_{23}|$ and $\theta_{13}$
for the input parameters
$\bar \tatm = 40^\circ$, $\Delta m^2_{23} = 2.5 \times 10^{-3}$ eV$^2$ and $\bar \tc = 7^\circ$.
We have then projected the 90\% three-parameters CL contour onto the $(\tatm,|\Delta m^2_{23}|)$ plane, 
Fig.~\ref{fig:sbdis}(right). In the absence of a complete simulation of the systematics and the background 
for the $\nu_\mu$ disappearance channel at the SPL Super-Beam \cite{Gomez-Cadenas:2001eu},
we have adopted as an estimate of the expected background and efficiency 
those used in \cite{Bouchez:2003fy,Burguet-Castell:2003vv} and \cite{Donini:2004hu} 
for the $\nu_e \to \nu_\mu$ appearance channel at the $\beta$-Beam facility. 
A 2\% systematic error has been assumed.
The solid line refers to the projection of the three-dimensional 90 \% CL contour on the 
($\theta_{23},|\Delta m^2_{23}|$) plane for $s_{atm} = \bar s_{atm}$, the dotted line to the projection 
of the 90 \% CL contour for $s_{atm} = - \bar s_{atm}$.
 
As expected, the three-parameters fit presents a second allowed region in the parameter space at $|\Delta m^2_{23}| > \Delta m^2_{atm}$ 
when the wrong $s_{atm}$ is considered. Notice that, performing a three-parameters fit
in $\theta_{23},\Delta m^2_{23}$ and $\delta$, the difference between the two- and three-parameters
contours is much smaller. In \cite{Donini:2004iv} it has been shown that a larger
spread in $\tatm$ is found for $\bar \theta_{23} \neq 45^\circ$.
We can perform fits with $\tatm$ non-maximal and for different input values for $\bar \tc \in [0, 10^\circ]$. 
The result of such an analysis is that the SPL Super-Beam will be able to measure $\tatm$ in the
interval $[36^\circ,55^\circ]$ and $\Delta m^2_{23}$ in $[2.3,2.9] \times 10^{-3}\,{\rm eV^2}$
for the input pair $\bar \theta_{23} = 40^\circ, \Delta m^2_{23} = \Delta m^2_{atm}$. 
Notice that the expected SPL precision on $\Delta m^2_{23}$ is comparable with what expected at T2K-I 
\cite{Itow:2001ee}. On the other hand, the expected SPL precision on $\theta_{23}$ is much worse
than the T2K-I one, a consequence of the fact that the considered 
SPL setup is a counting experiment and it has no energy resolution.

\section{Impact of parameter uncertainties on $\theta_{13}$ and $\delta$}
\label{sec:uncert}

In this section, we discuss the impact of the uncertainties in the solar and atmospheric 
parameters to the simultaneous measurement of $\theta_{13}$ and $\delta$ at the considered $\beta$-Beam, Super-Beam 
and Neutrino Factory. The three experiments will be discussed separately.

In all fits we have combined informations from all available channels for both polarities, 
\be
\label{eq:chi2atm}
\chi^2(\theta_{13}, \delta, x)= \sum_i \chi^2_i(\theta_{13},\delta,x) \, , 
\ee
where $\chi^2_i$ is the three-parameters $\chi^2$ function defined as 
in eq.~(\ref{chi2}) for a given channel and polarity.
All channels have been taken as independent measurement and no covariance matrix has been introduced, following 
the approach described in Sect.~\ref{sec:fit}. In general, a ``pessimistic'' systematic error, $\epsilon^\pm = 5$\%, has been used
in appearance channels. On the other hand, a 2\% systematic error has been used in disappearance channels. 

\subsection{The solar sector}
\label{sec:solar}

We study the effect of present uncertainties on the solar sector parameters in the measurement 
of $\theta_{13}$ and $\delta$ performing two distinct three-parameters fit in $\theta_{13},\delta$ and 
$\theta_{12}$ (for fixed $\Delta m^2_{12}$) or $\Delta m^2_{12}$ (for fixed $\theta_{12}$). 
The fits have been performed using 10 years of $\beta$-Beam running with both polarities.

\begin{figure}[t!]
\begin{center}
\begin{tabular}{cc}
\hspace{-1.0cm}
\epsfxsize8cm\epsffile{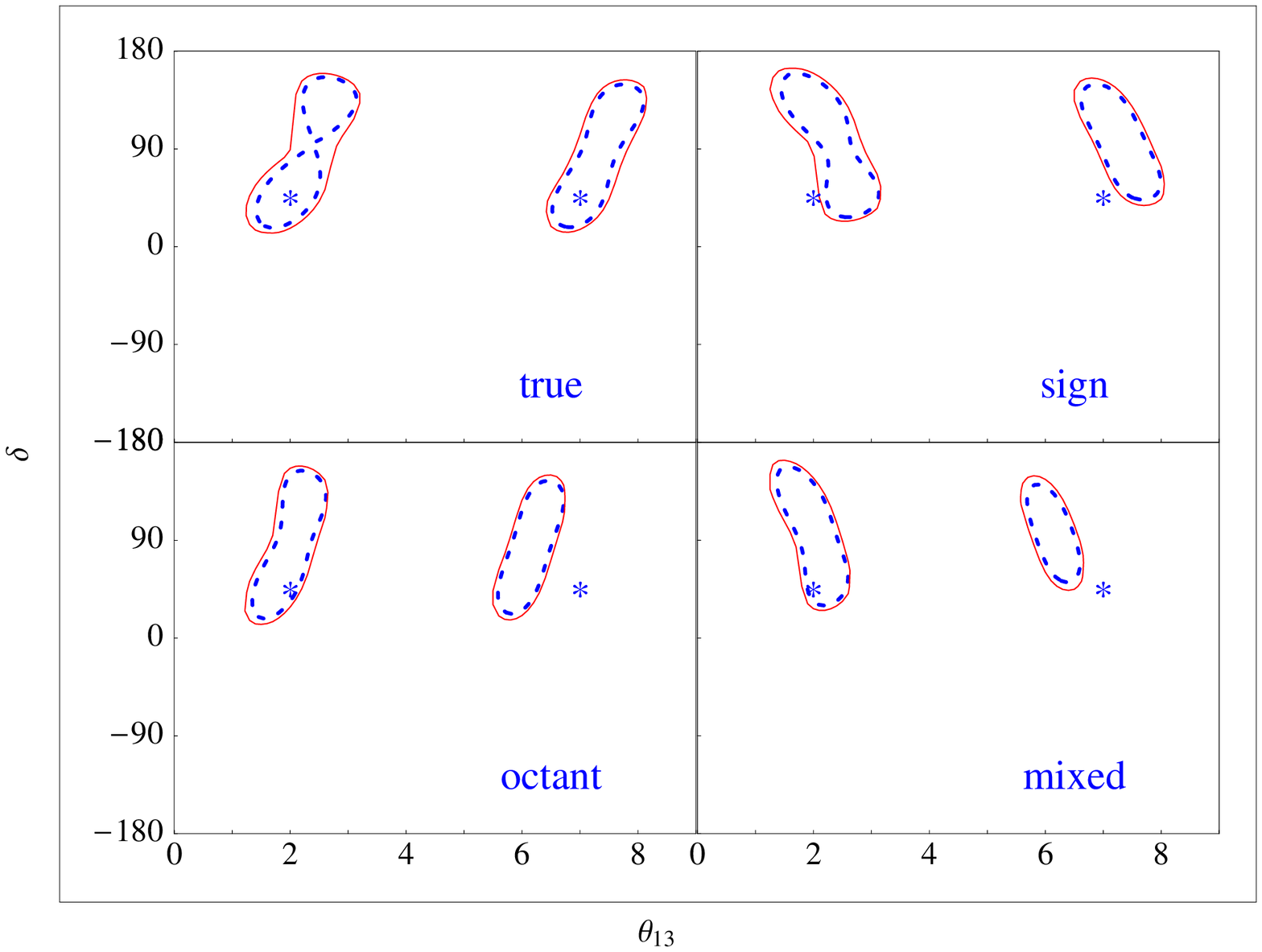} &
\epsfxsize8cm\epsffile{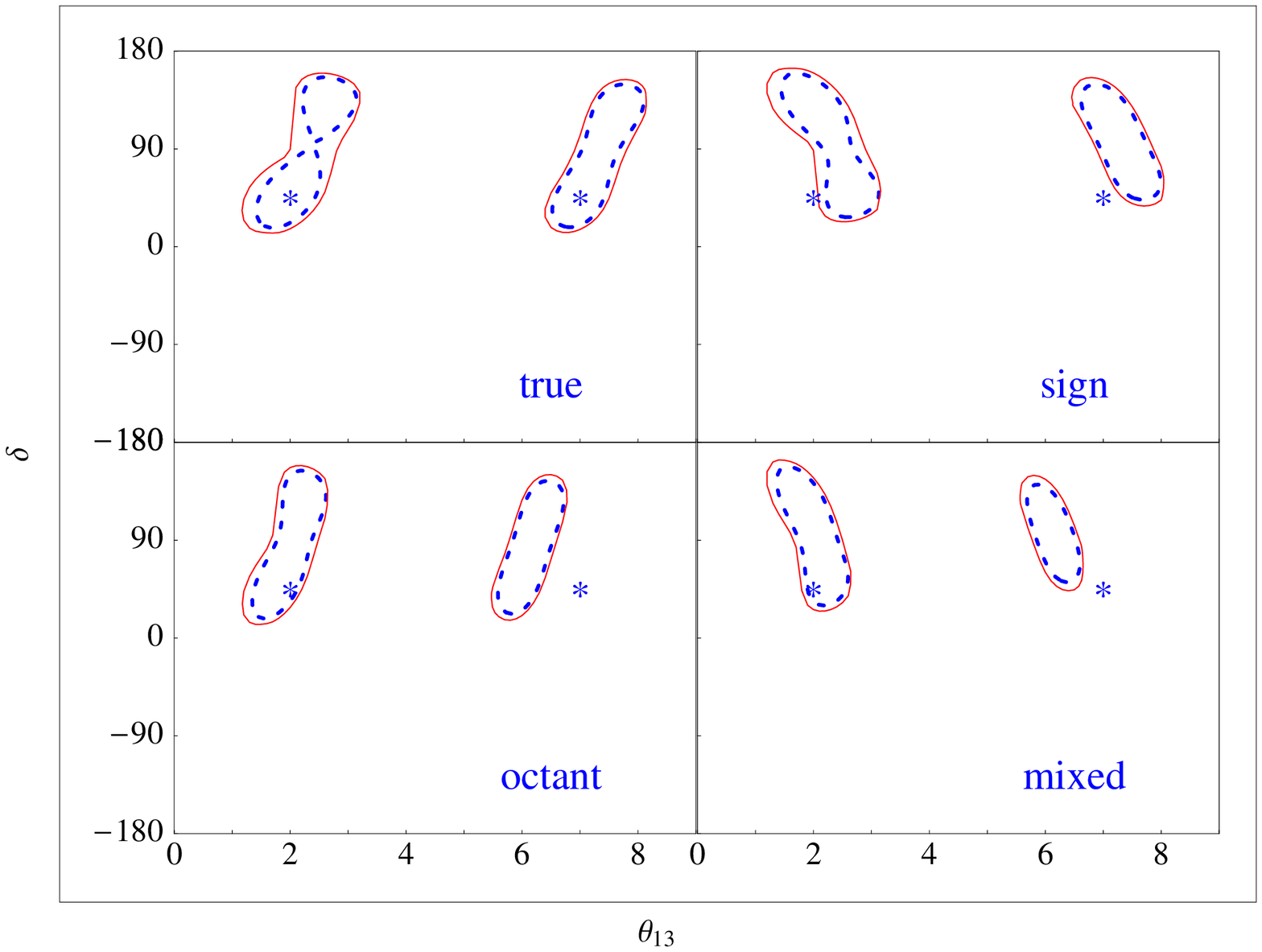}\\
\end{tabular}
\caption{\it Comparison of the projection of three-parameters 90\% CL contours onto the ($\theta_{13}$, $\delta$) plane 
(solid lines) with the corresponding two-parameters 90 \% CL contours (dashed lines) after a 10 years run at the $\beta$-Beam. 
Different choices of $s_{atm},s_{oct}$ are plotted separately.
The input parameters (represented by a star) are: $\bar \theta_{13} = 2^\circ,7^\circ$, $\bar \delta = 45^\circ$.
Left panel: $x = \theta_{12}$; right panel: $x = \Delta m^2_{12}$.}
\label{fig:solar}
\end{center}
\end{figure}

The projection of the three-parameters 90\% CL contours onto the ($\theta_{13},\delta$) plane are
presented in Fig.~\ref{fig:solar}. In the left panel we have fixed $\Delta m^2_{12} = 8.2 \times 10^{-5}$ eV$^2$ 
and drawn the projection of the three-dimensional contours for $\chi^2 (\theta_{13},\delta,\theta_{12})$, for 
$\theta_{12} \in [29^\circ,36^\circ]$. In the right panel we have fixed $\theta_{12}= 32^\circ$ and drawn 
the projection of the three-dimensional contours for $\chi^2 (\theta_{13},\delta,\Delta m^2_{12})$, for 
$\Delta m^2_{12} \in [7.5,9.1] \times 10^{-5}$ eV$^2$. In both cases, the atmospheric parameters have been fixed to 
$\Delta m^2_{23}= 2.5 \times 10^{-3}$ eV$^2$ and $\theta_{23}=40^\circ$.
The input values for the two unknowns are $\bar \theta_{13}= 2^\circ,7^\circ$ and $\bar \delta = 45^\circ$.
For each panel, the results for the four different choices of the two discrete variables, $s_{atm}$ and 
$s_{oct}$, are presented separately. Finally, the projection of the three-parameters 90\% CL contours (solid lines)
are directly compared with the two-parameters 90\% CL contours (dashed lines) obtained fixing the solar parameters 
to their present best fit values, $\theta_{12}=32^\circ, \Delta m^2_{12}=8.2 \times 10^{-5}$ eV$^2$.

As we can see in both panels, most of the plotted three-parameters contours coincide for any practical purpose 
with the corresponding two-parameters ones, with small deviations easily explained by the different CL  
in two- and three-parameters $\chi^2$. As a result, we claim that the impact of solar parameter uncertainties 
on the measurement of $\theta_{13}$ and $\delta$ is negligible for $\bar \theta_{13} \geq 2^\circ$.
This is indeed a consequence of the subleading dependence of the $\nu_e \to \nu_\mu$ oscillation probability
on the solar parameters (see, for example, Refs.~\cite{Cervera:2000kp,Akhmedov:2004ny}) for large 
values of $\bar \theta_{13}$. When $\bar \theta_{13}$ is large, we are in what has been called the 
``atmospheric regime'' in Ref.~\cite{Burguet-Castell:2001ez}; only for $\bar \theta_{13}$ below the verge
of the $\beta$-Beam $\theta_{13}$-sensitivity, i.e. for $\bar \theta_{13} > 2^\circ$, we enter in the 
so-called ``solar regime''. Clearly, no signal is expected at the $\beta$-Beam in this case: we can thus 
safely claim that the solar parameter uncertainties do not affect significantly the measurement of 
$\theta_{13}$ and $\delta$ at the considered facility. 

Similar conclusions can be drawn for the SPL Super-Beam and for different values of $\bar \delta$ and will therefore 
not be repeated here. For the rest of the paper, the solar parameters will be considered as fixed external inputs: 
$\theta_{12} = 32^\circ$ and $\Delta m^2_{12} = 8.2 \times 10^{-5}$ eV$^2$.

\subsection{The atmospheric sector at the $\beta$-Beam}
\label{sec:atmoBB}

As for the solar sector, we study the effect of present uncertainties on the atmospheric sector parameters in the measurement 
of $\theta_{13}$ and $\delta$ performing two distinct three-parameters fit in $\theta_{13},\delta$ and 
$\theta_{23}$ (for fixed $\Delta m^2_{23}$) or $\Delta m^2_{23}$ (for fixed $\theta_{23}$). 

\begin{figure}[t!]
\begin{center}
\begin{tabular}{cc}
\hspace{-1.0cm}
\epsfxsize8cm\epsffile{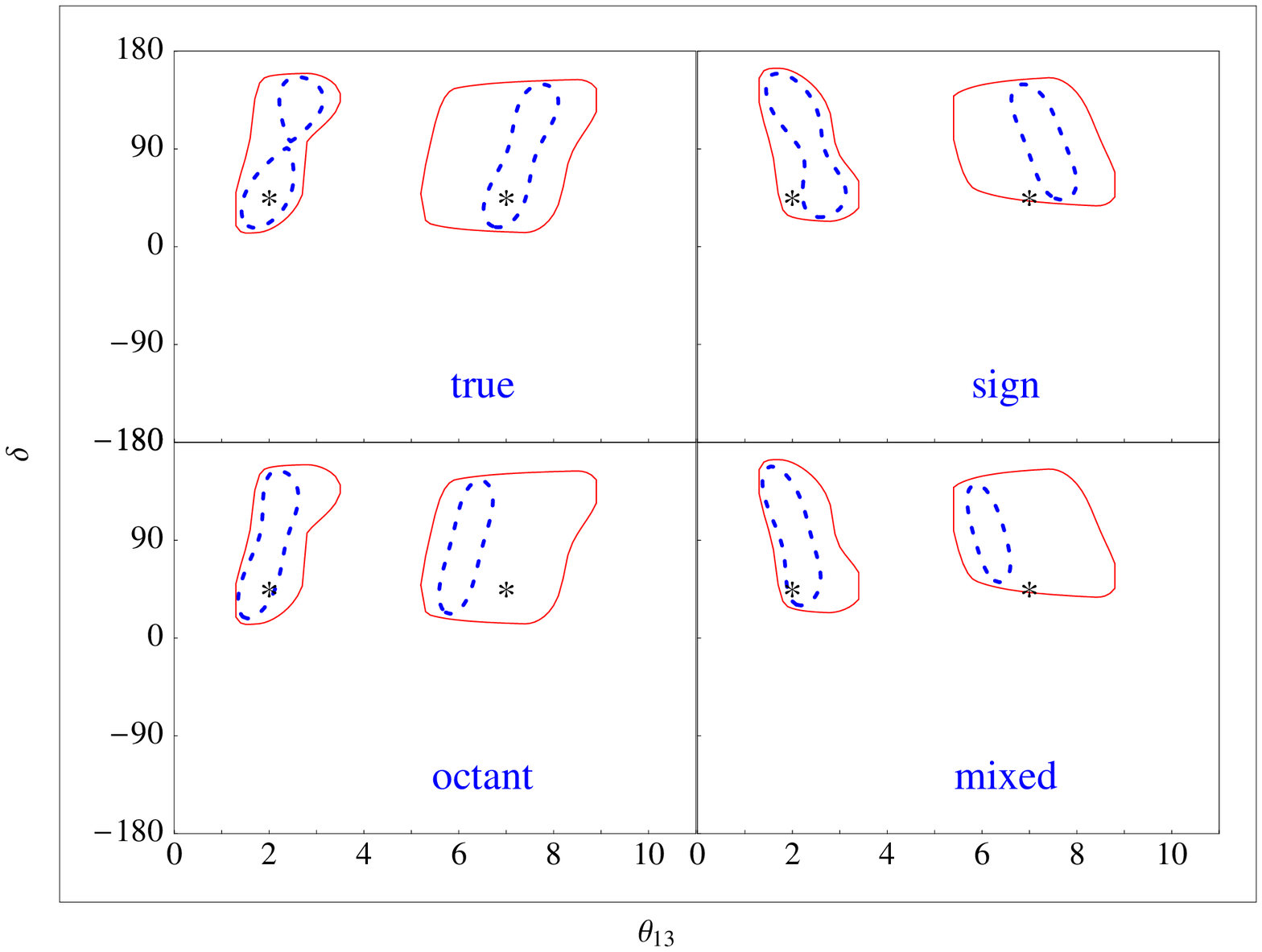} &
\epsfxsize8cm\epsffile{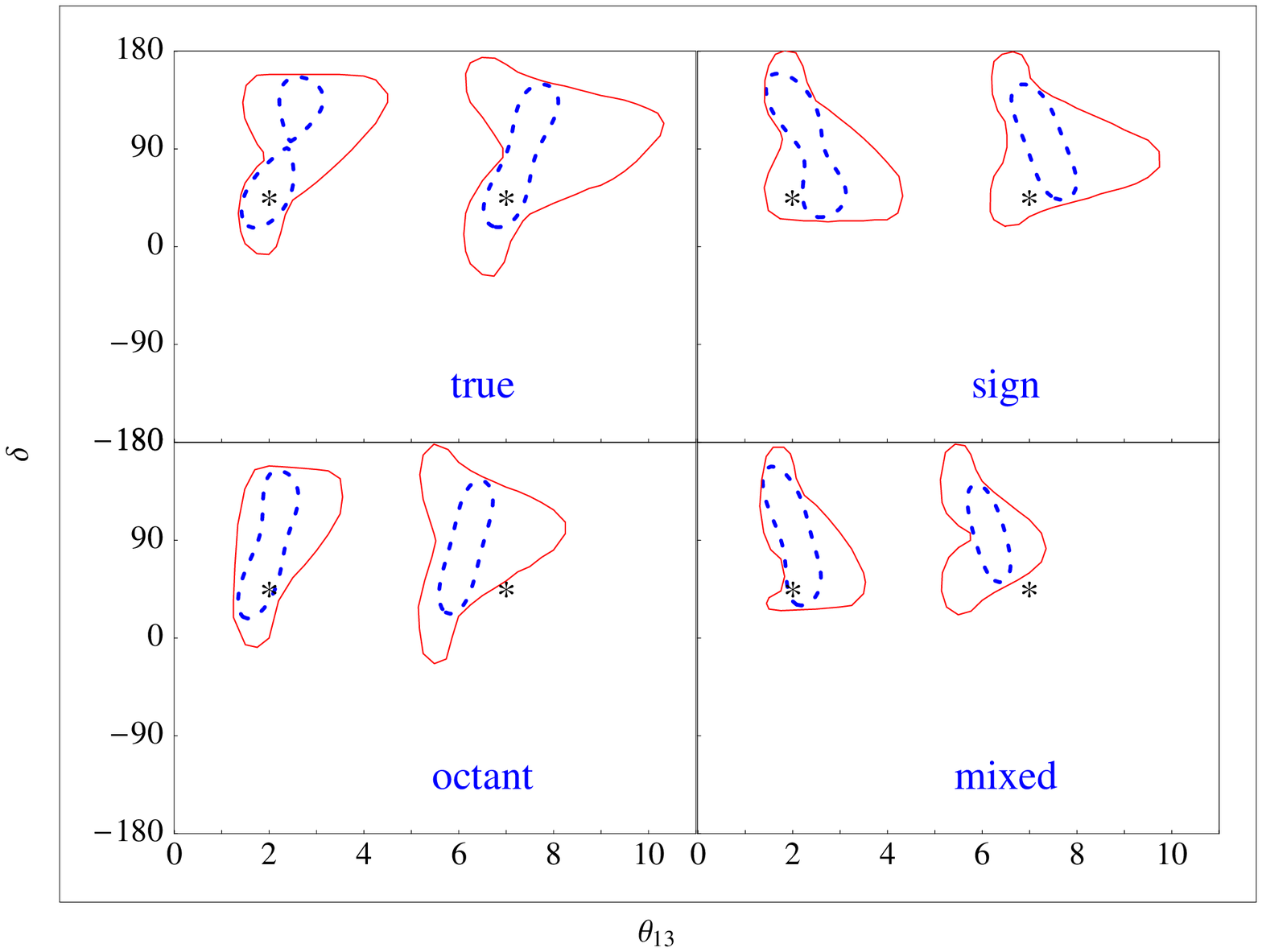}\\
\end{tabular}
\caption{\it Comparison of the projection of three-parameters 90\% CL contours onto the ($\theta_{13}$, $\delta$) plane 
(solid lines) with the corresponding two-parameters 90 \% CL contours (dashed lines) after a 10 years run at the $\beta$-Beam. 
Different choices of $s_{atm},s_{oct}$ are plotted separately.
The input parameters (represented by a star) are: $\bar \theta_{13} = 2^\circ,7^\circ$, $\bar \delta = 45^\circ$.
Left panel: $x = \theta_{23}$; right panel: $x = \Delta m^2_{23}$.}
\label{fig:atmo}
\end{center}
\end{figure}

The comparison between two- and three-parameters fits is presented in Fig.~\ref{fig:atmo}, where
the projection of the three-parameters 90\% CL contours onto the ($\theta_{13},\delta$) plane (solid lines)
and the two-parameters 90\% CL contours (dashed lines) have been plotted separately for each possible choice of the two 
discrete variables, $s_{atm}$ and $s_{oct}$. In the left panel we have fixed $\Delta m^2_{atm} = 2.5 \times 10^{-3}$ eV$^2$ 
and drawn the projection of the three-dimensional contours for $\chi^2 (\theta_{13},\delta,\theta_{23})$, for 
$\theta_{23} \in [36^\circ,55^\circ]$ (see Tab.~\ref{tab:input}). 
In the right panel we have fixed $\theta_{23}= 40^\circ$ and drawn the projection of the three-dimensional contours 
for $\chi^2 (\theta_{13},\delta,\Delta m^2_{23})$, for $\Delta m^2_{23} \in [1.7,3.5] \times 10^{-3}$ eV$^2$
(see Tab.~\ref{tab:input}). 
The input values for the two unknowns are $\bar \theta_{13}= 2^\circ,7^\circ$ and $\bar \delta = 45^\circ$.
The two-parameters contours have been drawn using fixed values for the atmospheric parameters, 
$\theta_{23}=40^\circ, \Delta m^2_{atm}=2.5 \times 10^{-3}$ eV$^2$. We have checked that no significant improvement is observed 
when the systematic error in the appearance channel is set to 2\%.

In Fig.~\ref{fig:atmo} it is manifest the impact of atmospheric parameter uncertainties on the measurement of $\theta_{13}$
and $\delta$, for both $\theta_{23}$ and $\Delta m^2_{23}$ fits. Both unknowns are measured with errors much larger than 
those expected from two-parameters contours. This must be compared with the results of the previous section, where it has been 
shown that solar parameter uncertainties have a negligible impact.

Consider first the left panel of Fig.~\ref{fig:atmo}: the results from a three-parameters fit in $(\theta_{13},\delta,\theta_{23})$.
Notice that, being $\theta_{23}$ a fitting variable in the whole range $\theta_{23} \in [36^\circ,55^\circ]$, 
the effect of the ``octant ambiguity'' is automatically taken into account by
the three-parameters $\chi^2$ function. 
For this reason the three-parameters contours labelled as ``true'' and ``octant'' are identical.
Notice also that this is not the case for the two-parameters contours, where the 
choice of $s_{oct}$ is reflected in contours located at different values of $\theta_{13}$ with respect to the
input $\bar \theta_{13}$.
For $\bar \theta_{13} = 7^\circ$ we can see that a large error in $\theta_{13}$ 
is induced by the uncertainty in $\theta_{23}$ (with $\Delta \theta_{13}$ as large as $4^\circ$). 
This is a consequence of the fact that the leading term in the $\nu_e \to \nu_\mu$ oscillation probability is proportional 
to the combination $\sin^2 (2 \theta_{13}) \sin^2 \theta_{23}$: 
to compensate a change in $\theta_{23}$, a change in $\theta_{13}$ is needed. For smaller values of $\bar \theta_{13}$
this effect is much smaller. The spread in $\delta$ is, on the other hand, extremely similar in two- and three-parameters 
contours. This is a consequence of the fact that a $\delta$-dependence would be induced in the fit through 
the subleading term in the oscillation probability, that is proportional to $\sin (2 \theta_{23})$ and thus less sensitive 
to changes in $\theta_{23}$ in the almost symmetric interval considered. Notice that three-parameters contours 
have a box-like shape, with no strong $\theta_{13}-\delta$ correlation. Finally, the largest values of $\theta_{13}$ 
are observed for both choices of $s_{atm}$ at lower values of $\theta_{23}$, 
whereas the smallest values of $\theta_{13}$ are reached for larger values of $\theta_{23}$. 

Consider now the right panel of Fig.~\ref{fig:atmo}: the results from a three-parameters fit in $(\theta_{13},\delta,\Delta m^2_{23})$. 
In this case to different choices of $s_{oct}$ correspond different contours ($\theta_{23}$ is a fixed external input and not 
a free parameter in the fit). As in the previous case, a large error in $\theta_{13}$ is induced by the error on the 
atmospheric parameter, especially for $\bar \theta_{13} = 7^\circ$. The largest value of $\theta_{13}$ is 
associated to the smallest value of $\Delta m^2_{23}$, in all plots. A characteristic feature of this fit is
the significant $\delta$-dependence that can be observed in all plots and that was not present in 
the fits in $\theta_{23}$. The three-parameters 90\% CL contours have a triangular shape
(the error in $\delta$ reduces for large values of $\theta_{13}$), pointing to a strong $\theta_{13}-\delta$ correlation.
It is important to stress that, when $\Delta m^2_{23}$ is the free parameter, the overall error in $\delta$ is significantly
larger in the three- than in the two-parameters contours. For both values of $\bar \theta_{13} = 2^\circ,7^\circ$ roughly 
half of the $\delta$-parameter space is covered. This result, considerably worse than what expected from two-parameters fit, 
can be compared with the CP-coverage expectation (explained in App.~A) for this particular input pair, 
Fig.~\ref{fig:coverage}. For both methods, a rather large error in $\delta$ is indeed expected.

It is clear from Fig.~\ref{fig:atmo} that both atmospheric parameter uncertainties are extremely important
in the measurement of $\theta_{13}$ and $\delta$: the three-parameters 90\% CL allowed regions are considerably
worse than those obtained with two-parameters fits. In particular, for the shown input pairs, $\delta$ would
remain completely unknown in the interval $\delta \in [0,\pi]$ and $\theta_{13}$ would be known only with a large
error.

As an example of how the situation can be improved when using reduced uncertainties on the atmospheric parameters,
in Fig.~\ref{fig:atmo2} we present the projection of the three-dimensional 90\% CL contours
onto the ($\theta_{13},\delta$) plane using the expected uncertainties on the atmospheric parameters after T2K-I
(last column of Tab.~\ref{tab:input}): $\theta_{23} \in [38^\circ,43^\circ]-[48^\circ,52^\circ]$ and 
$\Delta m^2_{23} \in [2.42,2.61] \times 10^{-3}$ eV$^2$ for $s_{atm} = +$ and 
$\Delta m^2_{23} \in [2.46,2.64] \times 10^{-3}$ eV$^2$ for $s_{atm} = -$, \cite{Enrique}. 
The octant-ambiguity (that will not be solved at T2K-I) recover its discrete nature: 
separate regions of the parameter space will be spanned by different choices of $s_{oct}$.
In this case, all choices of the two discrete variables $s_{atm}$ and $s_{oct}$
are presented together and no comparison with two-parameters contours is shown. 
In top panels $x = \theta_{23}$; in bottom panels $x=\Delta m^2_{23}$.
The results of the three-parameters fit with expected uncertainties (right panels), 
are directly compared with the results presented in Fig.~\ref{fig:atmo} computed with the present uncertainties
(left panels). The reduction of the uncertainties on the atmospheric parameters has indeed an important effect 
on the measurement of $\theta_{13}$ and $\delta$. As it can be seen in the right panels of Fig.~\ref{fig:atmo2}
a significant reduction of the $\theta_{13}$-spread is achieved, with plots resembling those obtained with 
standard two-parameters contours and fixed external atmospheric parameters (see Refs.~\cite{Donini:2004iv,Donini:2004hu}). 
The $\delta$-spread is also reduced considerably with respect to the results obtained with 
present uncertainties. These comments apply to both $\bar \theta_{13} = 2^\circ, 7^\circ$.
Notice that, as expected being $\theta_{23}$ restricted to one octant only, the octant- and mixed- ambiguities 
show themselves as separate contours in the ($\theta_{13},\delta$) plane, as for two-parameters fits.

A final comment on the impact of the uncertainties on the atmospheric parameters on the measurement of $\theta_{13}$ 
and $\delta$ at the low-gamma $\beta$-Beam is in order. We have shown that, with present uncertainties, the measurement 
of the two unknowns in the PMNS mixing matrix is severely spoiled. Errors as large as $\Delta \theta_{13} \simeq 4^\circ$
are found, and half of the parameter space in $\delta$ is spanned for different values of $\bar \delta$.
This corresponds to a CP-coverage $\xi \simeq 0.5$, a value that spoils completely the possibility to distinguish 
a CP-violating signal from a CP-conserving one at the considered facility (see App.~A).
A significant reduction in the uncertainties on the atmospheric parameters is mandatory if we plan to use such 
a facility to look for $\delta$. If $\theta_{23}$ and $\Delta m^2_{23}$ can be measured at 
the T2K-I experiment with the expected precision and for any value of $\bar \theta_{23}$, 
only then the results of present two-parameters studies \cite{Bouchez:2003fy,Donini:2004hu,Donini:2004iv,Burguet-Castell:2003vv} 
for facilities of this kind can be considered reliable.

In App.~B we present the results for different choices of $\bar \delta$, to illustrate the generality
of the results above. 

\subsection{The atmospheric sector at the Super-Beam}
\label{sec:atmoSB}

We now repeat the analysis of the impact of present and expected uncertainties on the atmospheric sector parameters 
in the measurement of $\theta_{13}$ and $\delta$ at a different facility: the SPL Super-Beam.
Again, two distinct three-parameters fits in $\theta_{13},\delta$ and 
$\theta_{23}$ (for fixed $\Delta m^2_{23}$) or $\Delta m^2_{23}$ (for fixed $\theta_{23}$) have been 
performed, with the Super-Beam running 2 years with $\pi^+$ and 8 years with $\pi^-$ to accumulate comparable 
statistics for neutrinos and antineutrinos. A significant difference between this facility and the low-gamma $\beta$-Beam 
considered previously is in that the $\nu_\mu$ disappearance channel at the Super-Beam reduces the uncertainties on the atmospheric parameters, 
as it can be seen in Fig.~\ref{fig:sbdis} (whereas the $\nu_e$ disappearance channel is useless to this purpose, 
see Ref.~\cite{Donini:2004iv}). In this case we therefore do not present results using ``present'' and ``expected'' 
uncertainties, but we just combine the results from the appearance and disappearance channel. 
We have checked that using a ``pessimistic'' 5\% systematic error in the disappearance channel 
does not change significantly our results.

\begin{figure}[t!]
\begin{center}
\begin{tabular}{cc}
\hspace{-1.0cm}
\epsfxsize8cm\epsffile{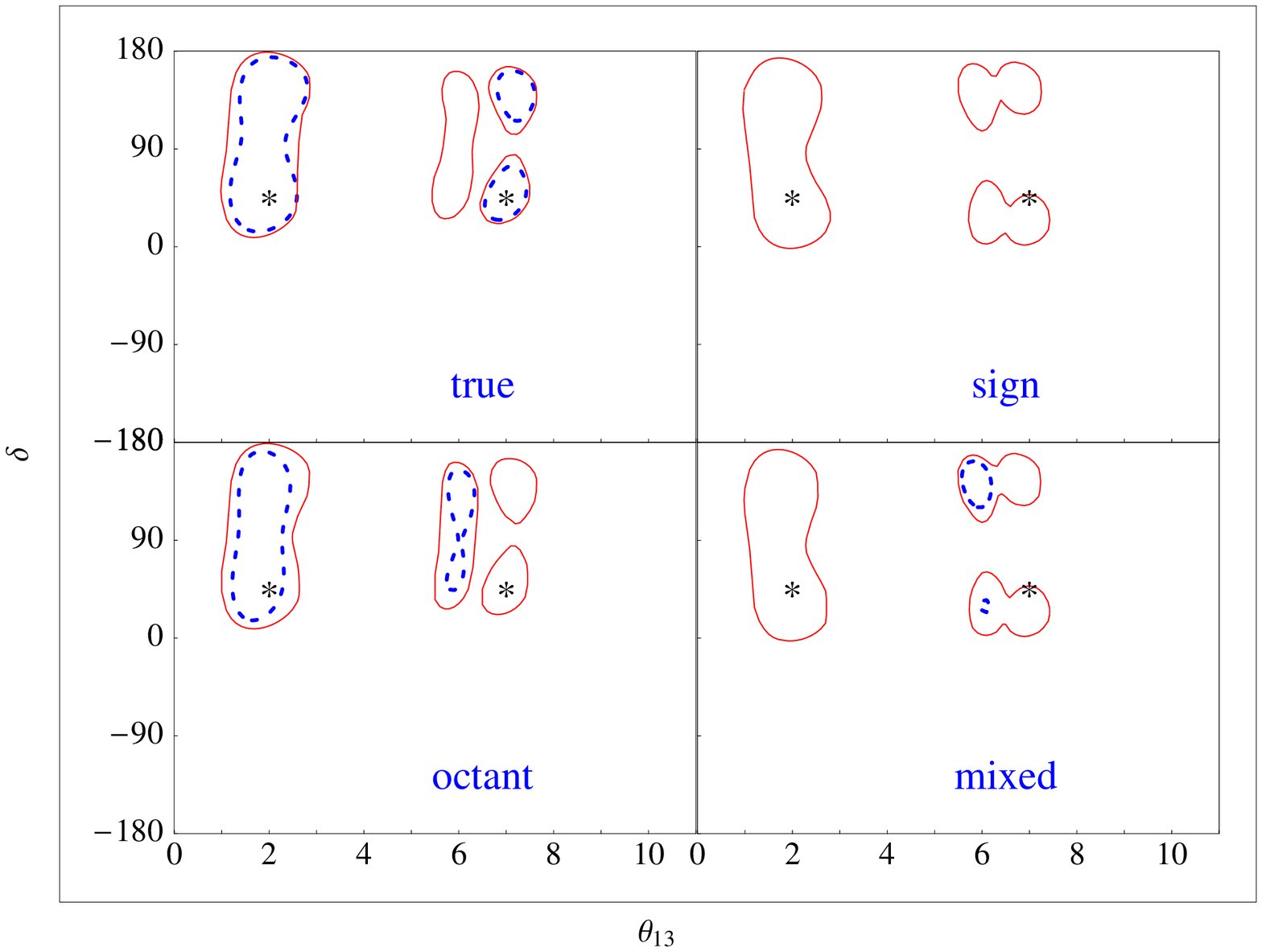} &
\epsfxsize8cm\epsffile{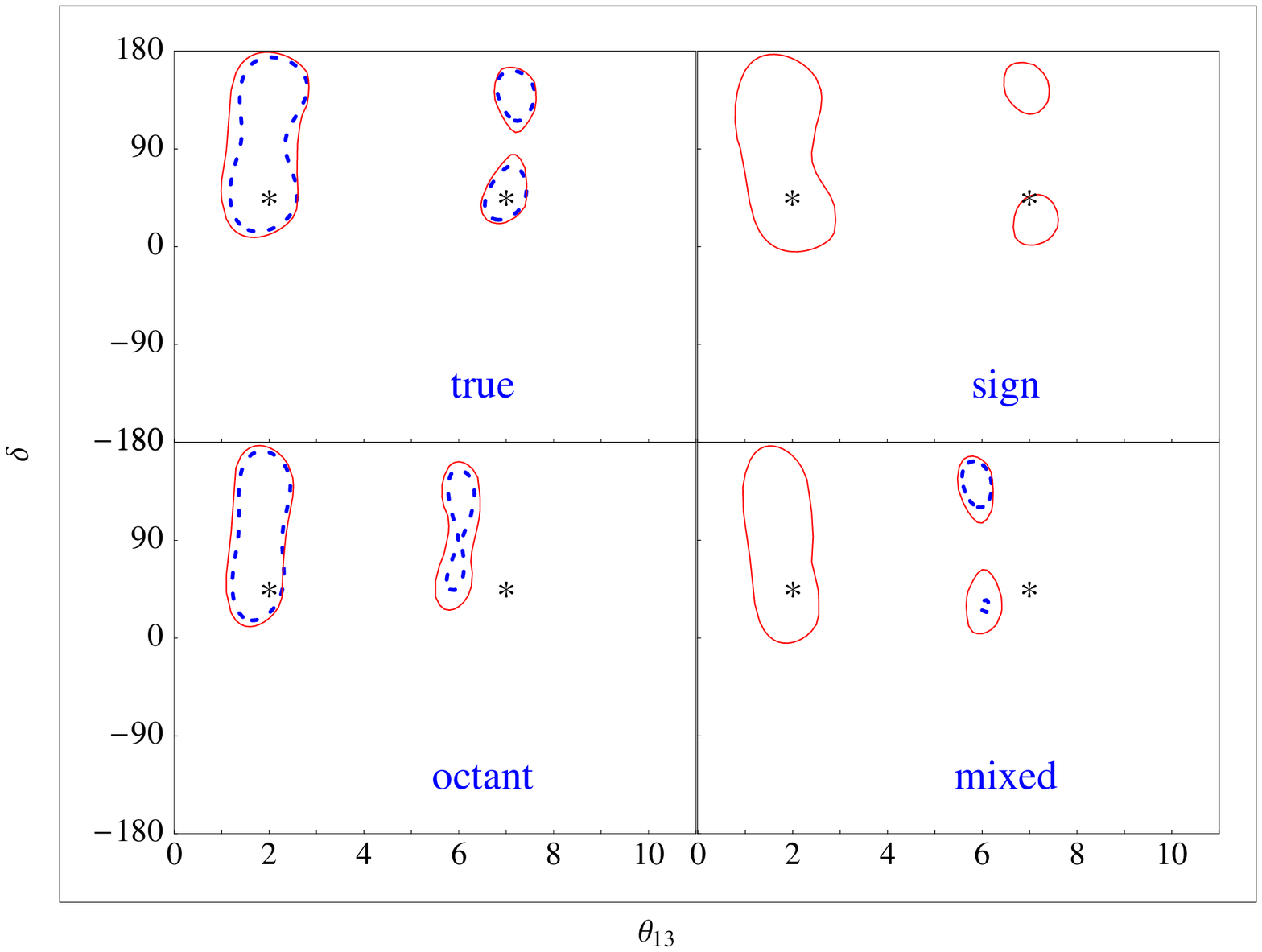}\\
\end{tabular}
\caption{\it Comparison of the projection of three-parameters 90\% CL contours onto the ($\theta_{13}$, $\delta$) plane 
(solid lines) with the corresponding two-parameters 90 \% CL contours (dashed lines) after a 2+8 years run at the Super-Beam. 
Different choices of $s_{atm},s_{oct}$ are plotted separately.
The input parameters (represented by a star) are: $\bar \theta_{13} = 2^\circ,7^\circ$, $\bar \delta = 45^\circ$.
Left panel: $x = \theta_{23}$; right panel: $x = \Delta m^2_{23}$.}
\label{fig:atmoSB}
\end{center}
\end{figure}

The comparison between two- and three-parameters fits is presented in Fig.~\ref{fig:atmoSB}.
In the left panel we have fixed $\Delta m^2_{atm} = 2.5 \times 10^{-3}$ eV$^2$ 
and drawn the projection of the three-dimensional contours for $\chi^2 (\theta_{13},\delta,\theta_{23})$, for 
$\theta_{23} \in [36^\circ,55^\circ]$ (see Tab.~\ref{tab:input}). 
In the right panel we have fixed $\theta_{23}= 40^\circ$ and drawn the projection of the three-dimensional contours 
for $\chi^2 (\theta_{13},\delta,\Delta m^2_{23})$, for $\Delta m^2_{23} \in [2.3,2.9] \times 10^{-3}$ eV$^2$
(see Tab.~\ref{tab:input} and Sect.~\ref{sec:disSB}). 
The input values for the two unknowns are $\bar \theta_{13}= 2^\circ,7^\circ$ and $\bar \delta = 45^\circ$.

The main difference between two- and three-parameters contours resides in that in the latter we observe some clones 
absent in the two-parameters plots. This is a consequence of the not satisfactory expected improvement on the error 
in $\theta_{23}$ and $\Delta m^2_{23}$ for $\theta_{23} = 40^\circ$. 

Also in this case the measurement of $\theta_{13}$ and $\delta$ is severely affected by the uncertainties on the atmospheric parameters.
Somewhat smaller errors are found in $\theta_{13}$ and $\delta$ with respect to the $\beta$-Beam case, but still almost half of the 
parameter space in $\delta$ is spanned for different values of $\bar \delta$. A crucial point is that it does not seem that the $\nu_\mu$ 
disappearance channel is capable of a significant reduction in the error on the atmospheric mixing angle $\theta_{23}$. 
The T2K-I experiment will therefore be crucial, if indeed the expected precision in the atmospheric angle can be met for any value 
of $\theta_{23}$.

In App.~B we present the results for different choices of $\bar \delta$, to illustrate the generality
of the results above. 

\subsection{The atmospheric sector at the Neutrino Factory}
\label{sec:atmoNF}

We repeat the analysis of the impact of atmospheric parameters uncertainties in the measurement of $\theta_{13}$ and $\delta$ 
at a third facility: the CERN-based SPL-fuelled 50 GeV Neutrino Factory.
We want to show in this way how the results of Sects.~\ref{sec:atmoBB} and \ref{sec:atmoSB} are quite general and must
be taken into account at any facility that is considered when looking for $\theta_{13}$ and $\delta$.

As before, two distinct three-parameters fit in $\theta_{13},\delta$ and 
$\theta_{23}$ (for fixed $\Delta m^2_{23}$) or $\Delta m^2_{23}$ (for fixed $\theta_{23}$) have been 
performed, with the Neutrino Factory running 5 years with $\mu^+$ and 5 years with $\mu^-$. 

In the absence of an updated analysis of the expected reduction of atmospheric parameters uncertainties at this facility through 
$\nu_\mu \to \nu_\mu, \nu_e \to \nu_e$ and $\nu_\mu \to \nu_\tau$ (see Ref.~\cite{Barger:1999fs,Bueno:2000fg} for old analyses), 
we only present results combining the two appearance channels $\nu_e \to \nu_\mu$ (i.e. the ``golden'' channel) and $\nu_e \to \nu_\tau$
(i.e. the ``silver'' channel) for both polarities. We use the expected uncertainties after T2K-I, in order to get a preliminar understanding of the 
impact of atmospheric parameter uncertainties at this facility.

\begin{figure}[t!]
\begin{center}
\begin{tabular}{cc}
\hspace{-1.0cm}
\epsfxsize5cm\epsffile{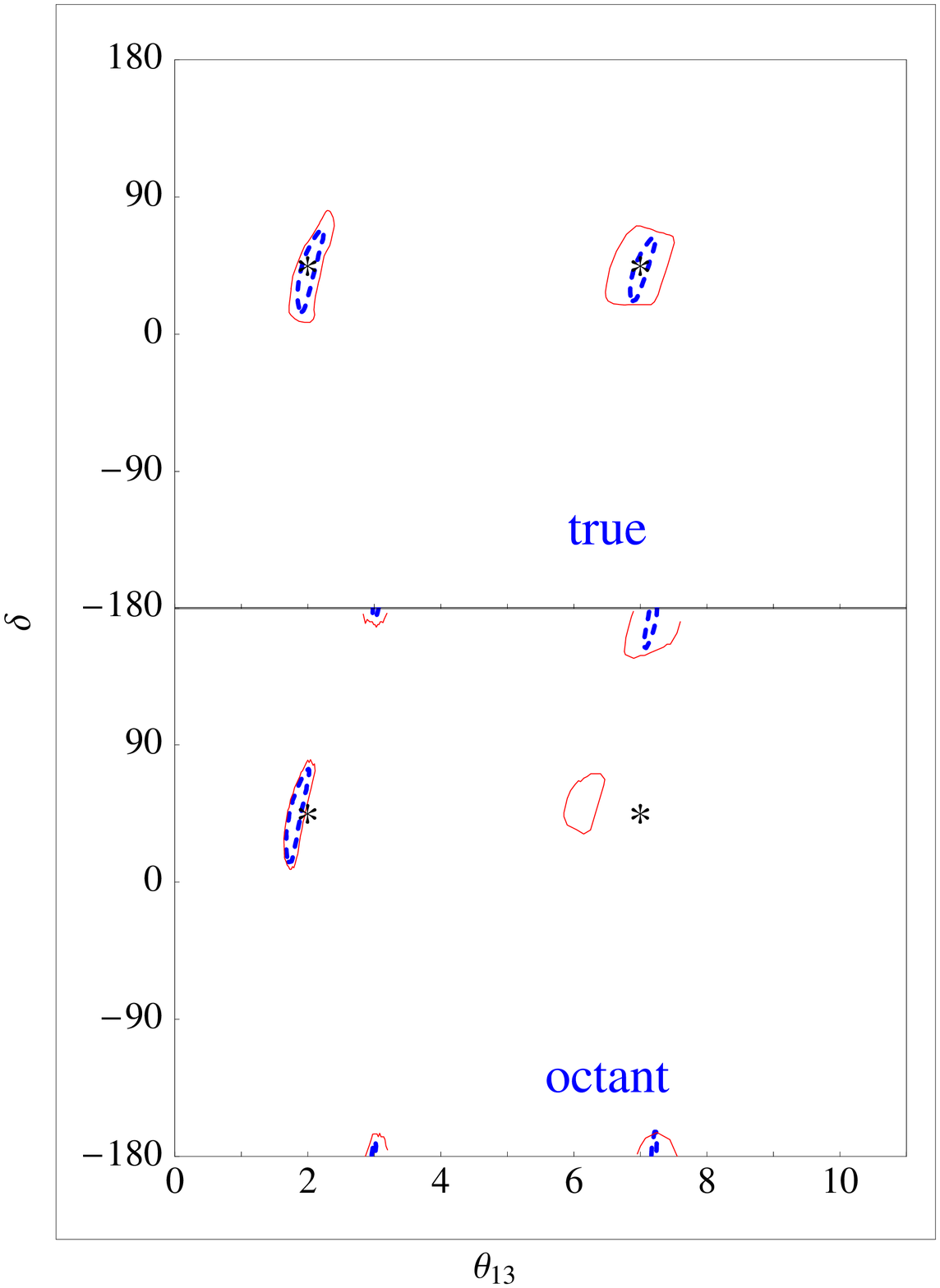} &
\epsfxsize5cm\epsffile{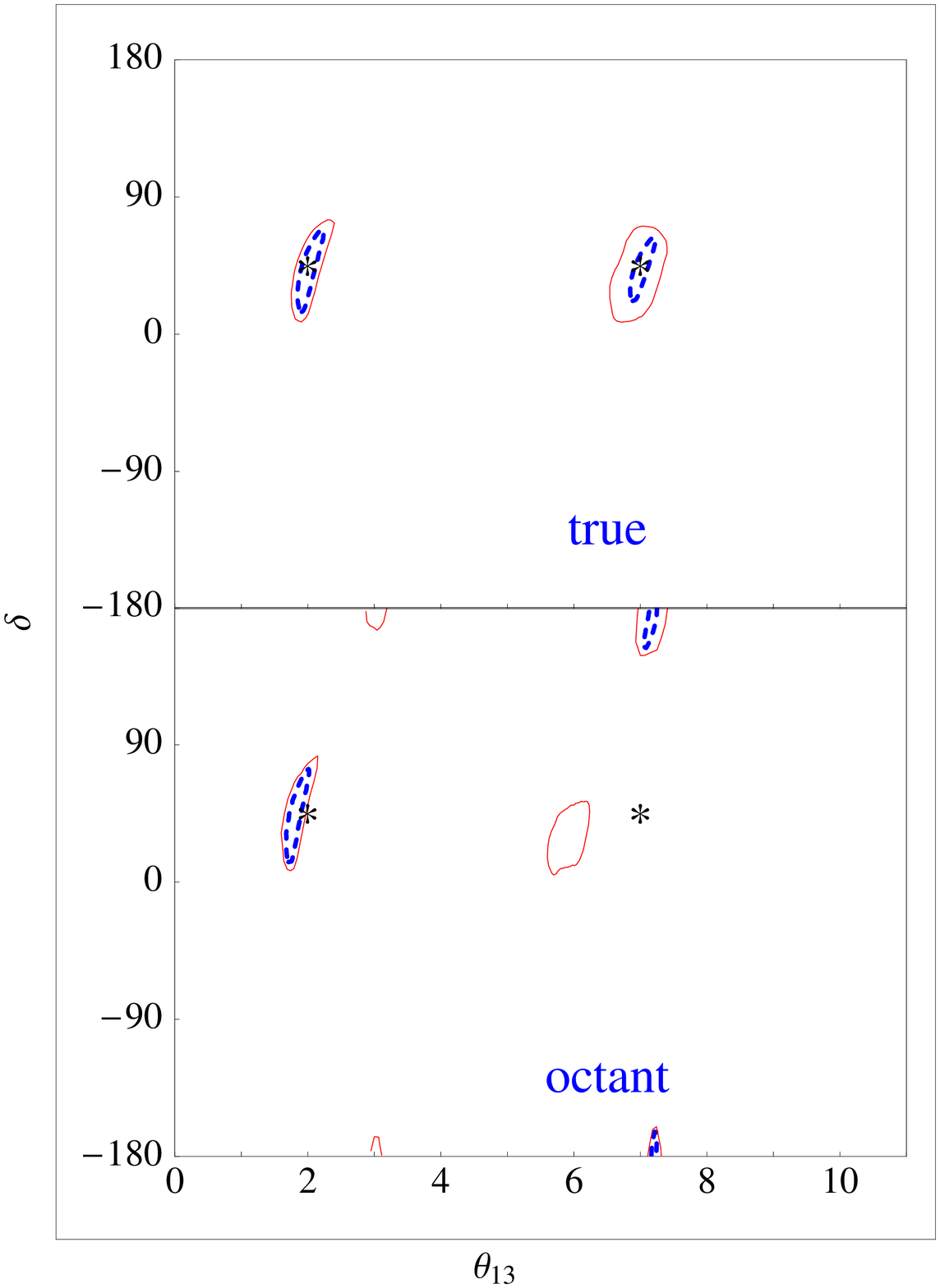}\\
\end{tabular}
\caption{\it Comparison of the projection of three-parameters 90\% CL contours onto the ($\theta_{13}$, $\delta$) plane 
(solid lines) with the corresponding two-parameters 90 \% CL contours (dashed lines) after a 5+5 years run at the Neutrino Factory. 
Different choices of $s_{oct}$ are plotted separately.
The input parameters (represented by a star) are: $\bar \theta_{13} = 2^\circ,7^\circ$, $\bar \delta = 42^\circ$.
Left panel: $x = \theta_{23}$; right panel: $x = \Delta m^2_{23}$.}
\label{fig:atmoNF}
\end{center}
\end{figure}

The comparison between two- and three-parameters fits is presented in Fig.~\ref{fig:atmoNF}.
In the left panel we have fixed $\Delta m^2_{atm} = 2.5 \times 10^{-3}$ eV$^2$ and drawn the projection of the three-dimensional contours 
for $\chi^2 (\theta_{13},\delta,\theta_{23})$, for $\theta_{23} \in [38^\circ,43^\circ]-[48^\circ,52^\circ]$  (see Tab.~\ref{tab:input}). 
In the right panel we have fixed $\theta_{23}= 40^\circ$ and drawn the projection of the three-dimensional contours 
for $\chi^2 (\theta_{13},\delta,\Delta m^2_{23})$, for 
$\Delta m^2_{23} \in [2.4,2.7] \times 10^{-3}$ eV$^2$ 
(see Tab.~\ref{tab:input} and Sect.~\ref{sec:disSB}). 
The input values for the two unknowns are $\bar \theta_{13}= 2^\circ,7^\circ$ and $\bar \delta = 42^\circ$.

First of all notice that at the Neutrino Factory the sign and mixed degeneracies are solved, being the magnetized iron detector with a $L=3000$ km 
baseline extremely sensitive to matter effects and thus capable to measure $s_{atm}$. For this reason we only present two panels, 
corresponding to two possible choices of the $\theta_{23}$-octant, $s_{oct} = \pm \bar s_{oct}$. As for the SPL Super-Beam, for small $\bar \theta_{13}$ 
the two- and three-parameters contours practically coincide. On the other hand, for $\bar \theta_{13}$ large we must make a distinction between 
$s_{oct} = \bar s_{oct}$ and $s_{oct} = - \bar s_{oct}$: whereas the impact of the atmospheric uncertainties for $s_{oct} = \bar s_{oct}$ is marginal 
(something already observed in \cite{Burguet-Castell:2001ez}, where the covariance matrix approach was adopted and only the right choice of the 
 $\theta_{23}$-octant was considered), we notice how extra octant clones are
 present in the three-parameters contours 
that are absent in the two-parameters ones when the wrong choice of $s_{oct}$ is taken. This happens because in the three-dimensional parameter
space $\theta_{23}$ cooperates with $\theta_{13}$ to identify a low $\chi^2$ region with $\delta \simeq \bar \delta$ but with $\theta_{13} < \bar \theta_{13}$.

As for the other facilities, we have seen that the impact of the uncertainties on the atmospheric parameters on the measurement of $\theta_{13}$ and 
$\delta$ at the Neutrino Factory is relevant (albeit perhaps not as important as for the $\beta$-Beam and the Super-Beam previously discussed). 
Again, we stress that the loss in precision is more important for large $\bar \theta_{13}$ than for small $\bar \theta_{13}$, 
a region of the PMNS parameter space that will be selected or excluded by the approaching T2K-I experiment. 
This is indeed a crucial problem for precision measurements of the PMNS matrix elements.

In App.~B we present the results for different choices of $\bar \delta$, to illustrate the generality
of the results above. 

\section{CP-violation discovery potential}
\label{sec:sensitivity}

\begin{figure}[h!]
\begin{center}
\begin{tabular}{cc}
\hspace{-1.0cm} \epsfxsize8cm\epsffile{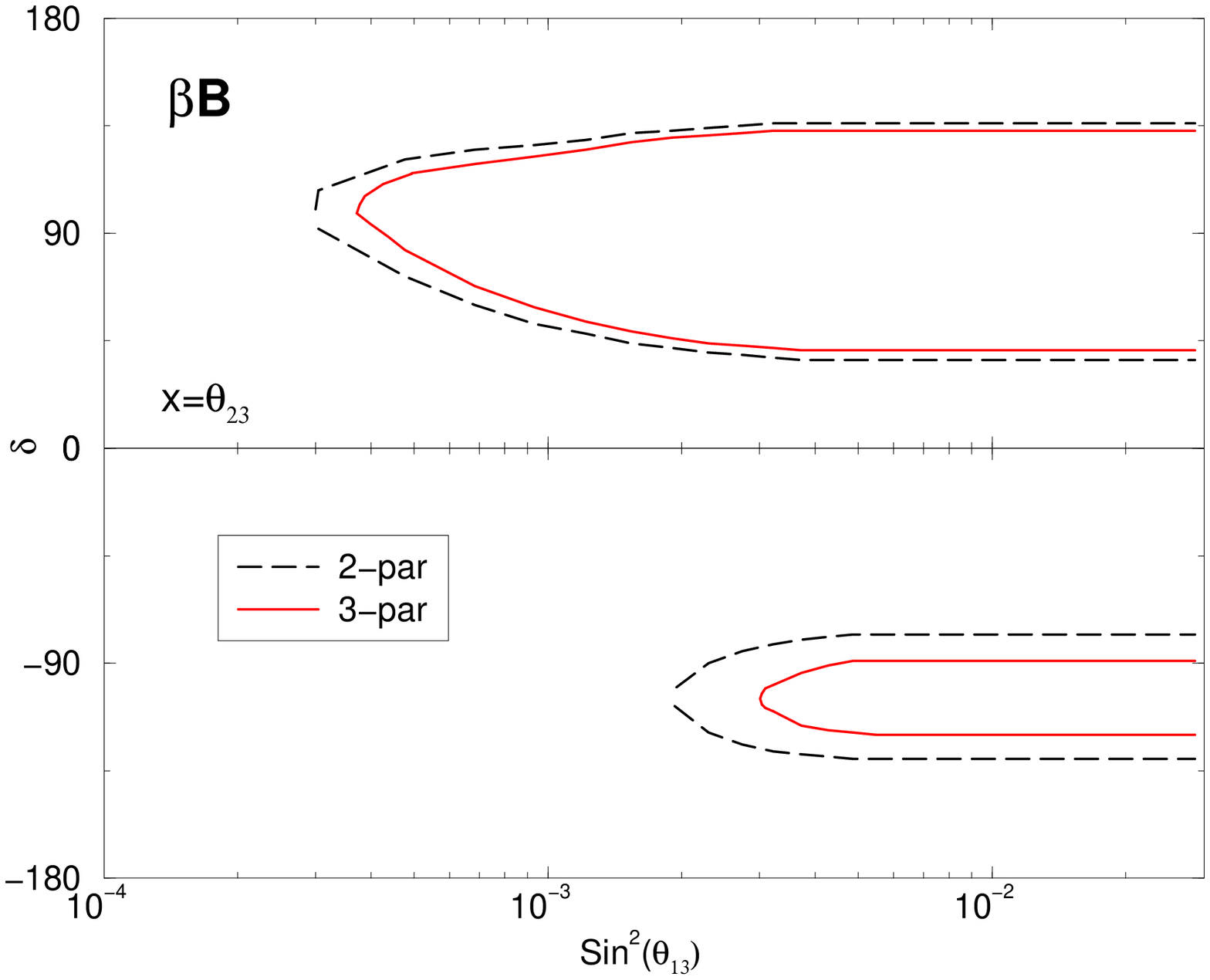} &
\hspace{-0.5cm} \epsfxsize8cm\epsffile{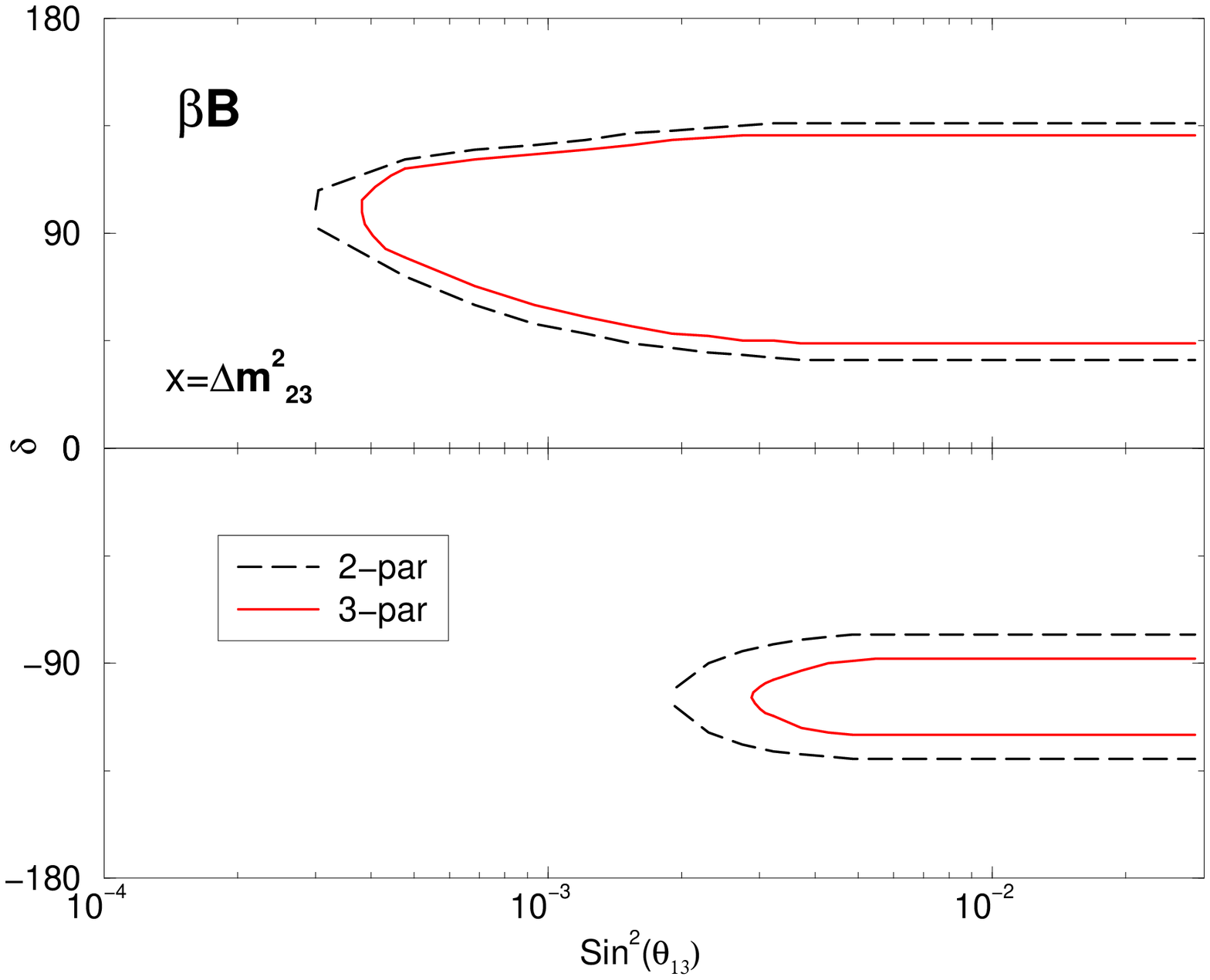} \\
\end{tabular}
\caption{\it CP-violation discovery potential after 10 years at the low-$\gamma$ $\beta$-Beam.}
\label{fig:sensitivityBB}
\end{center}
\end{figure}

\begin{figure}[ht!]
\begin{center}
\begin{tabular}{cc}
\hspace{-1.0cm} \epsfxsize8cm\epsffile{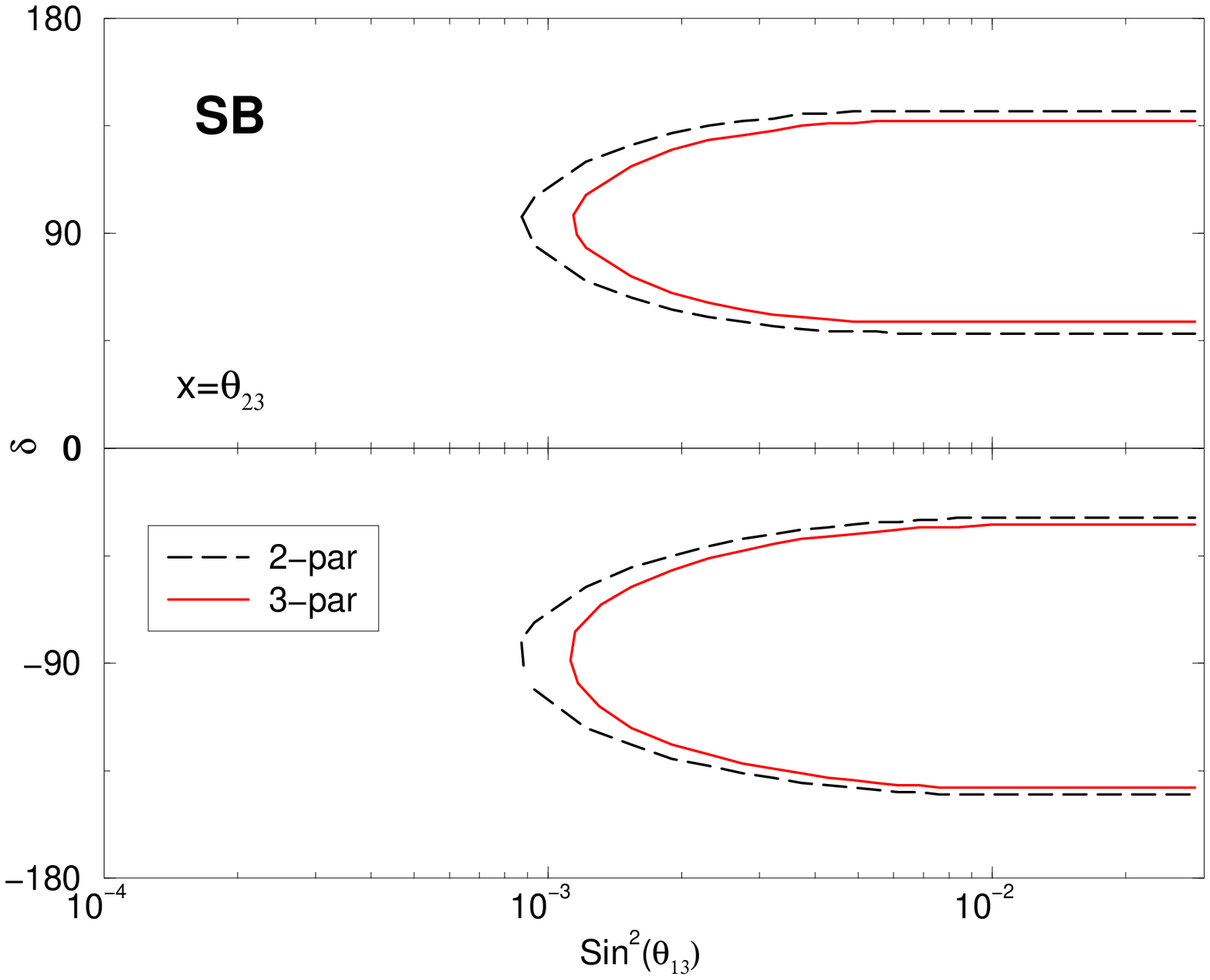} &
\hspace{-0.5cm} \epsfxsize8cm\epsffile{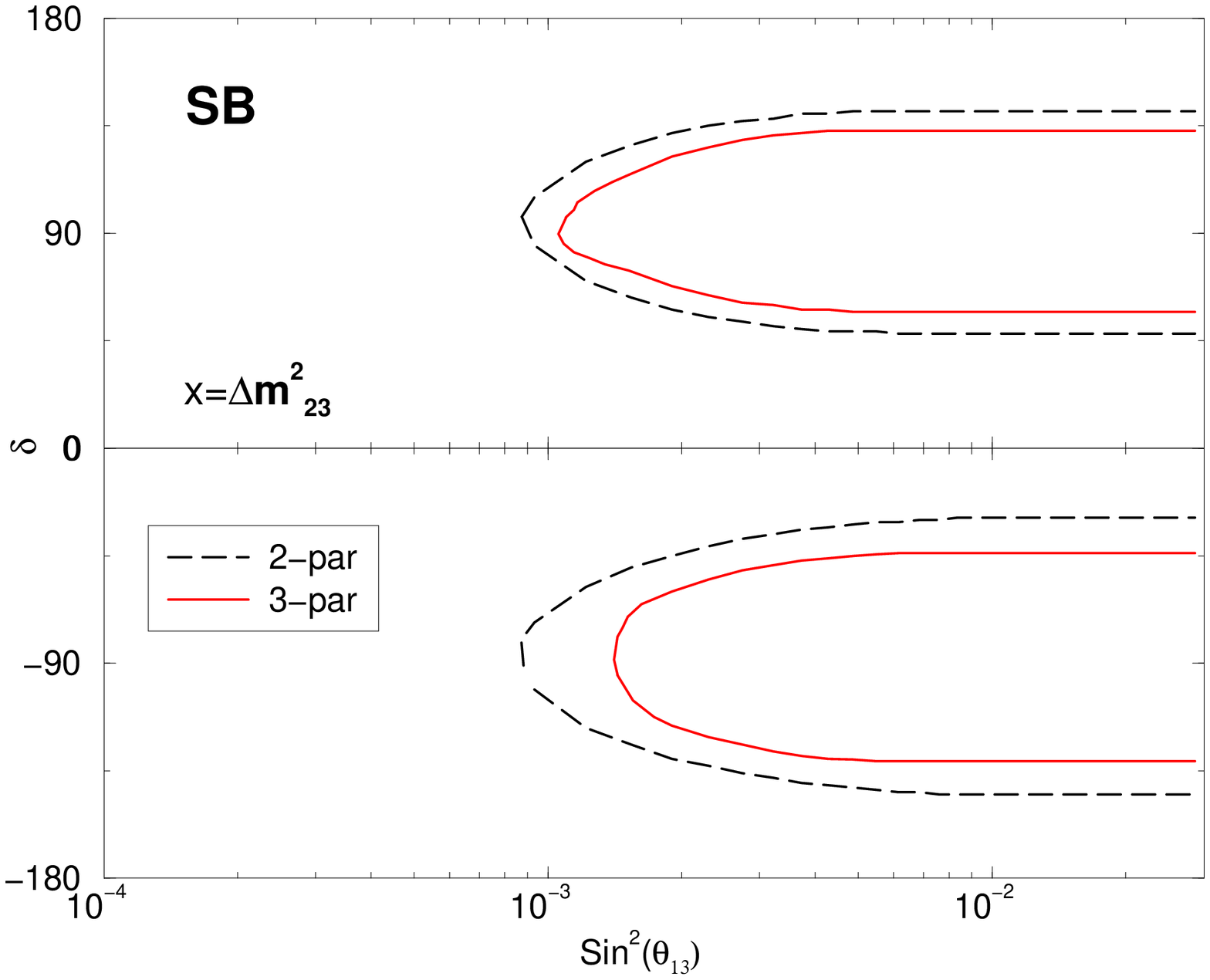} \\
\end{tabular}
\caption{\it CP-violation discovery potential after 2+8 years at the SPL Super-Beam.}
\label{fig:sensitivitySB}
\end{center}
\end{figure}

\begin{figure}[ht!]
\begin{center}
\begin{tabular}{cc}
\hspace{-1.0cm} \epsfxsize8cm\epsffile{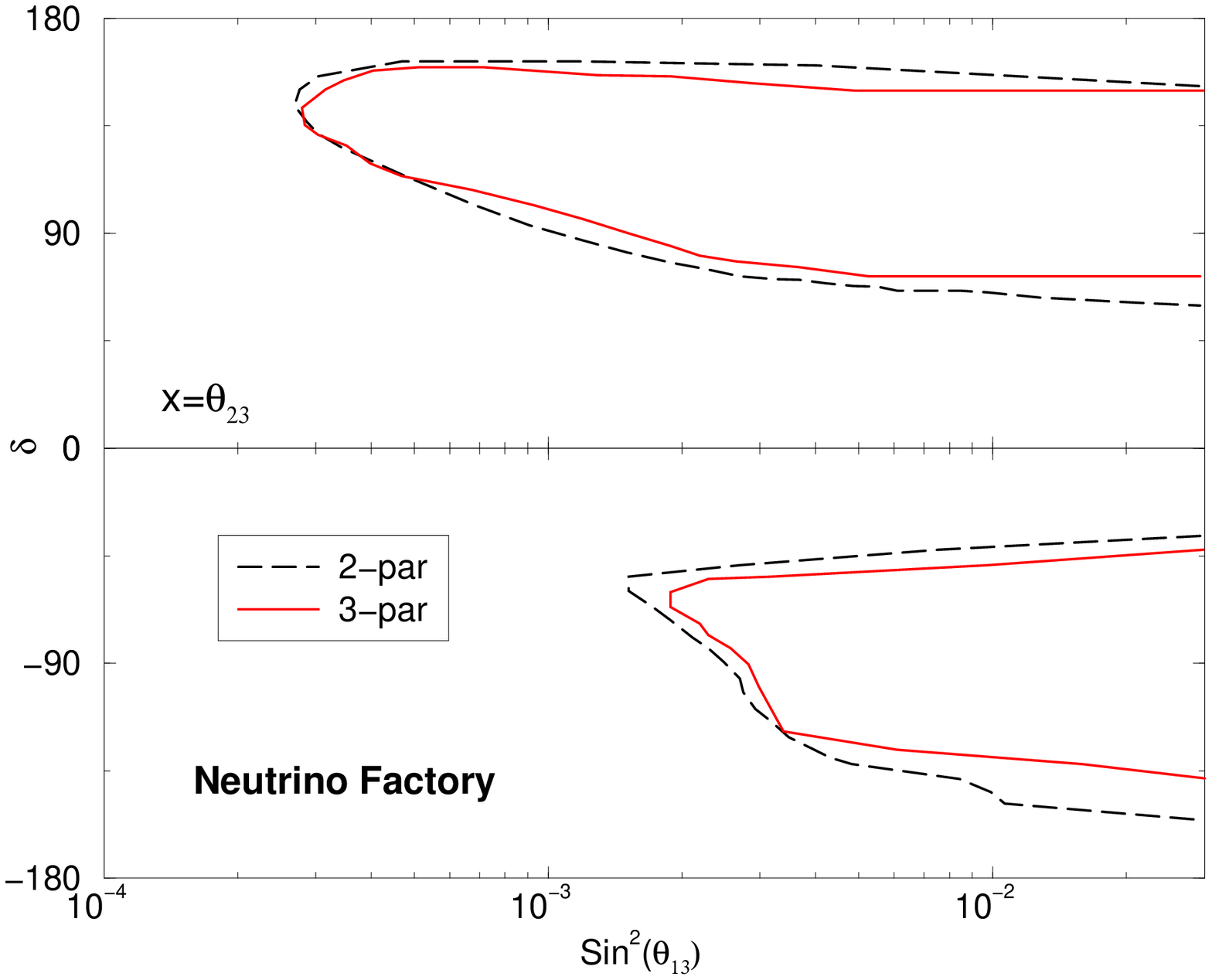} &
\hspace{-0.5cm} \epsfxsize8cm\epsffile{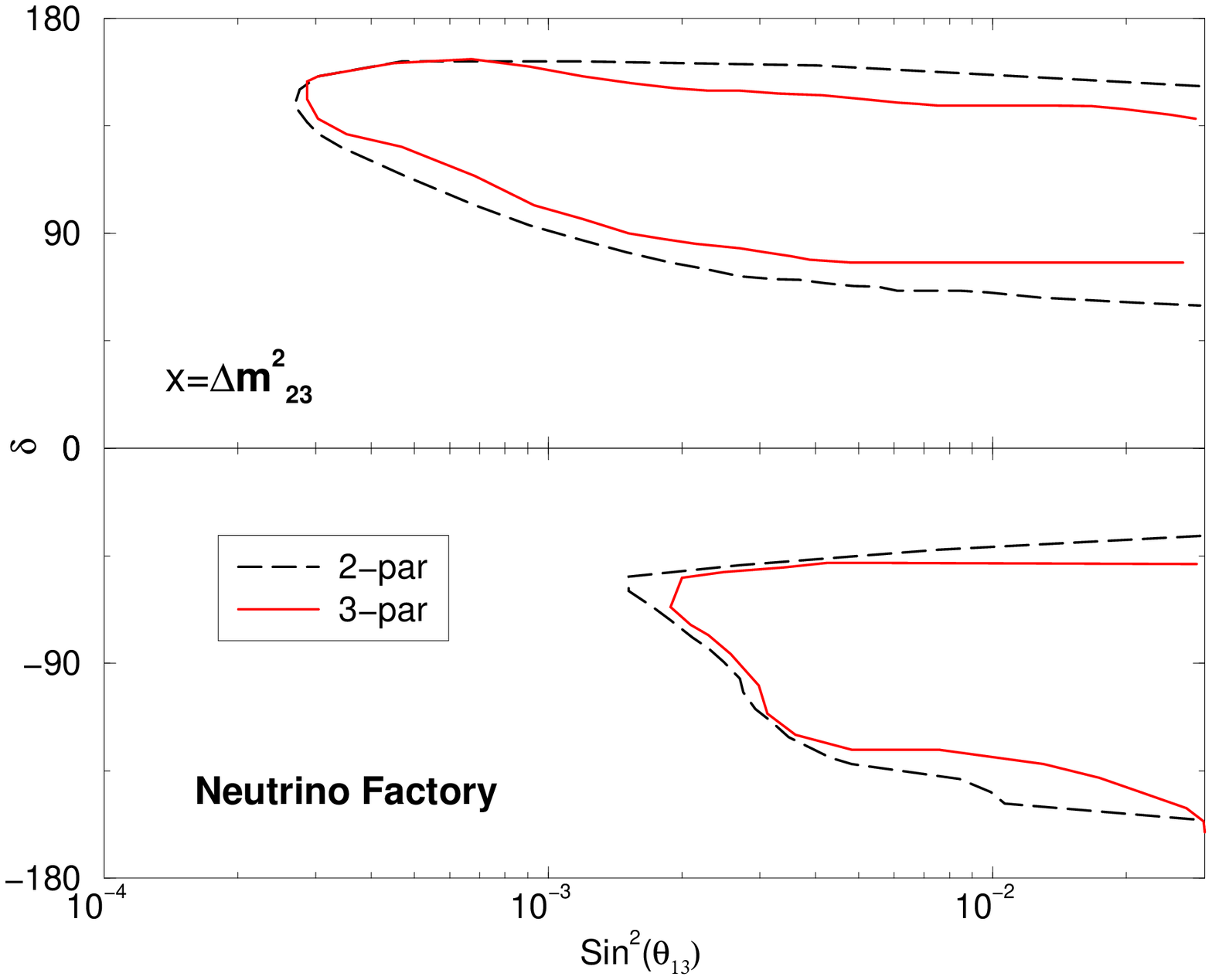} \\
\end{tabular}
\caption{\it CP-violation discovery potential after 5+5 years at the 50 GeV Neutrino Factory.}
\label{fig:sensitivityNF}
\end{center}
\end{figure}

Eventually, in Figs.~\ref{fig:sensitivityBB}-\ref{fig:sensitivityNF} we compare the sensitivity to $(\theta_{13},\delta)$ 
obtained with a two-parameters fit in ($\theta_{13},\delta$) or a
three-parameters fit in 
($\theta_{13},\delta,\theta_{23}$) or ($\theta_{13},\delta,\Delta m^2_{23}$) at the three considered facilities.
The 3$\sigma$ contours have been computed as in Ref.~\cite{Donini:2004iv}: 
at a fixed $\bar \theta_{13}$, we look for the smallest (largest) value of $|\bar \delta|$ for which the 
two- (three-) parameters 3$\sigma$ contours of any of the degenerate solutions (true, sign, octant and mixed) 
do not touch $\delta = 0^\circ$ nor $\delta = 180^\circ$. 
Notice that, although the input $\bar \theta_{13}$ value is fixed, the clones can touch $\delta = 0^\circ, 180^\circ$ 
at $\theta_{13} \neq \bar \theta_{13}$, also\footnote{This is not the case of Fig.~11 in Ref.~\cite{Donini:2004hu}, 
where the excluded region in $\delta$ at fixed $\bar \theta_{13}$ in the absence of a CP-violating signal at 90\% CL is presented.
In practice, in that figure we compare $N_\pm (\bar \theta_{13},\delta)$ with $N_\pm (\bar \theta_{13},0^\circ)$, 
thus obtaining a one-parameter sensitivity plot in $\delta$ only.}.
The outcome of this procedure is finally plotted, representing the region in the $(\theta_{13},\delta)$
parameter space for which a CP-violating signal is observed at 3$\sigma$. 
Within this approach we can thus take fully into account the impact of the parameter degeneracies in the CP-violation discovery 
potential of the three facilities. As for the previous section, we have applied a 2\% systematic error on 
disappearance channels and a 5\% systematic error on appearance channels.
As in the previous section we used the expected errors on the atmospheric parameters after T2K-I for the three-parameters fits 
at the $\beta$-Beam and the Neutrino Factory. The SPL Super-Beam analysis relies on SPL data, only (see Sect.~\ref{sec:disSB}).

Notice that results are given for the whole allowed range in $\delta$, $\delta \in \left [ -180^\circ,180^\circ \right ]$. 
This is particularly appropriate, since only an approximate symmetry is observed for 
$|\delta| \ge \pi/2$ and $|\delta| \le \pi/2$ and no symmetry at all between positive and negative $\delta$ in the 
case of the $\beta$-Beam and of the Neutrino Factory. 

Consider first Fig.~\ref{fig:sensitivityBB}, that refers to $\beta$-Beam results. Notice that the discovery 
potential is not symmetric for positive and negative values of $\delta$, as it has already been observed in 
Ref.~\cite{Donini:2004iv}. This asymmetric behaviour of the $\beta$-Beam is indeed a statistical mirage 
caused by the low background in the appearance antineutrino sample and the high background in the appearance neutrino one 
(see Ref.~\cite{Donini:2004hu}). A proper statistical treatment should be performed, following Ref.~\cite{Feldman:1997qc}, 
to get rid of this asymmetry for small $\sin^2 \theta_{13}$: the treatment, however, is extremely 
time-consuming and we do not consider meaningful applying it here. A further asymmetry can be observed in the
different behaviour of the three-parameters 3$\sigma$ contours projection onto the ($\theta_{13},\delta$) plane for 
positive and negative values of $\delta$: whereas for $\delta > 0$ we observe that the smallest value of $\sin^2 \theta_{13}$
for which a CP-violating phase can be distinguished from a null result goes from 
$[\sin^2 \theta_{13}]_{min} = 3 \times 10^{-4} \to 5 \times 10^{-4}$, for $\delta < 0$ we get 
$[\sin^2 \theta_{13}]_{min} = 2 \times 10^{-3} \to 4 \times 10^{-3}$ for both the $\theta_{23}$ and 
$|\Delta m^2_{23}|$ fits. A small loss in the discovery potential of this
facility with respect to the two-parameters fit is observed
in both three-parameters fits for negative $\delta$. In particular, the region in which a CP-violating signal can be distinguished
from a CP-conserving one goes from $\delta \in [-80^\circ,-130^\circ] \to \delta \in [-90^\circ,-120^\circ]$.

Consider now Fig.~\ref{fig:sensitivitySB}, that refers to Super-Beam results. The strong asymmetry 
for positive and negative $\delta$ is not observed in this case, both for two-
and three-parameters
fits. The impact of the third fitting variable, being it $\theta_{23}$ or $\Delta m^2_{23}$, is a rather small loss 
in the minimum value of $\sin^2 \theta_{13}$ for which a CP-violating phase is distinguished from a null result: 
$[\sin^2 \theta_{13}]_{min} = 9 \times 10^{-4} \to 1.2 \times 10^{-3}$
for $x = \theta_{23}$ and $[\sin^2 \theta_{13}]_{min} = 9 \times 10^{-4} \to 1.4 \times 10^{-3}$ for $x = \Delta m^2_{23}$.
The loss in the $\delta$-interval that is distinguishable from a null result is rather small for three-parameters fits
in $\Delta m^2_{23}$ and negligible when fitting in $\theta_{23}$.

For the Neutrino Factory, Fig.~\ref{fig:sensitivityNF}, we observe a mixed situation: a strong asymmetry between positive and negative
$\delta$ regions (as for the $\beta$-Beam), but a very small difference between two- and three-parameters fits (as for the Super-Beam).
The asymmetry, however, it is not a consequence of asymmetric signal-to-noise ratios\footnote{It must also be reminded that for 5 years 
of data taking in each polarity, a smaller statistics is accummulated in the wrong-sign muon sample for initial $\mu^-$ than for initial $\mu^+$, 
due to the different $\nu N$ and $\bar \nu N$ cross-sections. This reduces the sensitivity to $\theta_{13}$ for negative $\bar \delta$.}
as for the $\beta$-Beam but, rather, of a ``parametric
conspiracy'' that for the chosen values of energy and baseline results in clones that for many negative values of $\bar \delta$ move
toward $\delta = 0^\circ$ or $\delta = 180^\circ$ \cite{Donini:2003vz}, thus preventing a clean identification of a CP-violating signal
(see Figs.~\ref{fig:atmoNF_90}-\ref{fig:atmoNF_m0}). 
The impact of the third fitting variable, being it $\theta_{23}$ or $\Delta m^2_{23}$, is a rather small loss 
in the minimum value of $\sin^2 \theta_{13}$ for which a CP-violating phase is distinguished from a null result
for negative $\bar \delta$: $[\sin^2 \theta_{13}]_{min} = 1.5 \times 10^{-3} \to 2 \times 10^{-3}$. On the other hand, 
for positive values of $\bar \delta$, $[\sin^2 \theta_{13}]_{min} = 2.5 \times 10^{-4}$ both for two- and three-parameters fits.

\section{Conclusions}
\label{sec:concl}

The simultaneous measurement of $\theta_{13}$ and $\delta$ has been often performed in the literature 
considering the solar and atmospheric PMNS parameters as external quantities fixed to their best fit values. 
This is an approximation that has been adopted to get a first insight on the problems related to 
the ($\theta_{13},\delta$) measurement. The experimental uncertainties on these parameters can in principle 
affect the measurement of the unknowns, and it seemed important to us to perform an analysis that could go beyond 
the two-parameters fits presented in the literature.

In this paper we therefore have tried to study the impact that solar and atmospheric sector parameter uncertainties 
have on the measurement of $\theta_{13}$ and $\delta$ at three out of the many proposed setups, the standard low-$\gamma$
$\beta$-Beam, the 4 MWatt SPL Super-Beam and the 50 GeV SPL-fuelled Neutrino Factory. By doing this we wanted to catch 
the characteristic features of the inclusion of external parameters uncertainties in a ($\theta_{13},\delta$) measurement. 

Our first goal has been to identify which of the external parameters affects the most the 
results of two-parameters fits. To do so we have performed a series of three-parameters fits in $\theta_{13},\delta$
and one of the other parameters ($\theta_{12},\Delta m^2_{12},\theta_{23}$ and $\Delta m^2_{atm}$) 
in turn as the third fitting variable and compared our results with standard two-parameters fits. 
It turned out that the impact of solar parameters uncertainties on the measurement of ($\theta_{13},\delta$)
is negligible, in practice, whereas present uncertainties on the atmospheric parameters are large enough
to modify in a significant way the results of two-parameters fits. 
In particular, we have noticed that the main cause of the worsening from two- to three-parameters fits are
the wide displacements of the so-called {\it clones}, parametric degeneracies due to multiple solutions 
of eqs.~(\ref{eq:ene0int})-(\ref{eq:ene0t23sign}), as a consequence of small changes in the external parameters.
These results are general to all the considered facilities.

We have then focused our attention on how the reduction of the atmospheric parameters uncertainties
could ameliorate the previous results. To this respect, the three facilities we have considered are on different
footing. On one side, the $\nu_e$ disappearance channel at the standard low-$\gamma$ $\beta$-Beam 
cannot improve on its own the present measurement of the atmospheric parameters. This facility, therefore, 
must rely on other experiments to meet its goal on $\theta_{13}$ and $\delta$. Luckily enough, it turns out that
the precision on $\theta_{23}$ and $\Delta m^2_{23}$ expected at the approved T2K-phase I experiment, if met, 
would be enough to improve our three-parameters fits and reproduce the results of two-parameters fits in the literature
(that, however, were not so good). On the other hand, we have shown that the $\nu_\mu$ disappearance channel at the 4 MWatt SPL Super-Beam 
does improve the present errors on the atmospheric parameters. This facility, therefore, should not necessarily
rely on external inputs. The combination of appearance and disappearance data, indeed, improve significantly
our three-parameters fits. However, $\theta_{23}$ is not measured well enough and extra clones are still present 
in the ($\theta_{13},\delta$) that are absent in two-parameters contours. Finally, the Neutrino Factory has certainly
the potential to improve significantly the precision on the atmospheric parameters through $\nu_e$ and $\nu_\mu$ disappearance
and the $\nu_\mu \to \nu_\tau$  appearance channel, something that we have not studied in this paper. Using the
errors on $\theta_{23}$ and $\Delta m^2_{23}$ expected at the T2K-I experiment we have checked that extra clones
are present in three-parameters fits that were absent in the two-parameters analysis. This is a clear indication of the 
fact that the problem we are addressing is common to all the facilities, not only to the low-$\gamma$ $\beta$-Beam or the SPL Super-Beam. 
It is not sufficient to just wait and see, but it must be taken into account when envisaging future facilities to look for
$\theta_{13}$ and $\delta$.

To include the impact of external parameters uncertainties, other methods than direct multi-parameter fits have been 
proposed in the literature. For this reason, we have presented a direct
comparison of our three-parameters fit results
with the so-called {\it CP-coverage} introduced in Ref.~\cite{Huber:2002mx}. We have shown that in both methods 
a significant worsening of two-parameters fits arise as a consequence of the inclusion of errors on the external parameters
in the fit. Whereas the {\it CP-coverage} method, however, can be quite useful to condense informations about the 
CP-sensitivity of a facility irrespectively of the specific input pair ($\bar \theta_{13},\bar \delta$) considered, 
direct three-parameters fits offer a detailed information for both $\theta_{13}$ and $\delta$ for specific points 
in the parameter space. We believe that the two methods are, in some sense, complementary and should be combined 
to get a thorough view of the performance of a specific facility designed to measure the ($\theta_{13},\delta$) pair.
To this scope we have presented in App.~B the results of a series of three-parameters fits for the 
standard low-$\gamma$ $\beta$-Beam, the 4 MWatt SPL Super-Beam and the 50 GeV Neutrino Factory for different choices of the input parameters. 

Eventually, we have studied the impact of the atmospheric parameters uncertainties in the CP-violation discovery
potential of the three considered facilities. Our results show that the discovery potential at
the standard low-$\gamma$ $\beta$-Beam and at the SPL Super-Beam is somewhat reduced for negative values of $\delta$
when uncertainties on $\theta_{23}$ and $\Delta m^2_{23}$ are taken into account.
On the other hand the Neutrino Factory appears less affected by the inclusion of external parameter errors. 

In conclusion, we think that this paper shows that present uncertainties on atmospheric parameters are 
indeed too large so that the widely adopted approximation of fixing $\theta_{23}$ and $\Delta m^2_{23}$ to 
their present best fit values be reliable. A new phase of experiments that could improve these uncertainties
are needed. The precision that is expected on the atmospheric parameters at the T2K-I experiment is
shown to be such that three-parameters fits could reproduce the results of two-parameters fits
presented in the literature. This experiment is therefore a crucial step in the way to the measurement of 
the two PMNS unknowns, if the precision goals can indeed be met.

The same kind of analysis we presented here must be clearly repeated at all of the proposed setups, 
something that goes beyond the scope of this paper but is extremely important to establish on solid grounds 
the quest for $\theta_{13}$ and leptonic CP violation in the near future. 

\section*{Acknowledgements}

We would like to thank E.~Fern\'andez-Mart\'{\i}nez for extremely useful discussions and comments, 
D.~Autiero, Y.~Declais, B.~Gavela, J.J.~Gomez-Cadenas, P.~Hernandez, P.~Huber, P.~Lipari, E.~Lisi, M.~Lusignoli, O.~Mena, 
P.~Migliozzi, T.~Schwetz and W.~Winter for discussions. The authors acknowledge the financial support of 
MEC through project FPA2003-04597, of CICYT-INFN through the ``Neutrinos and others windows to new physics beyond the SM'' 
agreement and of the European Union through the networking activity BENE and the RTN European Program MRTN-CT-2004-503369.

\section*{Appendix A}
\label{app:coverage}

In this Appendix we review some of the statistical methods that have been proposed in the literature to take
into account uncertainties on the external parameters in the measurement of ($\theta_{13},\delta$).

\noindent {\bf 1) Comparison between naive two- and three-parameters fit} \\
The obvious difference is in the CL contours that can be drawn in the two cases: for two-parameters fit 
the 90\% CL corresponds to $\Delta \chi^2 = 4.61$, whereas for three-parameters fit is $\Delta \chi^2 = 6.25$. 
As a consequence, when a single minimum is found the projection of a three-parameters fit onto the two-dimensional contour is, 
in general, a bit larger. The second, not obvious, difference resides in that in the three-parameters fit the 
three-dimensional manifold automatically allows for a displacement of the clones solutions arranging 
for a lower $\chi^2$ at the relative minima. This is indeed the case for the clones corresponding to wrong choices of $s_{atm}$
and of the $\theta_{23}$-octant (see \cite{Donini:2003vz}, also). If the clones location moves in the three-dimensional 
manifold, the resulting projection onto the plane can be much larger than the
two-parameters contour. 
This is indeed the main result of this paper and is discussed at length in Sect.~\ref{sec:uncert}.

\noindent {\bf 2) Inclusion of a covariance matrix in the two-parameters fit} \\
A fixed error range for any non-fitted parameter can be taken into account introducing a covariance matrix
in the $\chi^2$ function as follows:
\bea
\chi^2_{\{ \bar \alpha \}}(\theta_{13}, \delta) &=& \sum_{i,j} 
\left \{ \left[  N_i (\theta_{13}, \delta) - N_i (\bar \theta_{13}, \bar \delta) \right] 
C^{-1}_{ij}
\left[  N_j (\theta_{13}, \delta) - N_j (\bar \theta_{13}, \bar \delta) \right] 
\right \}_{\{ \bar \alpha \}}
\nn \\
&&
\label{eq:chi2cov}
\eea
where $C_{ij}$ is the covariance matrix, $i,j$ refer to different channels at the same experiment or to different experiments
and $\{ \bar \alpha \}$ is a given set of external parameters. 
If the errors on the entries $i$ and $j$ of the covariance matrix are statistically independent, $C$ is 
\be
C_{ij} = \delta_{ij} \, \delta N_i^2 + 
\sum_{\alpha=1}^{N_\alpha} \frac{\partial N_i}{\partial \alpha} \frac{\partial N_j}{\partial \alpha}
\sigma^2(\alpha)
\ee
where $\sigma (\alpha)$ is the 1$\sigma$ error on the parameter $\alpha$.
This procedure, followed in \cite{Burguet-Castell:2001ez} for the Neutrino Factory and in \cite{Bouchez:2003fy}
for the facilities considered in this paper, reproduces the enlargement of the
two-parameters CL contours 
observed from a multi-parameter fit projected onto the ($\theta_{13},\delta$) plane. However, within this approach, 
the clones locations are not free to move in the multi-dimensional manifold to arrange for a lower $\chi^2$: 
they are indeed stucked to the location in the ($\theta_{13},\delta$) plane that can be computed once known 
the external (fixed) parameters (see \cite{Donini:2003vz}, again, and \cite{Huber:2004gg}, Sect.~3.3).
The displacement of the relative minima is indeed the characteristic feature of the multi-parameter fit, where 
$N_{\alpha}$ external parameters cooperate with $\theta_{13}$ and $\delta$ to locate lower $\chi^2$ regions 
than those found in a two-parameters fit with fixed external parameters.
The resulting regions are therefore large than those obtained including the
covariance matrix in a two-parameters fit.

\noindent {\bf 3) CP-coverage and marginalization over $N_\alpha$ external parameters} \\
A parameter that can be used to compare in a condensed way the capability of different setups 
to measure the CP-violating phase $\delta$ has been proposed in \cite{Huber:2002mx}. 
The CP-coverage is defined as follows: 
\be
\xi (\bar \delta )= {\rm Coverage\ in}\, \delta = \frac{1}{2 \pi} {\large U}_{I = 1}^{N_{deg}} \Delta_I (\bar \delta) \, ,
\ee
is the fraction of the $\delta$-parameter space (i.e., $2 \pi$) that is allowed at a given CL as a result 
of a measure when the input parameter is $\bar \delta$. The sum goes over $N_{deg}$ possible allowed regions 
induced by parameter degeneracies, each of them spanning an interval $\Delta_I (\bar \delta)$ of the parameter space. 
We take the union of these intervals to take into account possible partial overlaps of the $\Delta_I (\bar \delta)$.
The smaller the CP-coverage, the better the measurement of $\delta$ at a given experiment. 
In particular, to distinguish a maximally CP-violating signal (i.e., $\bar \delta = \pm 90^\circ$)
at a given experiment from $\delta = 0^\circ$ or $\delta = 180^\circ$, the CP-coverage must be less than 0.5. 

The definition of the CP-coverage above is, however, incomplete: we must still define how the $\Delta_I (\bar \delta)$
intervals are computed and which is the dependence of $\xi (\bar \delta )$ on other parameters. 
Indeed, if all the parameters of the PMNS matrix but $\delta$ were measured, $\xi (\bar \delta)$ would be just an involute
way to express the expected error of an experiment for a certain value of $\bar \delta$. This has been called 
the ``CP-pattern'', see Fig.~4(right) in Ref.~\cite{Huber:2004gg}. If, on the other hand, $\theta_{13}$ is also an 
unknown parameter, we can think of defining a function $\xi (\bar \theta_{13},\bar \delta)$ and to plot it as a function 
of different values of $\bar \theta_{13}$ (called ``CP-scaling'' in Ref.~\cite{Huber:2004gg}) for a fixed value of $\bar \delta$. 
In this second case, for the particular value $\bar \delta = 0^\circ$, the ``CP-scaling'' as a function of $\bar \theta_{13}$ is 
nothing else that the CP-sensitivity. For example, Fig.~5 in Ref.~\cite{Bouchez:2003fy} or Fig.~11 of Ref.~\cite{Donini:2004hu}
represent the sensitivity to $\delta$ for varying $\bar \theta_{13}$ defined as a one-parameter fit where all mixing parameters
have been measured but $\delta$ and $\theta_{13}$. The same idea is presented in Fig.~6 of Ref.~\cite{Bouchez:2003fy} and
Fig.~13 of Ref.~\cite{Burguet-Castell:2003vv}, where the plot represents the capability to distinguish a non-vanishing $\delta$ from 
$\delta = 0^\circ$ or $\delta = 180^\circ$ at a given one-parameter CL\footnote{It should be stressed that it is not completely 
correct to present this ``CP discovery potential'' with one-parameter CL contours: being $\theta_{13}$ a parameter to be 
fitted together with $\delta$, we should consider two-parameters CL contours, instead. In this case the CP discovery potential 
can be smaller than when only $\delta$ is left as a free parameter, as a result
of the fact that the two-parameters 
$\chi^2 (\theta_{13},\delta)$ can be lower than $\chi^2(\bar \theta_{13},\delta)$ for specific
values of $\theta_{13} \neq \bar \theta_{13}$. This is indeed what reported in Fig.~7 of Ref.~\cite{Donini:2004iv}.}.

To take into account the fact that, in general, the parameters of the mixing matrix are known only with a finite precision
and that $\theta_{13}$ is completely unknown at present (and thus a one-parameter $\delta$-sensitivity plot 
is generally overestimating the performance of a given experiment), the authors of Ref.~\cite{Huber:2002mx} have proposed 
to compute the CP-coverage $\xi (\bar \delta)$ by first minimizing a ($N_\alpha + 2$)-parameter $\chi^2$ over $N_\alpha$ 
external parameters. In this way, for any input pair $\bar \theta_{13}$ and $\bar\delta$, 
a two-dimensional surface of the $\chi^2$ minimum as a function of $\theta_{13}$ and $\delta$ is generated. 
If we then minimize the resulting function in $\theta_{13}$, also, we can deduce a one-dimensional function of $\delta$ 
and of the input parameters representing the minimum value of the ($N_\alpha + 2$)-dimensional $\chi^2$ for a given value of 
$\bar \delta$. From this marginalized $\chi^2$ we can finally compute the allowed $\Delta_I (\bar \delta)$ intervals imposing 
that $\chi^2_{min} (\delta, \bar \delta)$ be equal to a given one-parameter CL. 

\begin{figure}[h!]
\begin{center}
\begin{tabular}{cc}
\hspace{-1.5cm}\epsfxsize8cm\epsffile{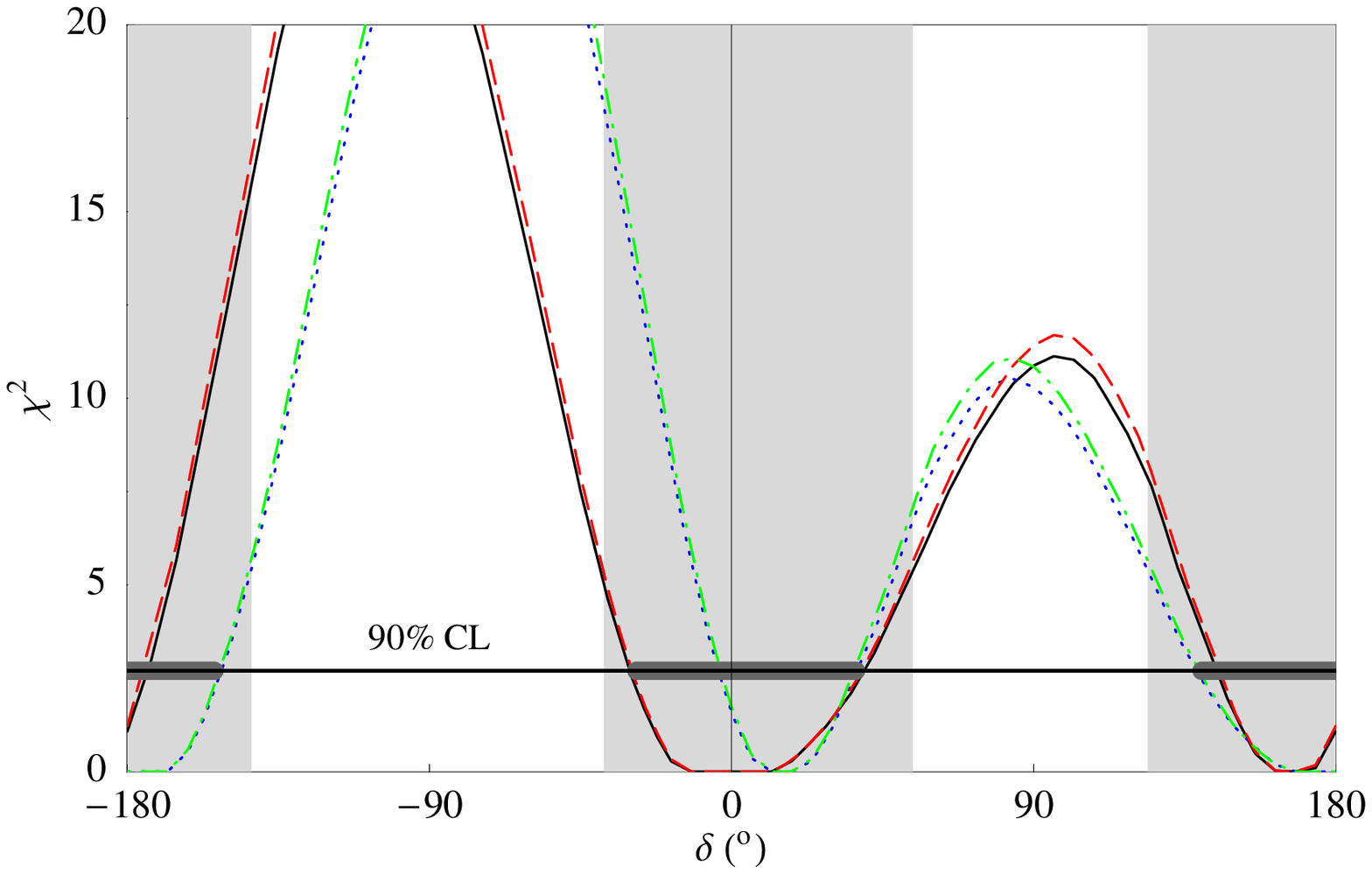} & 
               \epsfxsize8cm\epsffile{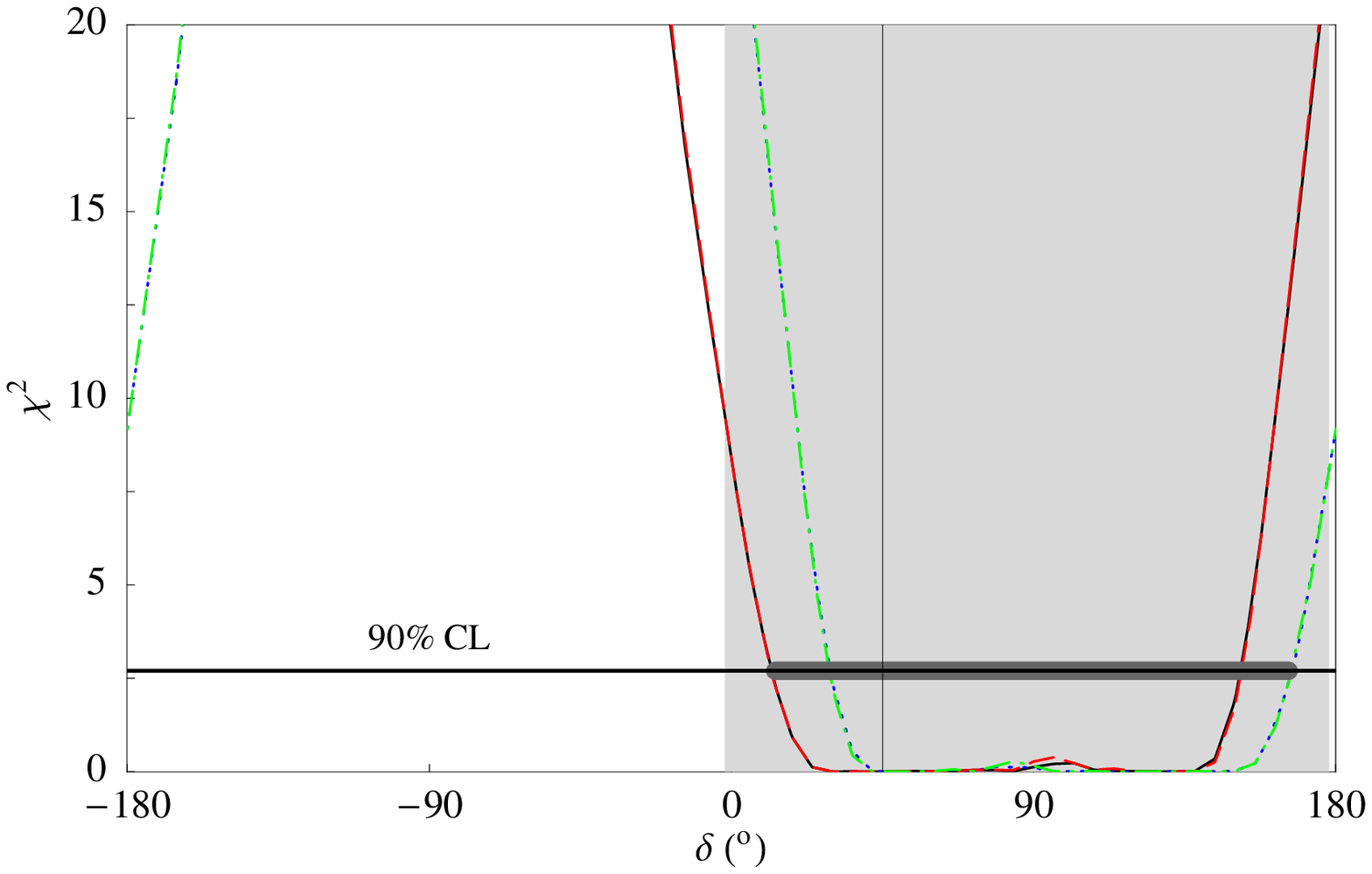} \\
\hspace{-1.5cm}\epsfxsize8cm\epsffile{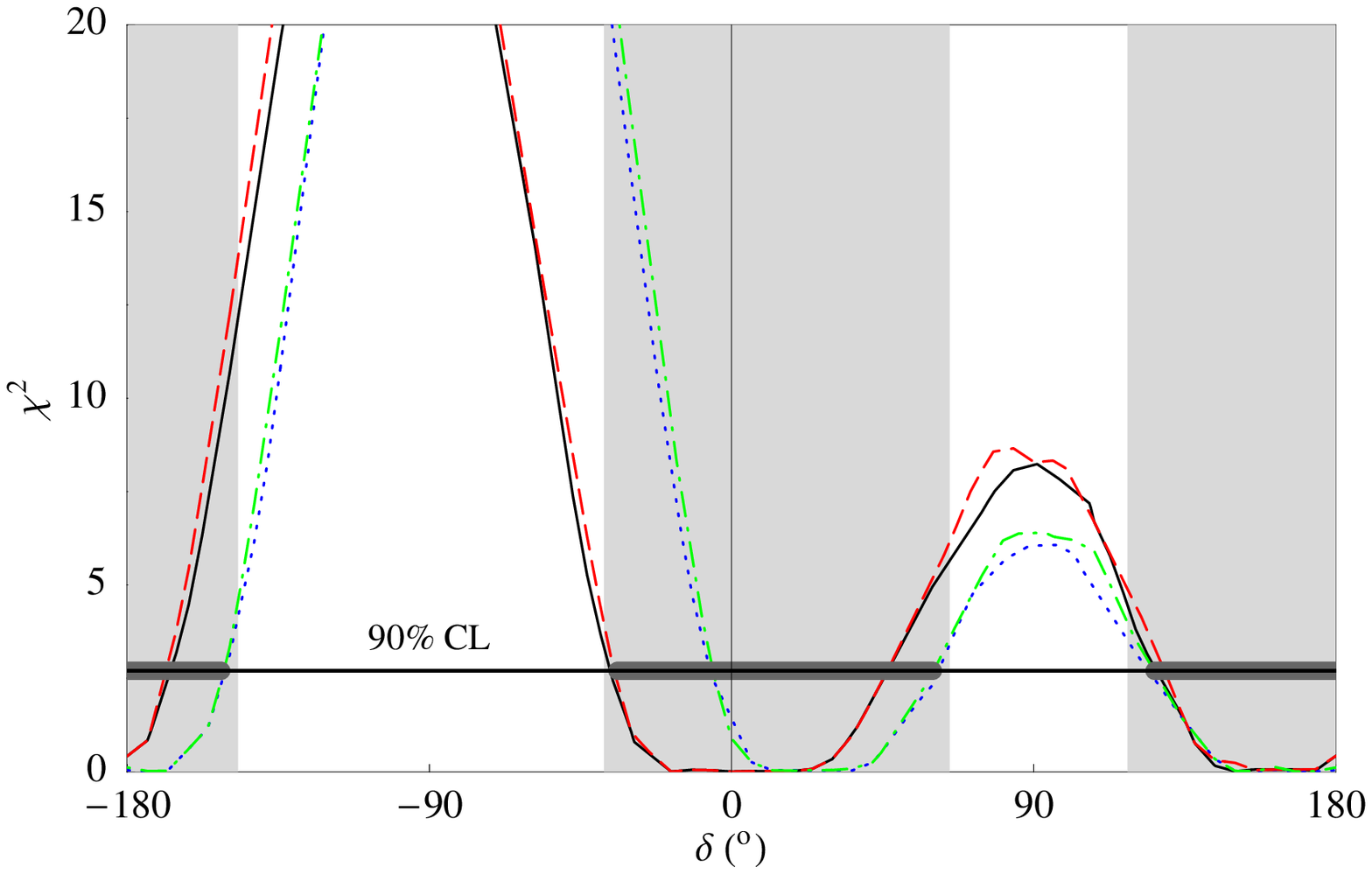} & 
               \epsfxsize8cm\epsffile{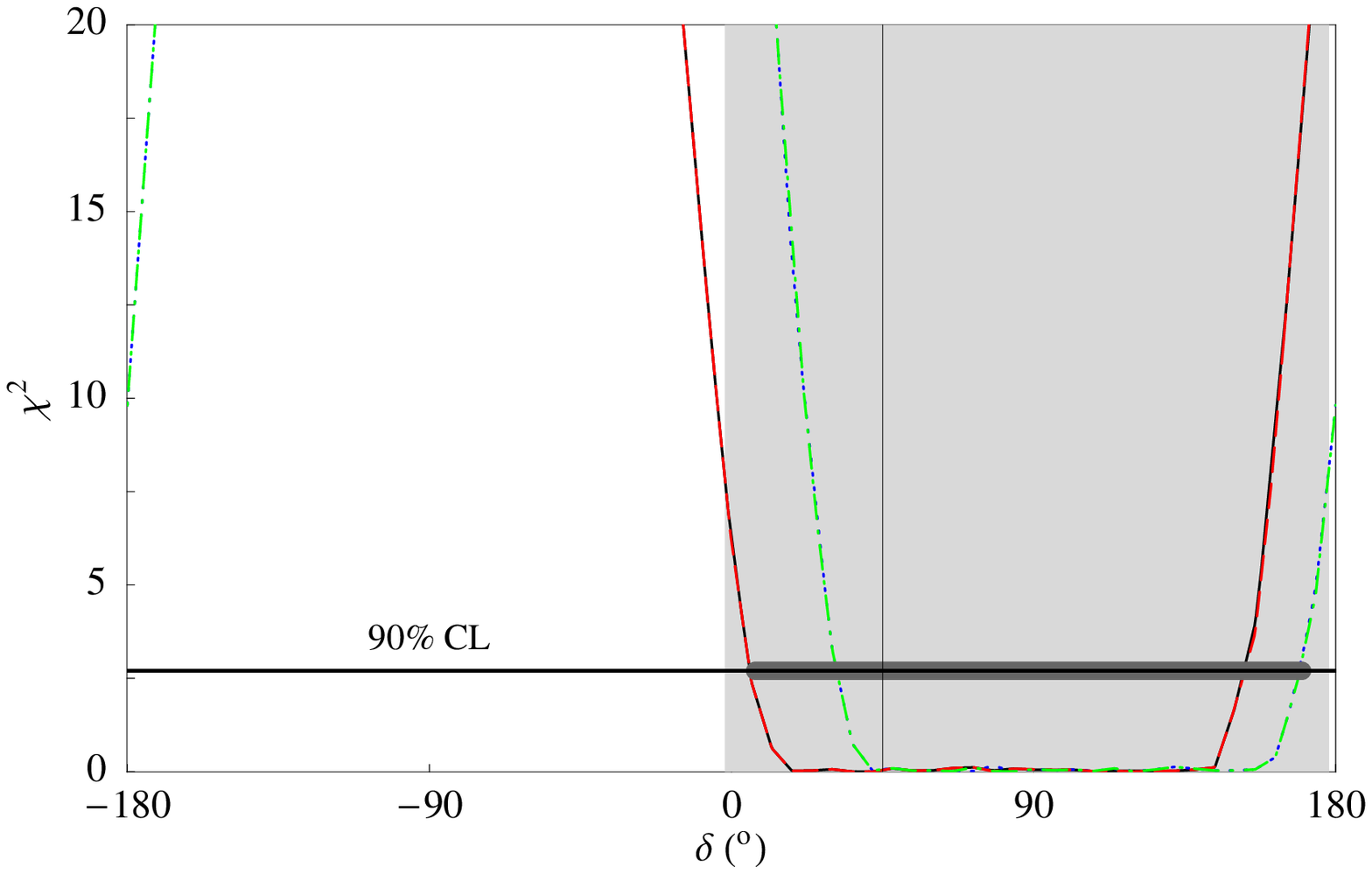} \\
\end{tabular}
\caption{\it  \label{fig:coverage}
Comparison of the CP-coverage $\xi$ and the $\delta$-axis projection 
of the three-parameters 90\% CL contour to the same input parameters at the $\beta$-Beam. 
The marginalized $\chi^2_{min} (\delta,\bar\delta)$ function is plotted for different choices of $s_{atm}$ and $s_{oct}$:
$s_{atm} = \bar s_{atm}, s_{oct} = \bar s_{oct}$ (solid); $s_{atm} = -\bar s_{atm}, s_{oct} = \bar s_{oct}$ (dotted);
$s_{atm} = \bar s_{atm}, s_{oct} = -\bar s_{oct}$ (dashed); $s_{atm} = -\bar s_{atm}, s_{oct} = -\bar s_{oct}$ (dot-dashed). 
Horizontal thick black lines are the $\Delta_I (\bar \delta)$ intervals and 
the gray-shaded region are the corresponding three-parameters 90\% CL contours. 
Of the two possible three-parameters fit (i.e., in $\theta_{23}$ or $\Delta m^2_{23}$), 
that with the largest error in $\delta$ is reported. 
The thin vertical line displays the value of $\bar \delta$ for each plot.
The different plots refer to: 
$\bar \theta_{13} = 2^\circ, \bar \delta = 0^\circ$ (top left);
$\bar \theta_{13} = 2^\circ, \bar \delta = 45^\circ$ (top right);
$\bar \theta_{13} = 7^\circ, \bar \delta = 0^\circ$ (bottom left);
$\bar \theta_{13} = 7^\circ, \bar \delta = 45^\circ$ (bottom right).
The atmospheric input parameters are $\bar \theta_{23} = 40^\circ$, $\Delta m^2_{atm} = 2.5 \times 10^{-3}$ eV$^2$.
}
\end{center}
\end{figure}

Clearly, this procedure can fail when multiple minima are present at each marginalization step. When multiple minima are 
present, the minimization procedure will generally look for the absolute minimum. In this way, the information on other 
relative minima in the $\chi^2$ can be lost. This is particularly problematic since we know that, due to parametric 
degeneracies, multiple minima should be present. Marginalization near a second minimum will give a second function 
$\chi^2_{min} (\delta, \bar \delta)$, from which a new set of $\Delta_I (\bar \delta)$ intervals can be found.
The procedure suggested in \cite{Huber:2004gg} is indeed to marginalize around each of the expected minima in the 
multi-dimensional $\chi^2$ and to draw several distinct one-dimensional functions $\chi^2_{min} (\delta, \bar \delta)$. 
Imposing on any of them the constraint $\chi^2_{min} (\delta, \bar \delta) = CL$, the full set of allowed regions 
in $\delta$ at a given one-parameter CL for a fixed input $\bar \delta$ is deduced and the $\xi (\bar \delta)$ can 
be finally computed. Since the minimization procedure must be repeated several times (once per expected minima), 
it is useful to choose the starting point of the minimization algorithm in a clever way. In \cite{Huber:2004gg}
it is suggested to solve eqs.~(\ref{eq:ene0int})-(\ref{eq:ene0t23sign}) as it has been done in Ref.~\cite{Donini:2003vz}
to find the expected clone locations and to use the latter as starting points for the minimization. 
Applying the previous algorithm it is possible to deduce a $\xi (\bar \delta, \bar \theta_{13})$ parameter that reduce
significantly the overestimation of the experiment performances in the measurement of $\delta$ that is typical of the 
one-parameter $\delta$-sensitivity plots. We should therefore compare this procedure with the projection onto the 
$\delta$-axis of our three-parameters fits to understand if a residual overestimation is still present. 
To this purpose, in Fig.~\ref{fig:coverage} we present the CP-coverage $\xi (\bar \theta_{13},\bar \delta)$
and the projection onto the $\delta$-axis of the three-parameters CL contours for different values of $\bar \theta_{13}$ 
and $\bar \delta$. The results in the figure have been obtained using the considered $\beta$-Beam facility, for definiteness. 

As it can be seen in the figure, some underestimation of the error on $\delta$  at the considered experiment 
is still present when computing $\xi (\bar\delta)$ and comparing it with the
$\delta$-axis projection of the three-parameters 
fits. The main interest of the CP-coverage parameter is that the algorithm described above can be iterated for any value of 
$\bar \theta_{13}$ to obtain a ``CP-scaling'' for any given value of $\bar \delta$, thus replacing the (poorly reliable) 
one-parameter $\delta$-sensitivity plots. This will give the general picture of the expected error on $\bar \delta$ at
a given facility, to be complemented in our opinion with multi-parameter fits to particular values of $\theta_{13}$ and $\delta$ 
to get a robust estimate of the facility performance.

\newpage


\newpage

\section*{Appendix B: three-parameters fits}
\label{app:plots}

In this Appendix we present the projection of the three-dimensional 90\% CL contours onto the ($\theta_{13},\delta$) plane 
for the three reference setups (the low-$\gamma$ $\beta$-Beam, the SPL Super-Beam and the 50 GeV Neutrino Factory) for different choices of 
the input pair ($\bar \theta_{13},\bar \delta$). The external input parameters in all plots are: $\theta_{12} = 32^\circ$, 
$\Delta m^2_{12} = 8.2 \times 10^{-5}$ eV$^2$; $\theta_{23} = 40^\circ, \Delta m^2_{23} = 2.5 \times 10^{-3}$ eV$^2$.
In this case, all choices of the two discrete variables $s_{atm}$ and $s_{oct}$
are presented together and no comparison with two-parameters contours is shown. 

For the $\beta$-Beam plots we compare the impact of the present uncertainties on the atmospheric parameters (third column of Tab.~\ref{tab:input})
with that of the expected uncertainties after T2K-I (last column of Tab.~\ref{tab:input}): $\theta_{23} \in [38^\circ,43^\circ]-[48^\circ,52^\circ]$ 
and $\Delta m^2_{23} \in [2.42,2.61] \times 10^{-3}$ eV$^2$ for $s_{atm} = +$ and 
$\Delta m^2_{23} \in [2.46,2.64] \times 10^{-3}$ eV$^2$ for $s_{atm} = -$, \cite{Enrique}. 
Notice that the error on $\theta_{23}$ is just the expected
error at T2K-I, \cite{Itow:2001ee}, shifted around $\theta_{23} = 40^\circ$. It has been shown in Sect.~\ref{sec:disSB}
that the $\nu_\mu$ disappearance channel at the Super-Beam is rather effective in reducing the uncertainties on the atmospheric parameters
(whereas the $\nu_e$ disappearance channel at the low-$\gamma$ $\beta$-Beam is useless to this purpose, see Ref.~\cite{Donini:2004iv}). 
In this case we therefore do not present results using ``present'' and ``expected'' uncertainties, but we just combine the results 
from the appearance and disappearance channel. 
Finally, for the Neutrino Factory we have considered only the two appearance channels $\nu_e \to \nu_\mu,\nu_\tau$ with the atmospheric 
parameters with the expected uncertainties after T2K-I. 

In general, a ``pessimistic'' systematic error, $\epsilon^\pm = 5$\%, has been used in appearance channels. 
A 2\% systematic error has been used in disappearance channels. However, we checked that using a ``pessimistic'' 5\% systematic 
error in the disappearance channel does not change significantly our results.

The input parameters are: $\bar \theta_{13} = 2^\circ, 7^\circ$; $\bar \delta = 90^\circ, 45^\circ, 0^\circ, -45^\circ, -90^\circ$.


\begin{figure}[h!]
\vspace{-0.35cm}
\begin{center}
\begin{tabular}{c c}
\hspace{-2cm} 
\epsfxsize9cm\epsffile{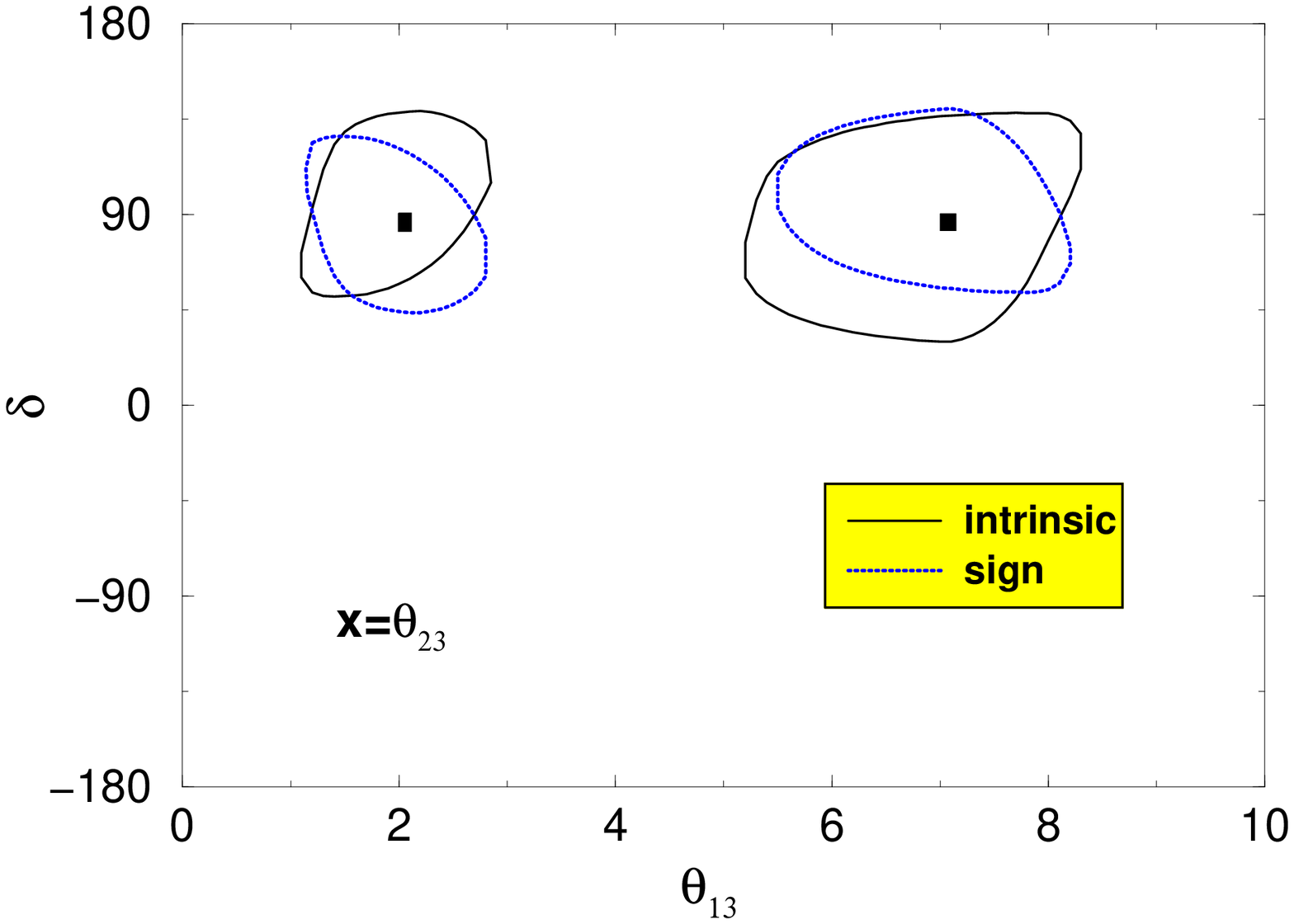} & 
\hspace{-1cm} 
\epsfxsize9cm\epsffile{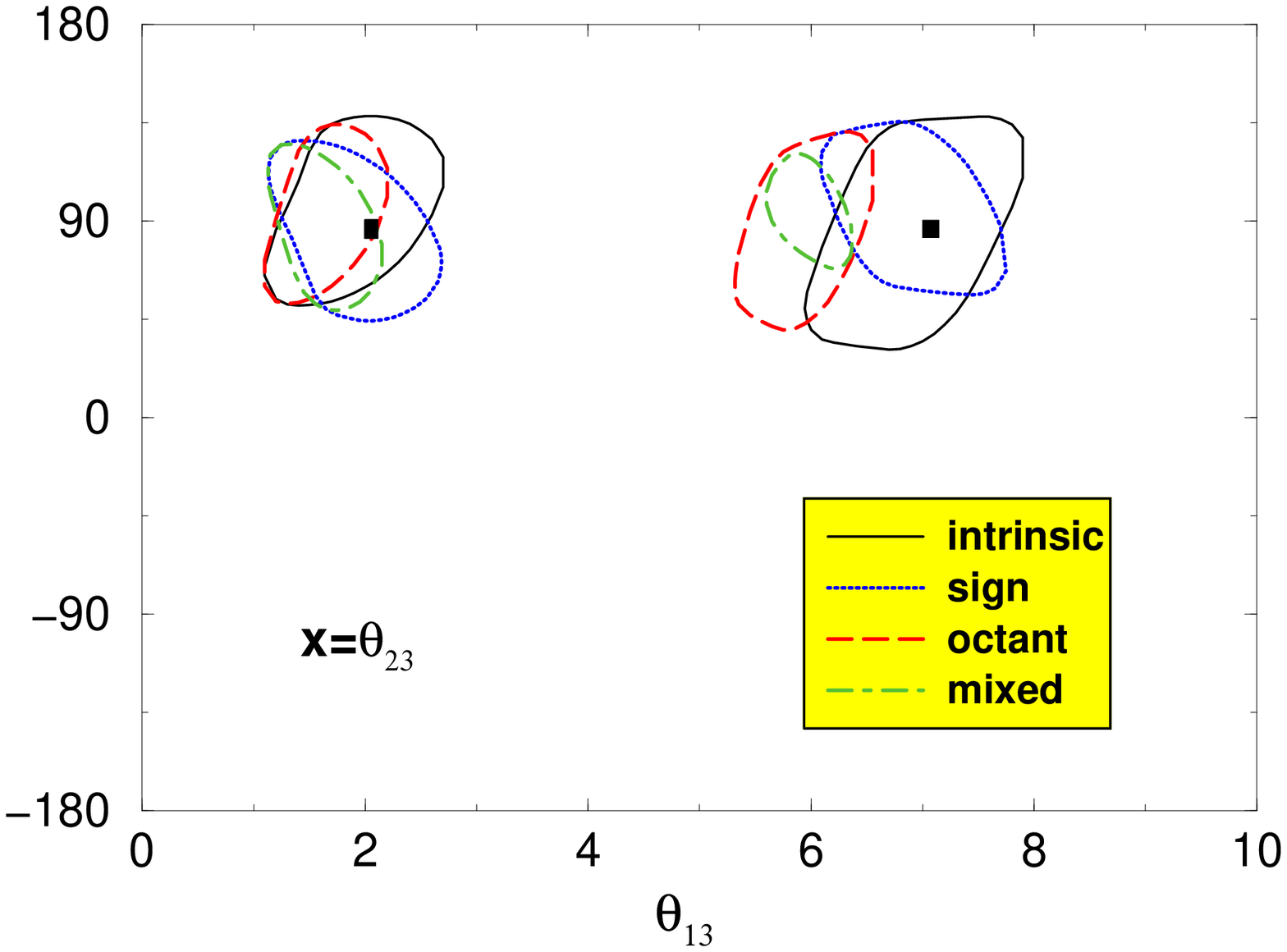} \\
\hspace{-2.5cm} 
\epsfxsize9cm\epsffile{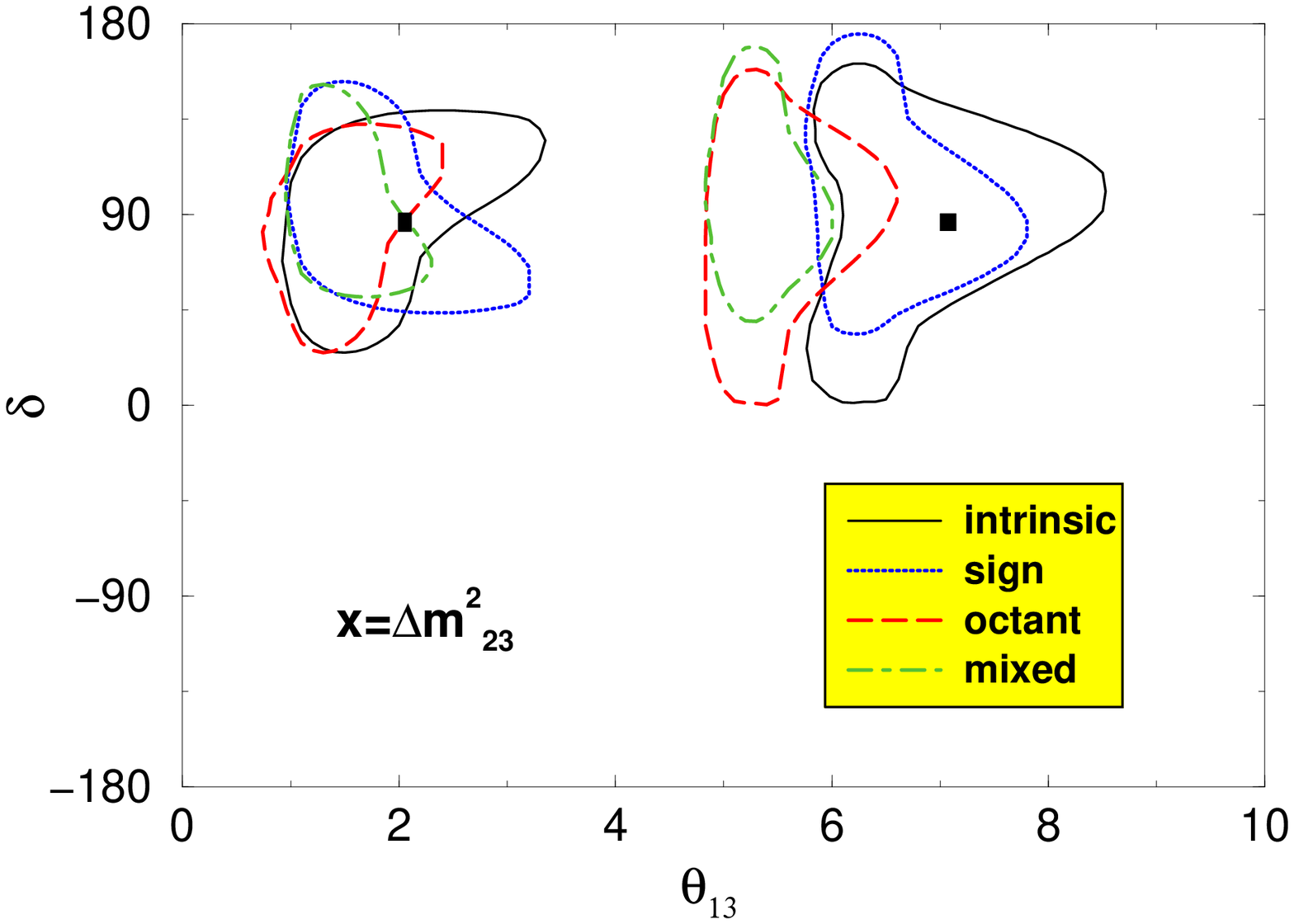} &
\hspace{-1cm} 
\epsfxsize9cm\epsffile{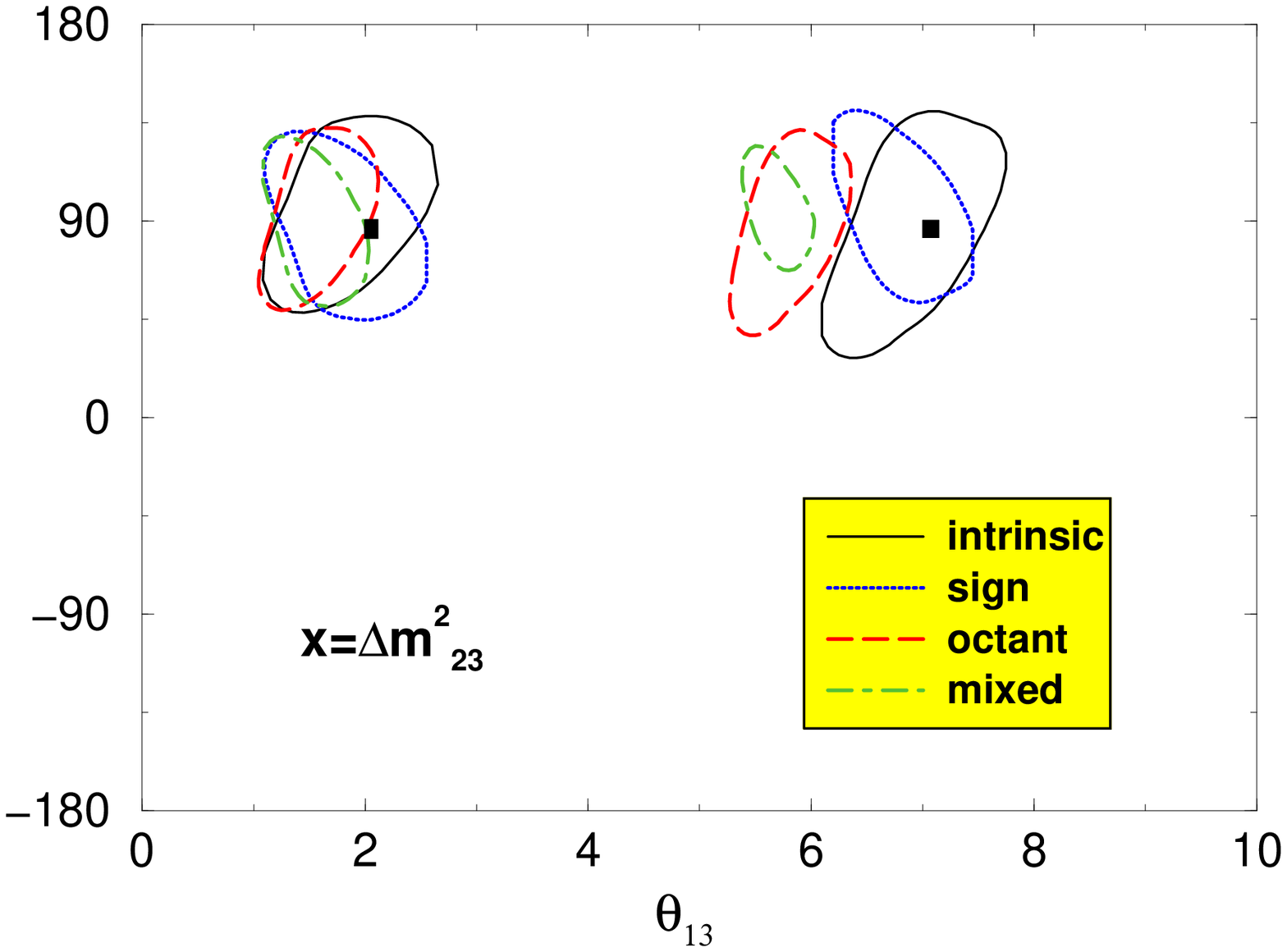} \\
\end{tabular}
\caption{\it Three-parameters 90 \% CL contours after a 10 years run at the $\beta$-Beam. 
Input parameters: $\bar \theta_{13} = 2^\circ, 7^\circ$; $\bar \delta = 90^\circ$.
Top panels: $x = \theta_{23}$; bottom panels: $x = \Delta m^2_{23}$. 
Left panels: present uncertaintes; right panels: after T2K-I.}
\label{fig:atmo2_90}
\end{center}
\end{figure}

\begin{figure}[h!]
\vspace{-0.35cm}
\begin{center}
\begin{tabular}{c c}
\hspace{-2cm} 
\epsfxsize9cm\epsffile{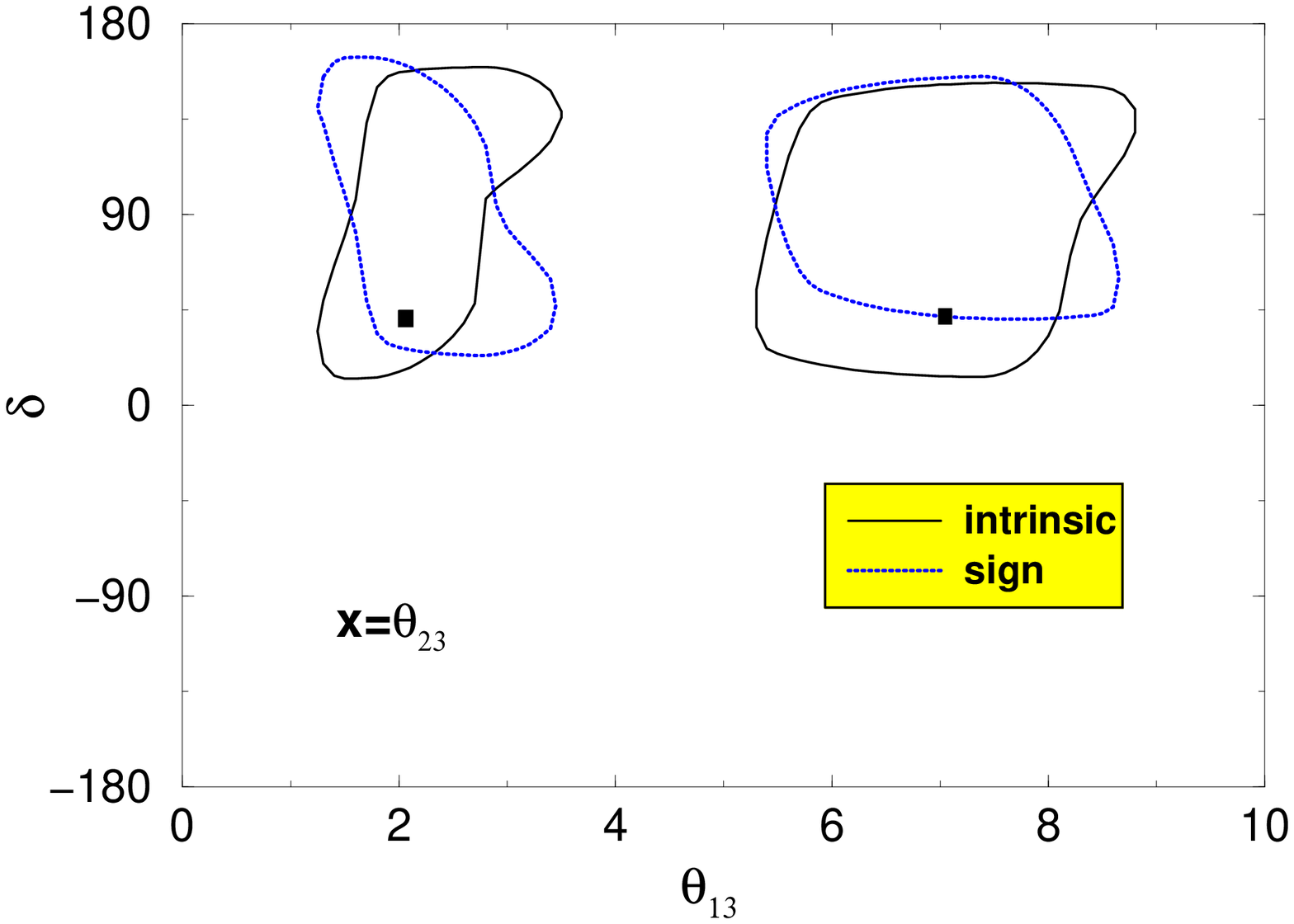} & 
\hspace{-1cm} 
\epsfxsize9cm\epsffile{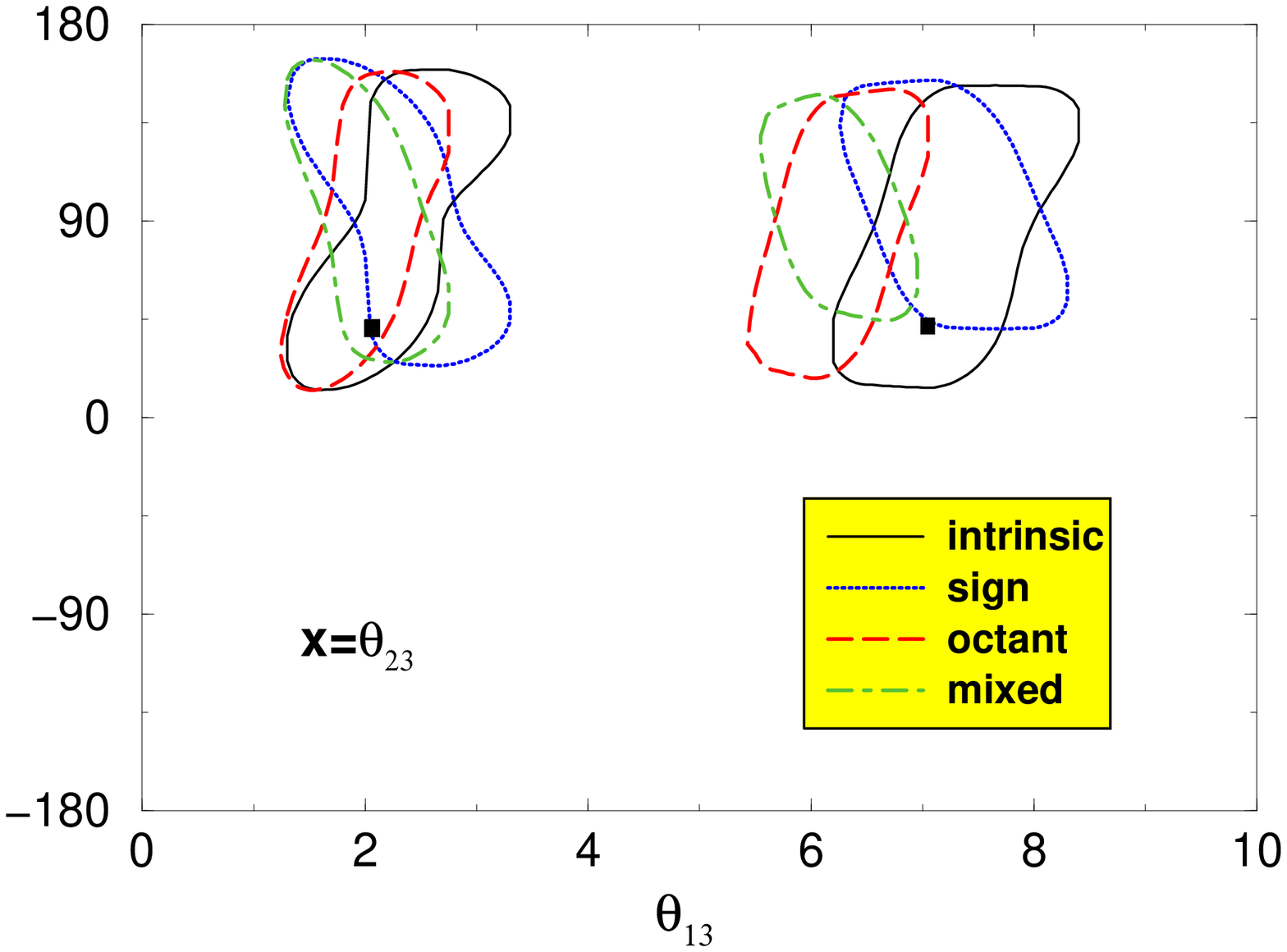} \\
\hspace{-2.5cm} 
\epsfxsize9cm\epsffile{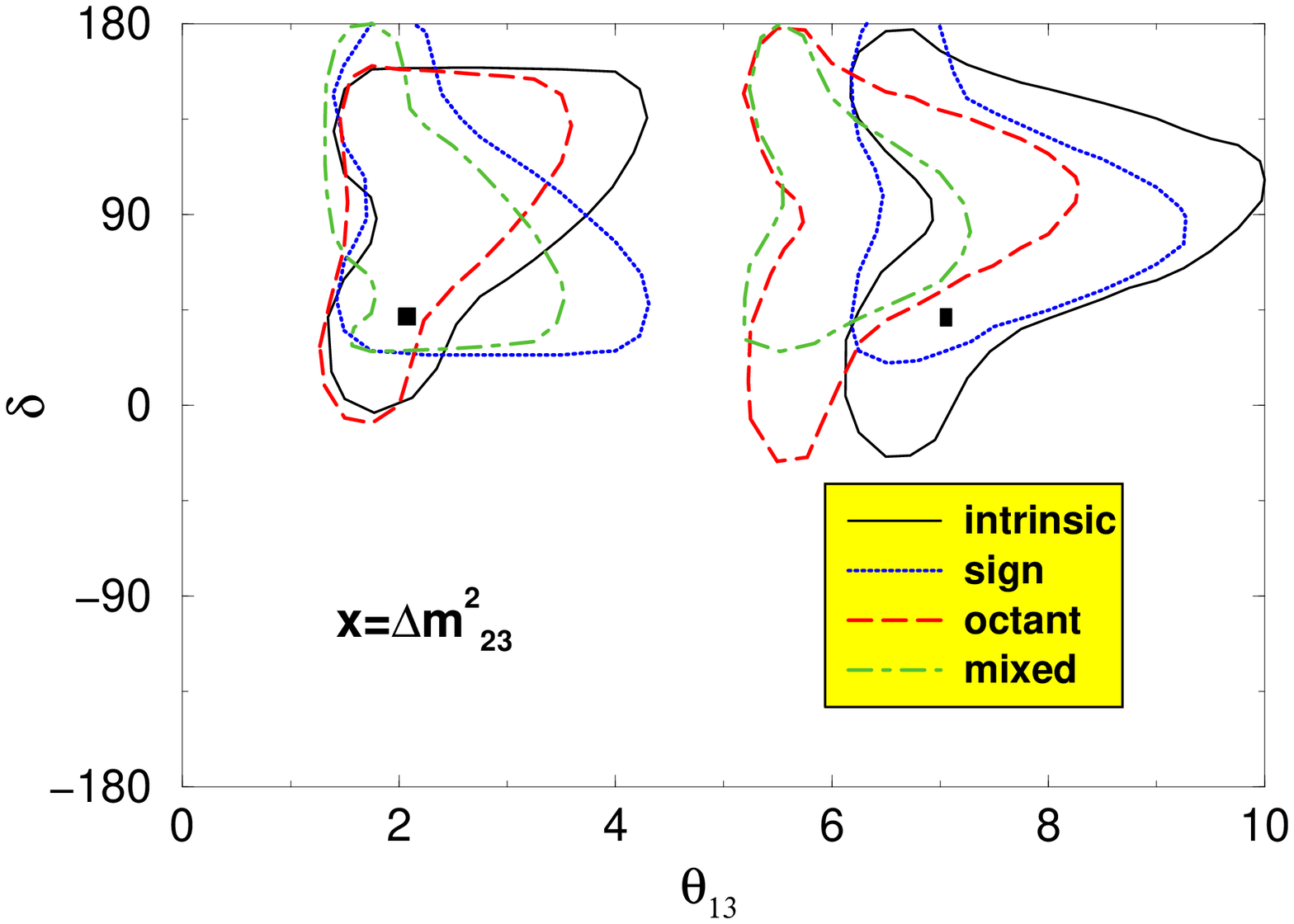} &
\hspace{-1cm} 
\epsfxsize9cm\epsffile{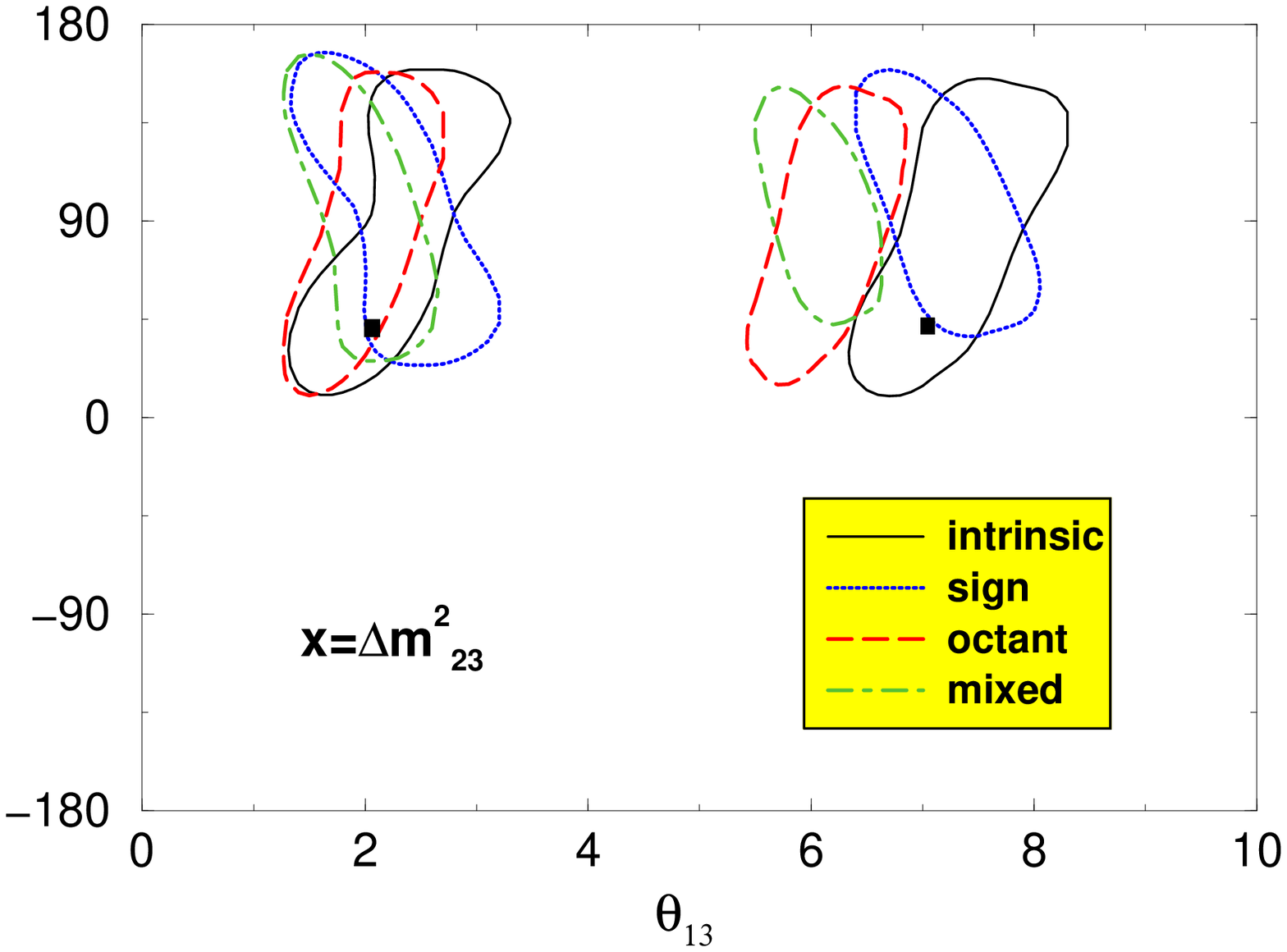} \\
\end{tabular}
\caption{\it Three-parameters 90 \% CL contours after a 10 years run at the $\beta$-Beam. 
Input parameters: $\bar \theta_{13} = 2^\circ, 7^\circ$; $\bar \delta = 45^\circ$.
Top panels: $x = \theta_{23}$; bottom panels: $x = \Delta m^2_{23}$. 
Left panels: present uncertaintes; right panels: after T2K-I.}
\label{fig:atmo2}
\end{center}
\end{figure}

\begin{figure}[h!]
\vspace{-0.35cm}
\begin{center}
\begin{tabular}{c c}
\hspace{-2cm} 
\epsfxsize9cm\epsffile{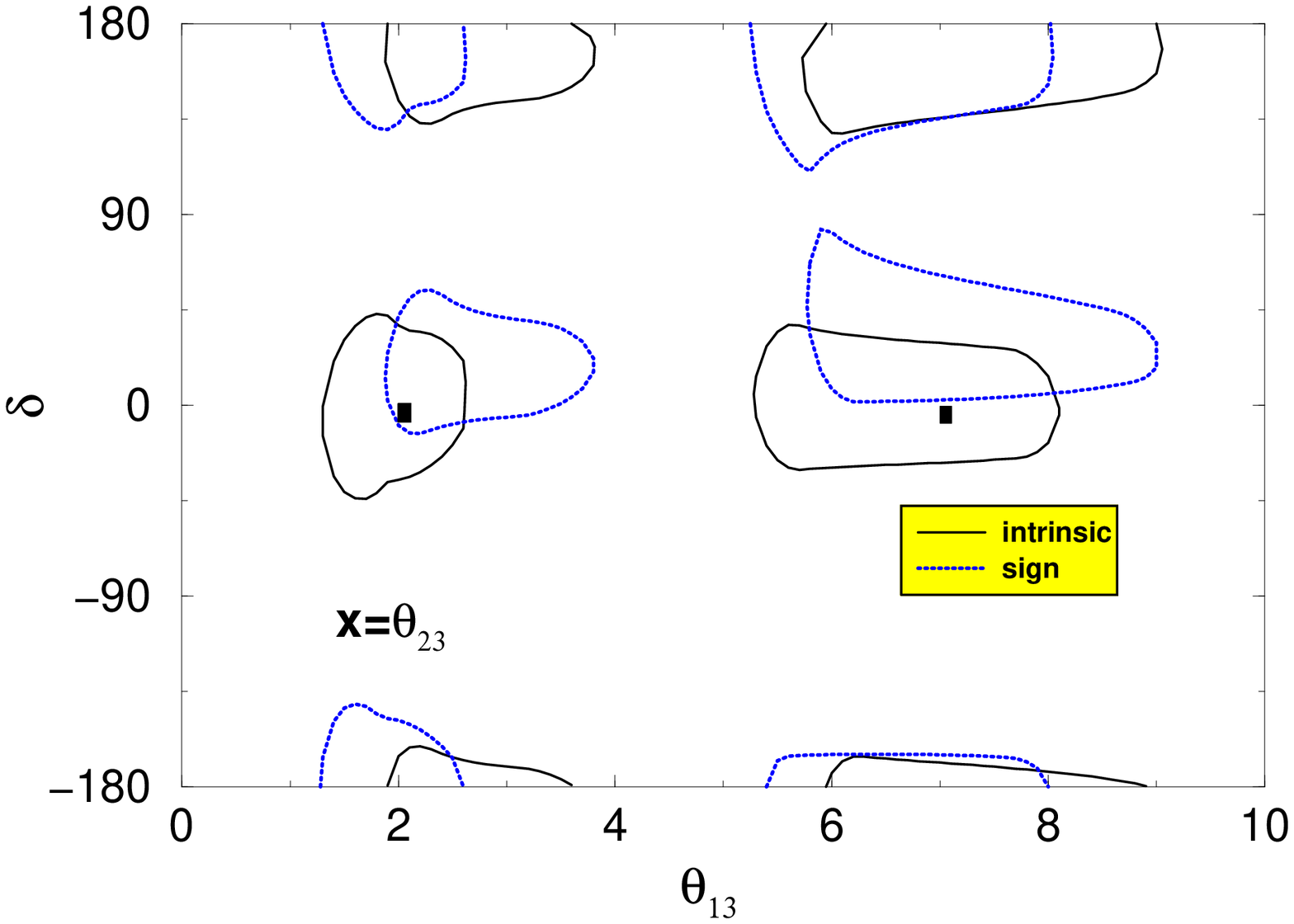} & 
\hspace{-1cm} 
\epsfxsize9cm\epsffile{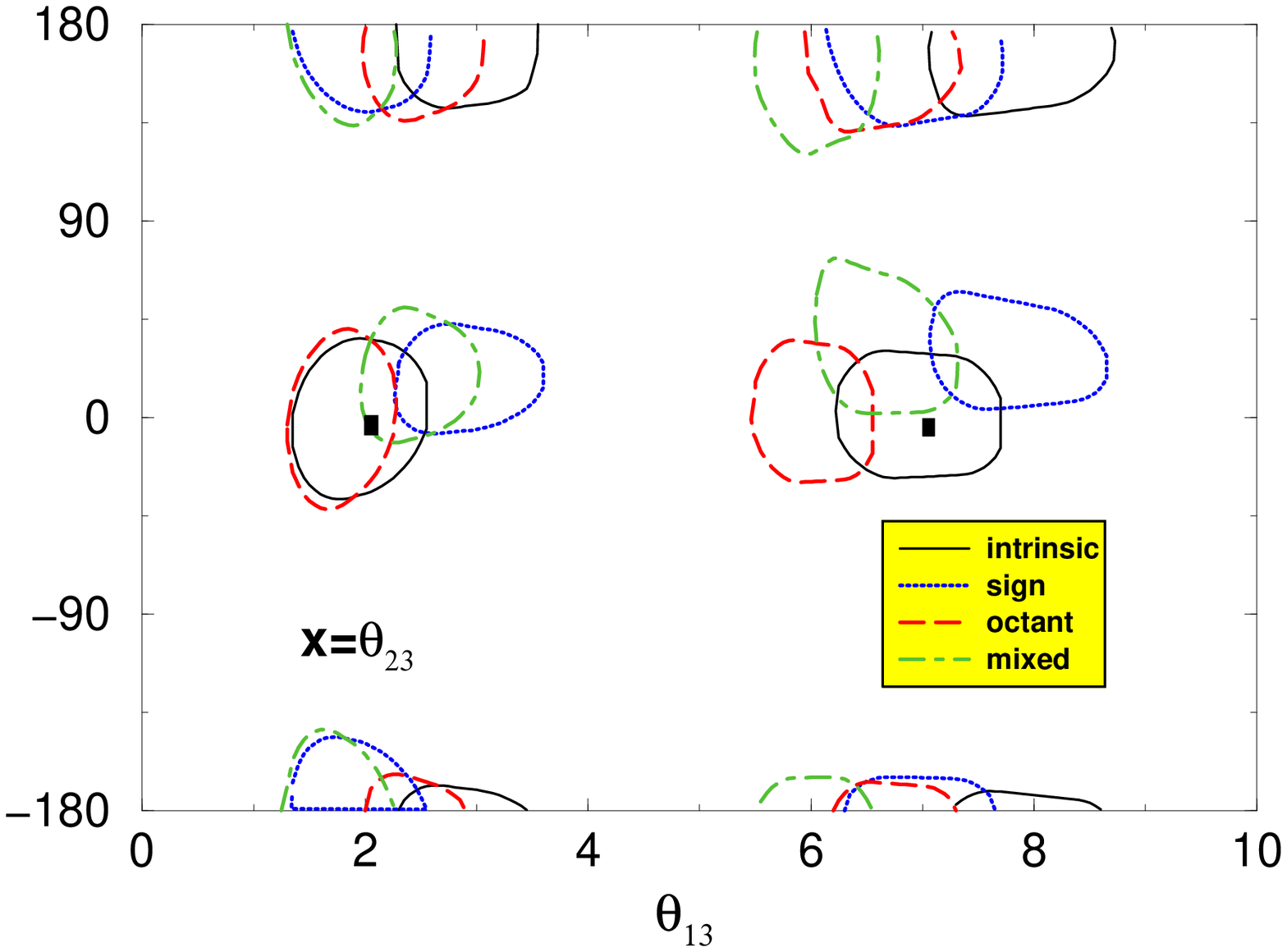} \\
\hspace{-2.5cm} 
\epsfxsize9cm\epsffile{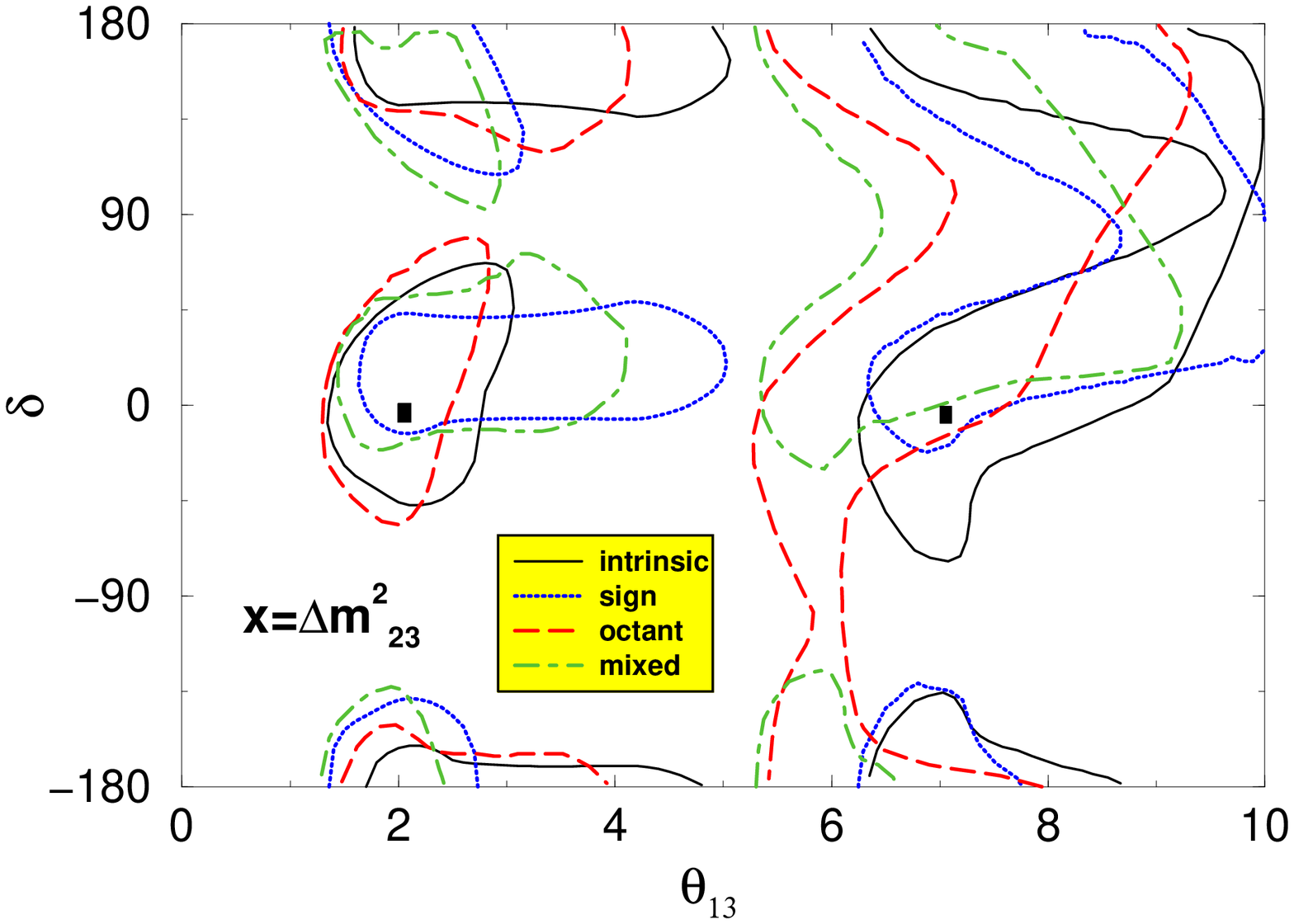} &
\hspace{-1cm} 
\epsfxsize9cm\epsffile{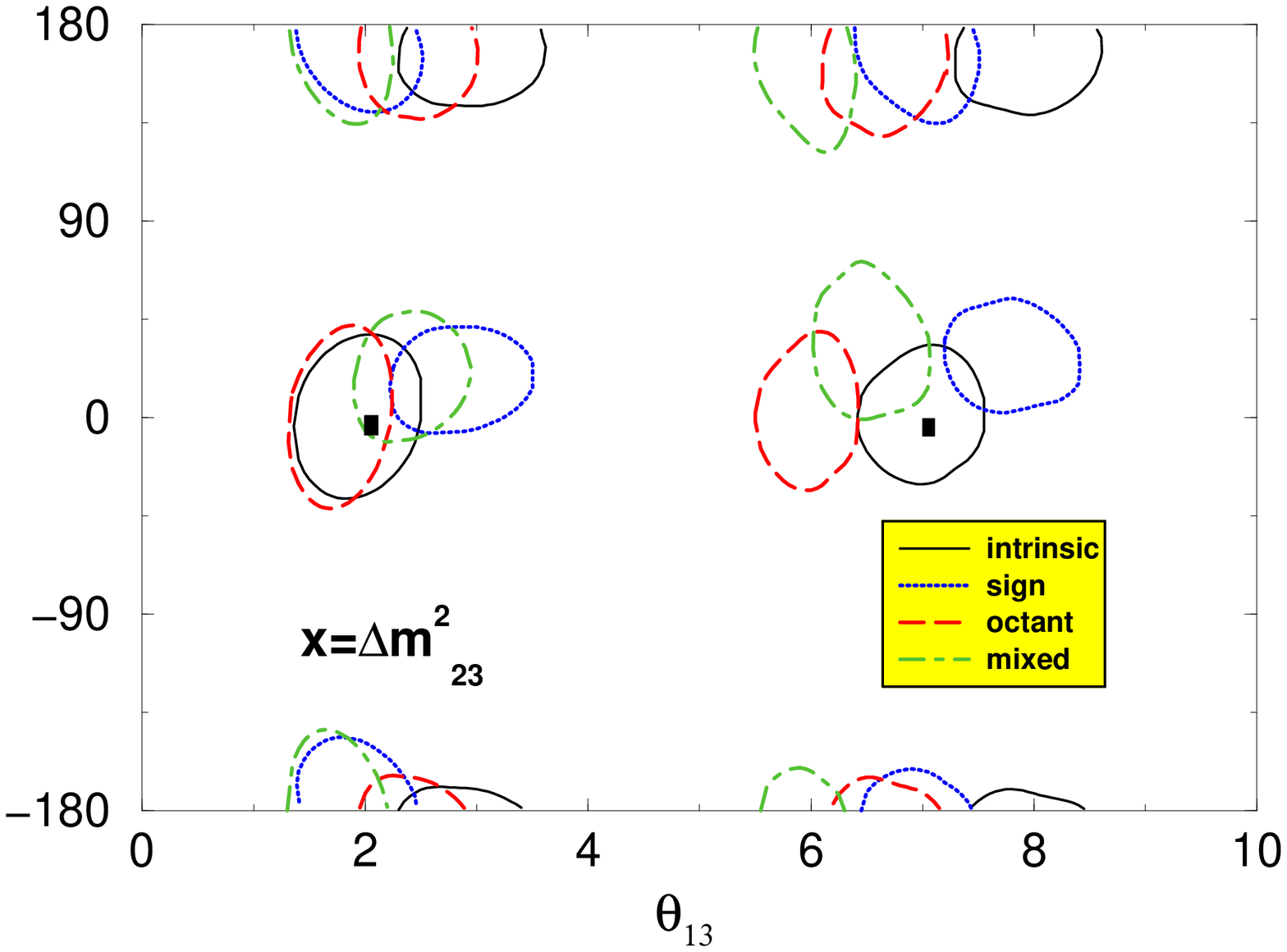} \\
\end{tabular}
\caption{\it Three-parameters 90 \% CL contours after a 10 years run at the $\beta$-Beam. 
Input parameters: $\bar \theta_{13} = 2^\circ, 7^\circ$; $\bar \delta = 0^\circ$.
Top panels: $x = \theta_{23}$; bottom panels: $x = \Delta m^2_{23}$. 
Left panels: present uncertaintes; right panels: after T2K- I.}
\label{fig:atmo2_00}
\end{center}
\end{figure}

\begin{figure}[h!]
\vspace{-0.35cm}
\begin{center}
\begin{tabular}{c c}
\hspace{-2cm} 
\epsfxsize9cm\epsffile{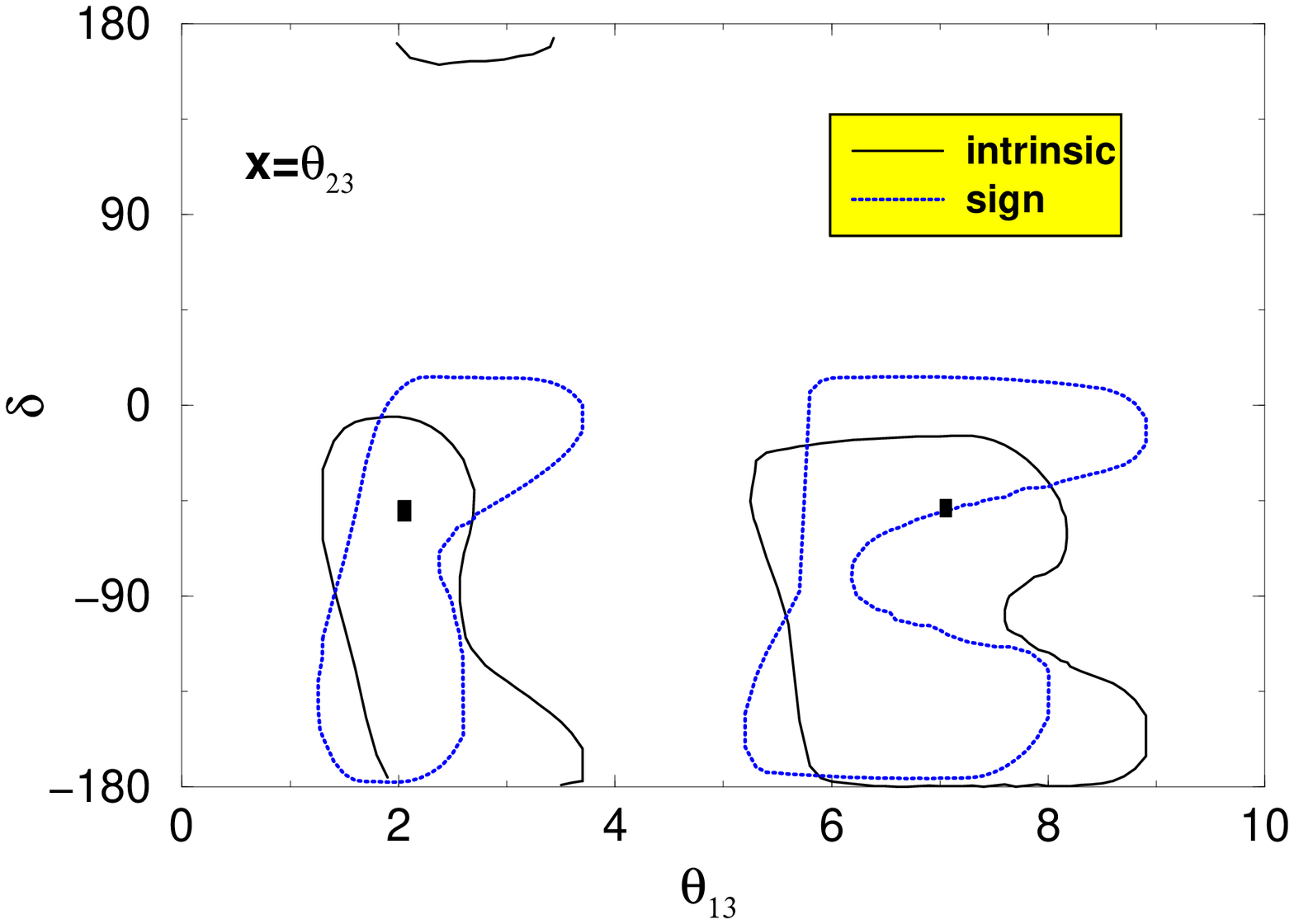} & 
\hspace{-1cm} 
\epsfxsize9cm\epsffile{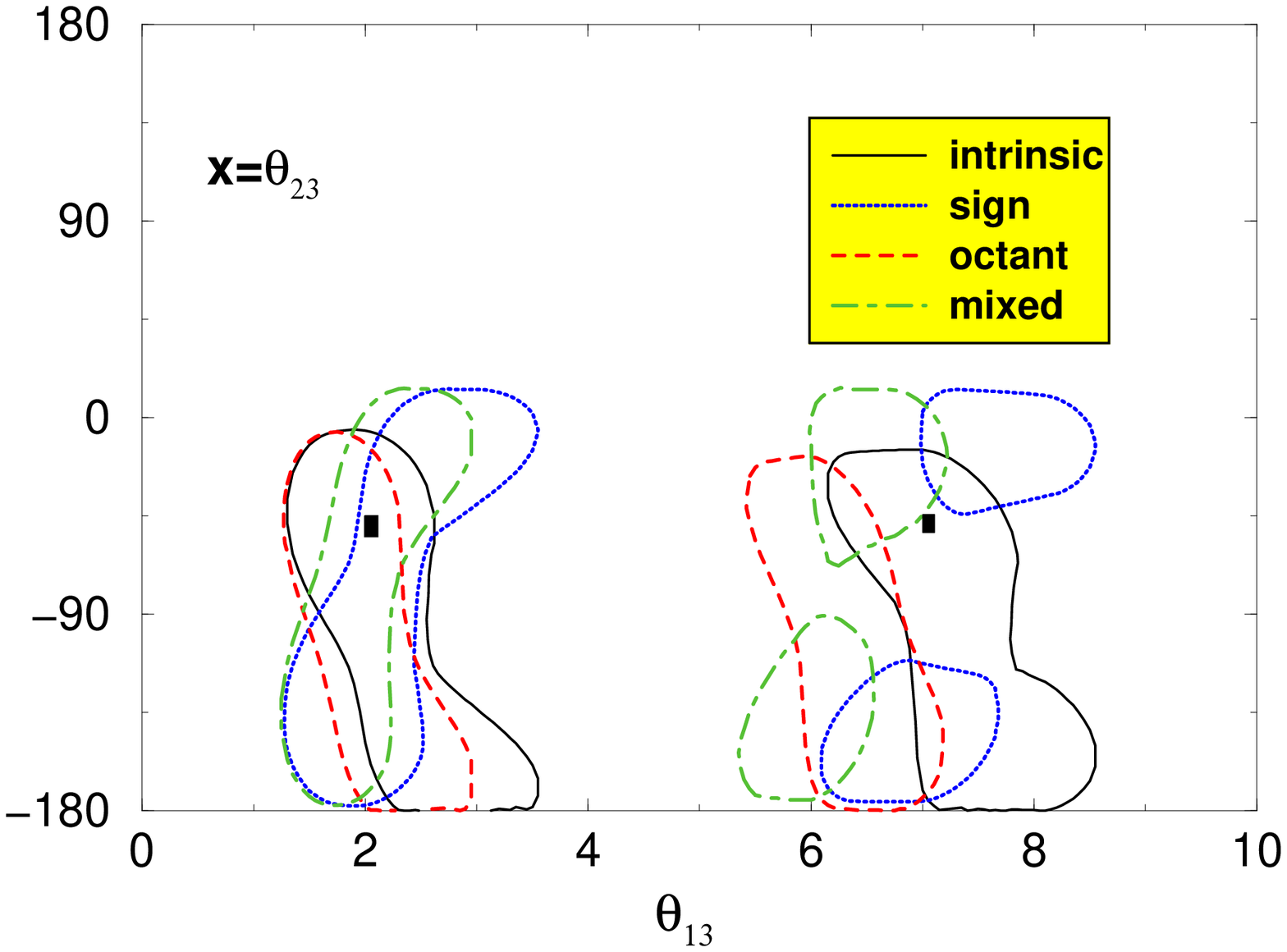} \\
\hspace{-2.5cm} 
\epsfxsize9cm\epsffile{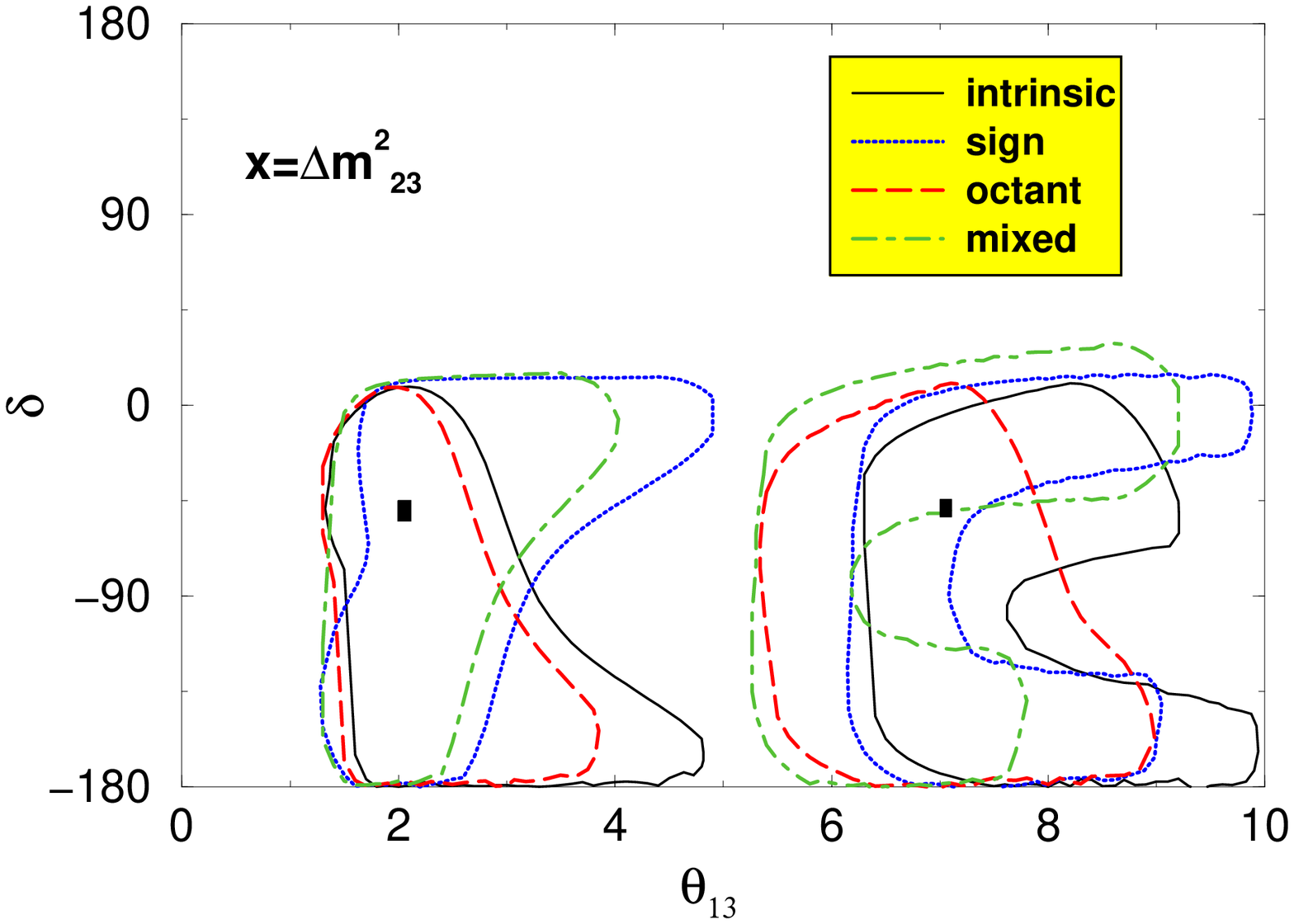} &
\hspace{-1cm} 
\epsfxsize9cm\epsffile{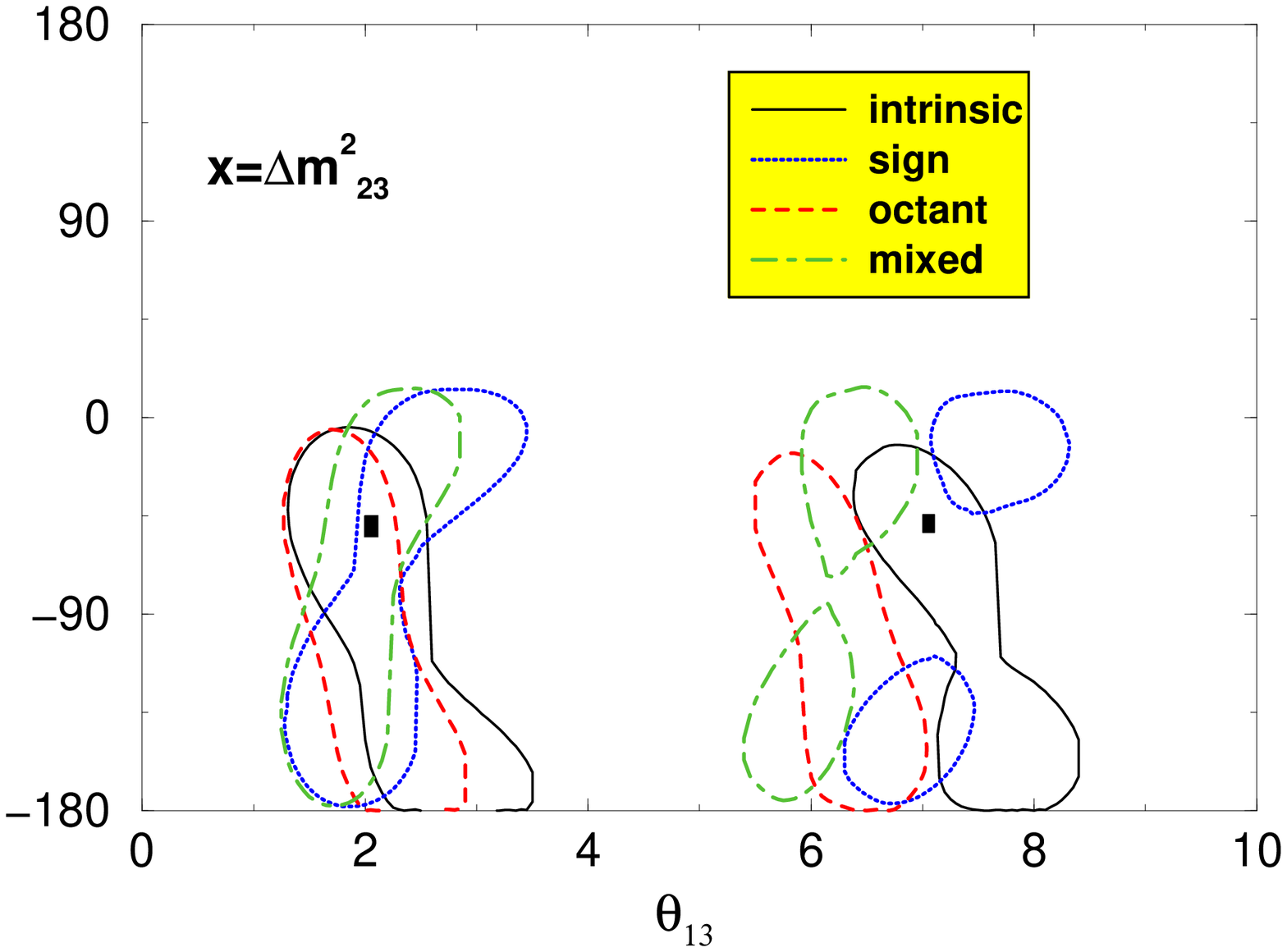} \\
\end{tabular}
\caption{\it Three-parameters 90 \% CL contours after a 10 years run at the $\beta$-Beam. 
Input parameters: $\bar \theta_{13} = 2^\circ, 7^\circ$; $\bar \delta = -45^\circ$.
Top panels: $x = \theta_{23}$; bottom panels: $x = \Delta m^2_{23}$. 
Left panels: present uncertaintes; right panels: after T2K-I.}
\label{fig:atmo2_m5}
\end{center}
\end{figure}

\begin{figure}[h!]
\vspace{-0.35cm}
\begin{center}
\begin{tabular}{c c}
\hspace{-2cm} 
\epsfxsize9cm\epsffile{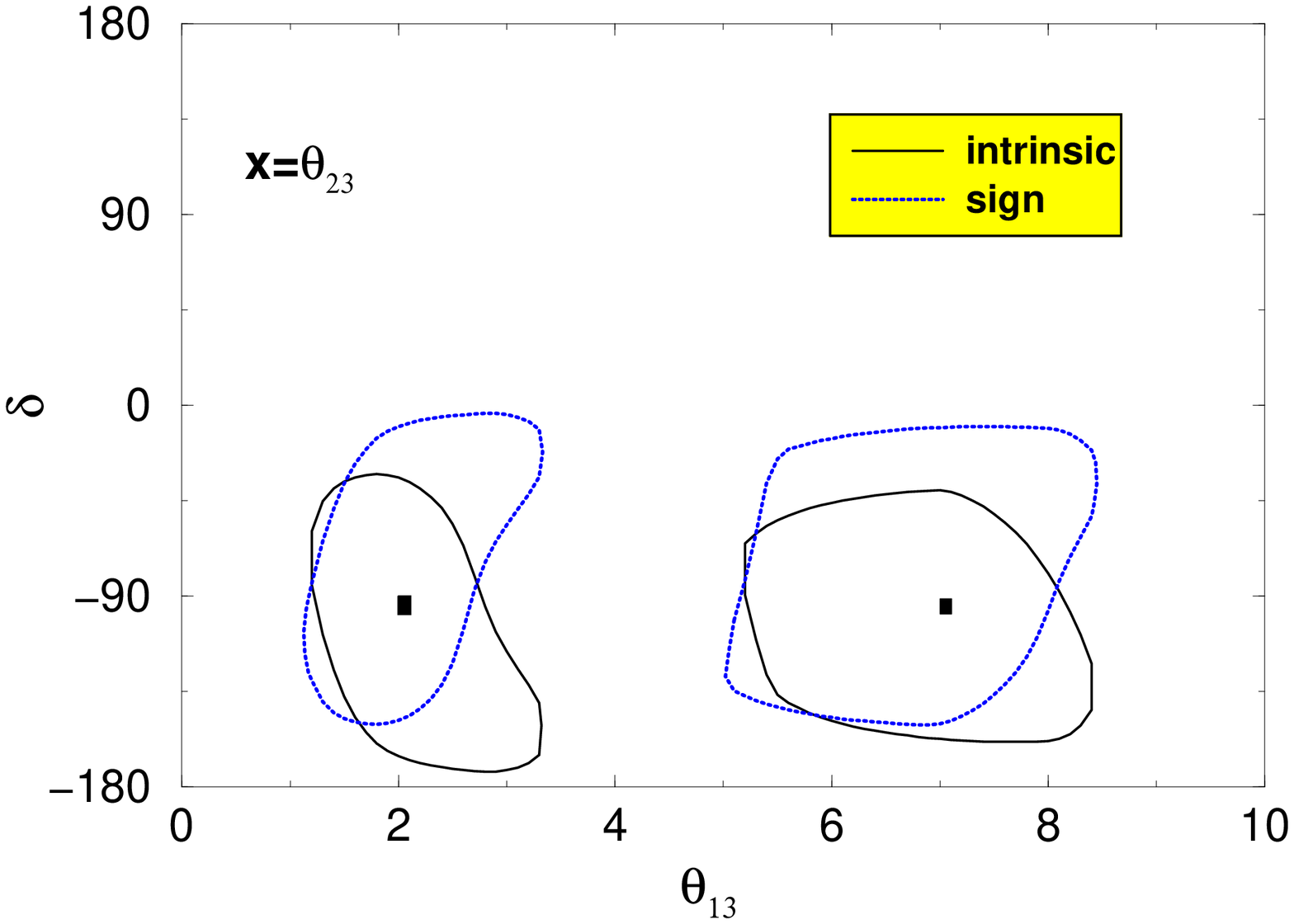} & 
\hspace{-1cm} 
\epsfxsize9cm\epsffile{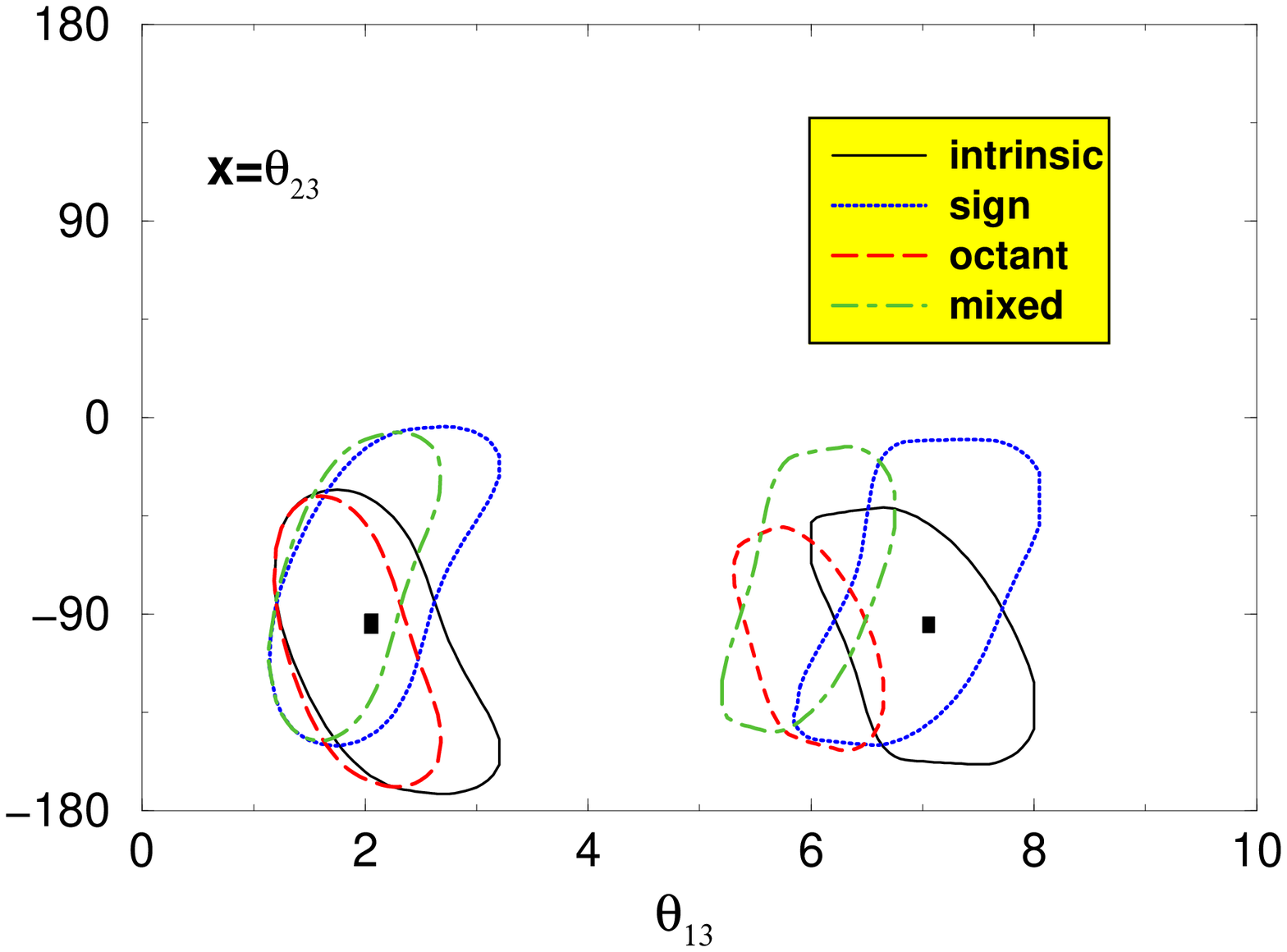} \\
\hspace{-2.5cm} 
\epsfxsize9cm\epsffile{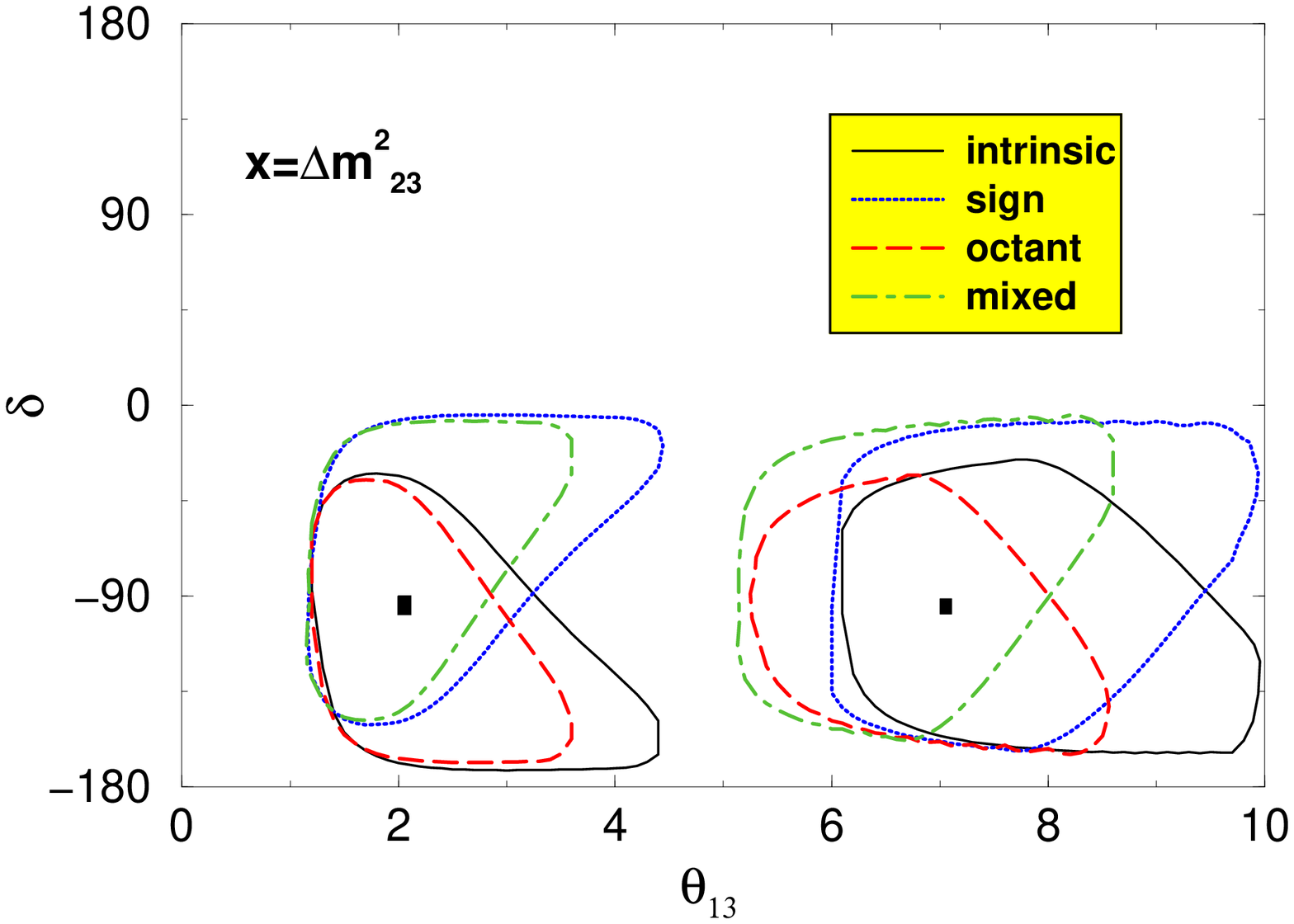} &
\hspace{-1cm} 
\epsfxsize9cm\epsffile{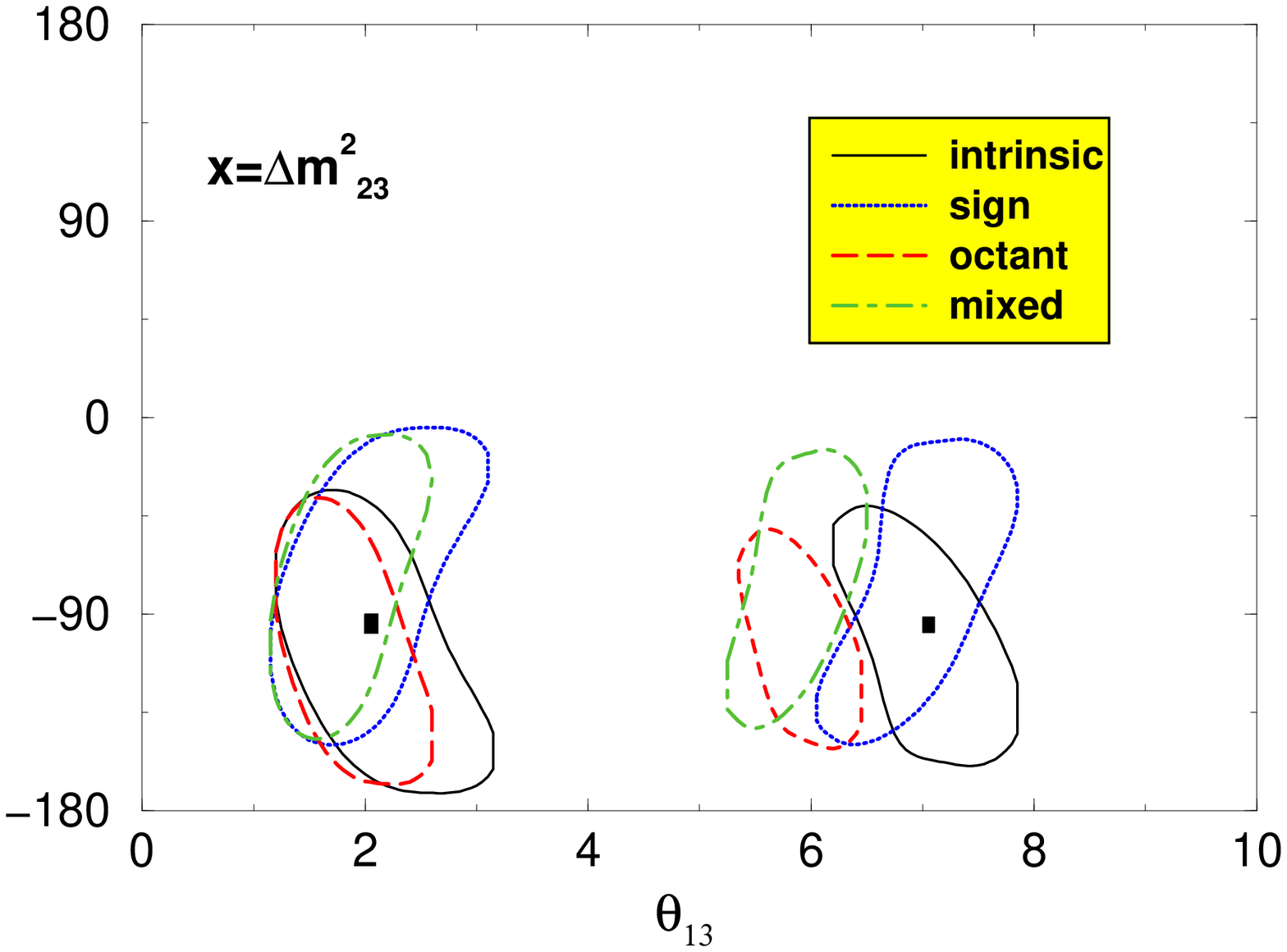} \\
\end{tabular}
\caption{\it Three-parameters 90 \% CL contours after a 10 years run at the $\beta$-Beam. 
Input parameters: $\bar \theta_{13} = 2^\circ, 7^\circ$; $\bar \delta = -90^\circ$.
Top panels: $x = \theta_{23}$; bottom panels: $x = \Delta m^2_{23}$. 
Left panels: present uncertaintes; right panels: after T2K-I.}
\label{fig:atmo2_m0}
\end{center}
\end{figure}


\begin{figure}[h!]
\vspace{-0.35cm}
\begin{center}
\begin{tabular}{c c}
\hspace{-2cm} 
\epsfxsize9cm\epsffile{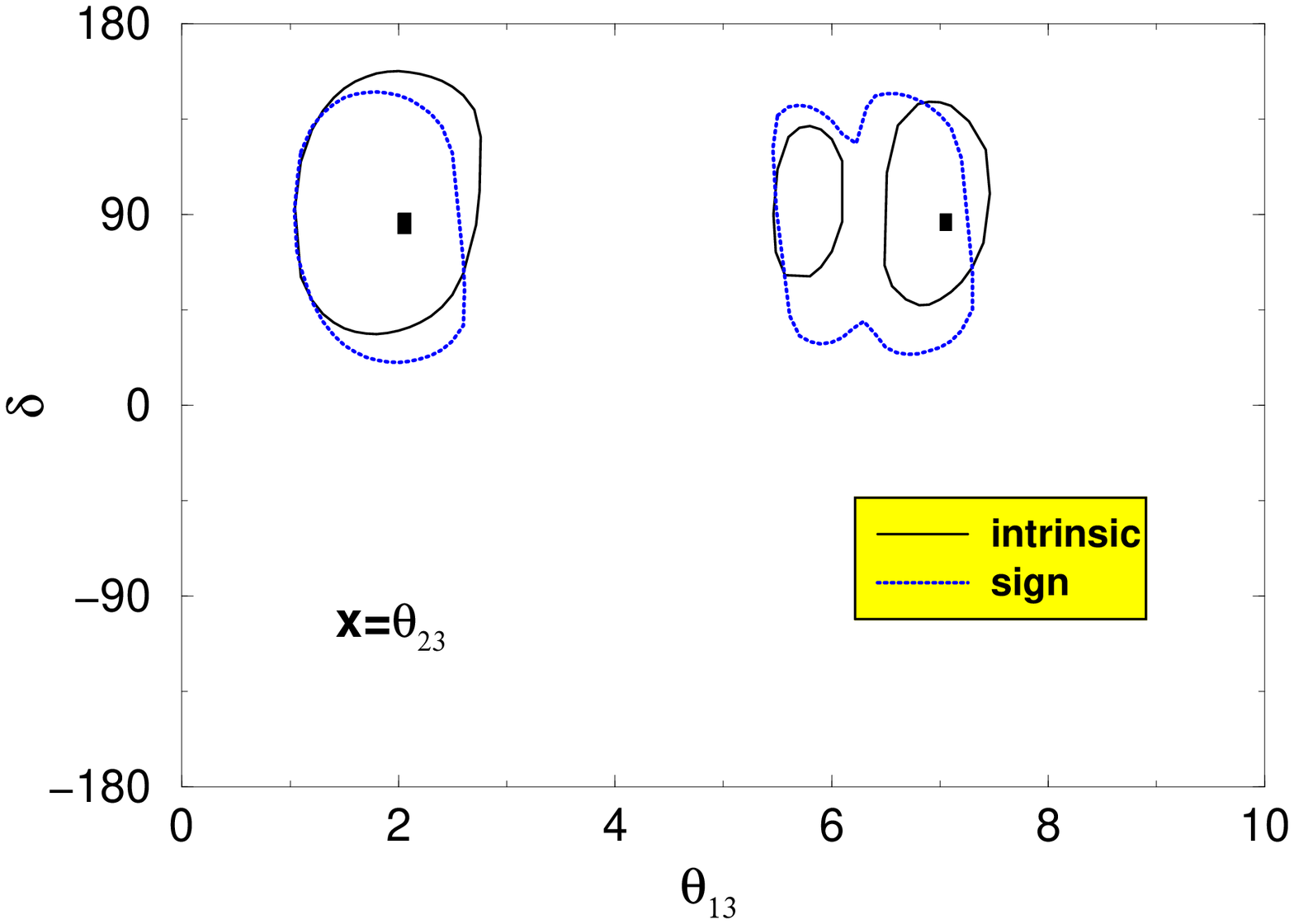} & 
\epsfxsize9cm\epsffile{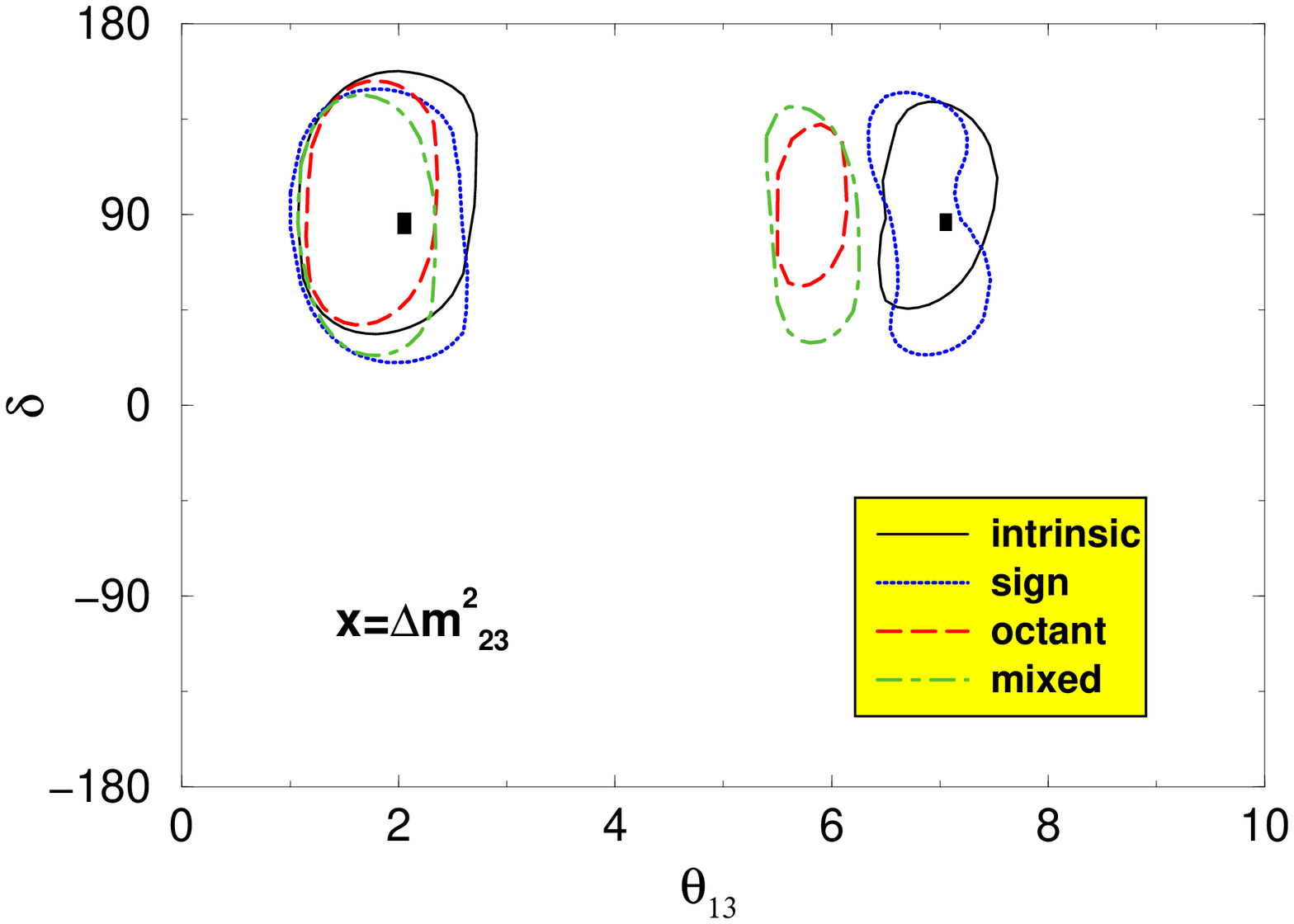} \\
\end{tabular}
\caption{\it Three-parameters 90 \% CL contours after a 2+8 years run at the Super-Beam. 
Input parameters: $\bar \theta_{13} = 2^\circ, 7^\circ$; $\bar \delta = 90^\circ$.
Left: $x = \theta_{23}$; right: $x = \Delta m^2_{23}$.}
\label{fig:atmoSB2_90}
\end{center}
\end{figure}

\begin{figure}[h!]
\vspace{-0.35cm}
\begin{center}
\begin{tabular}{c c}
\hspace{-2cm} 
\epsfxsize9cm\epsffile{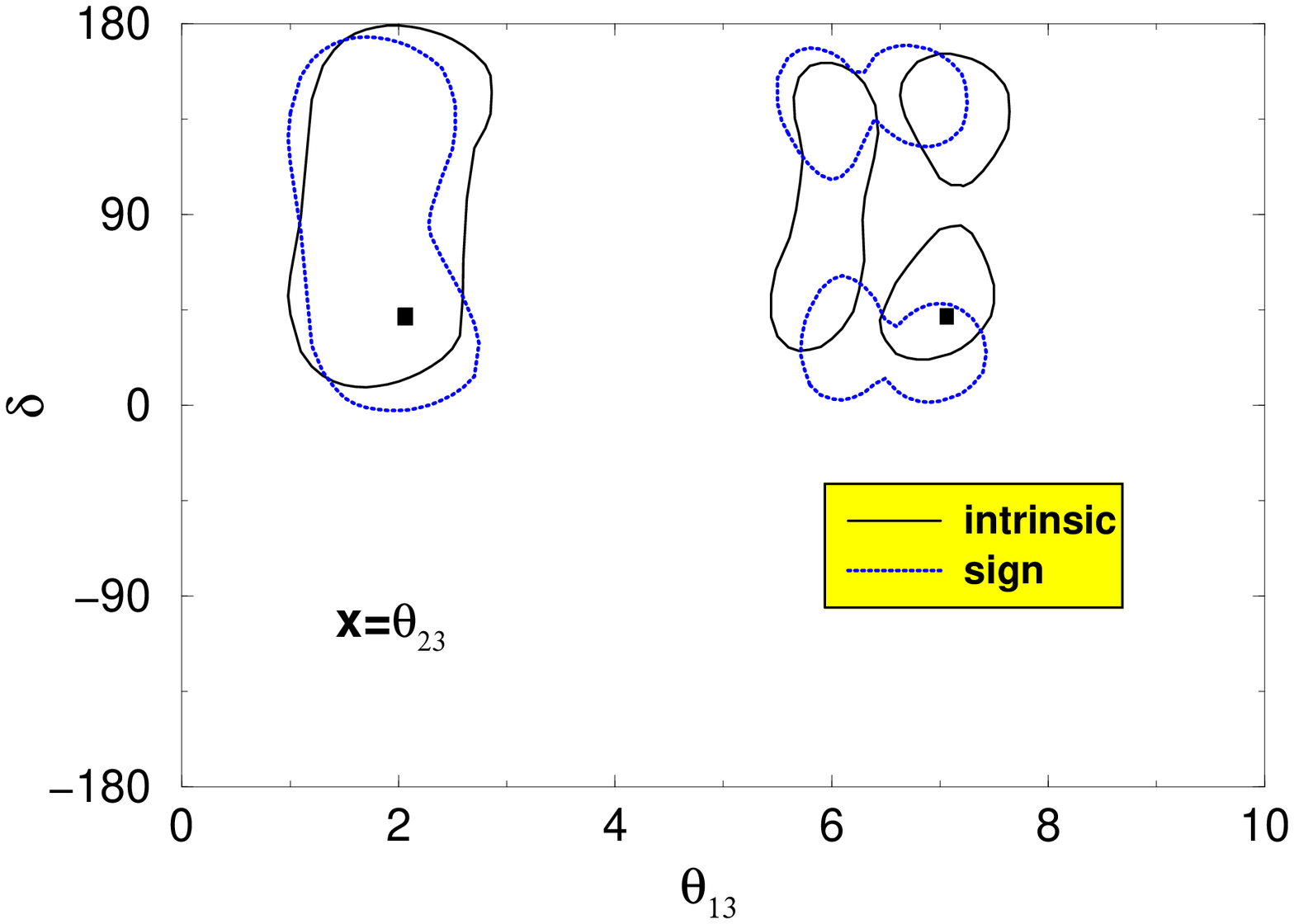} & 
\epsfxsize9cm\epsffile{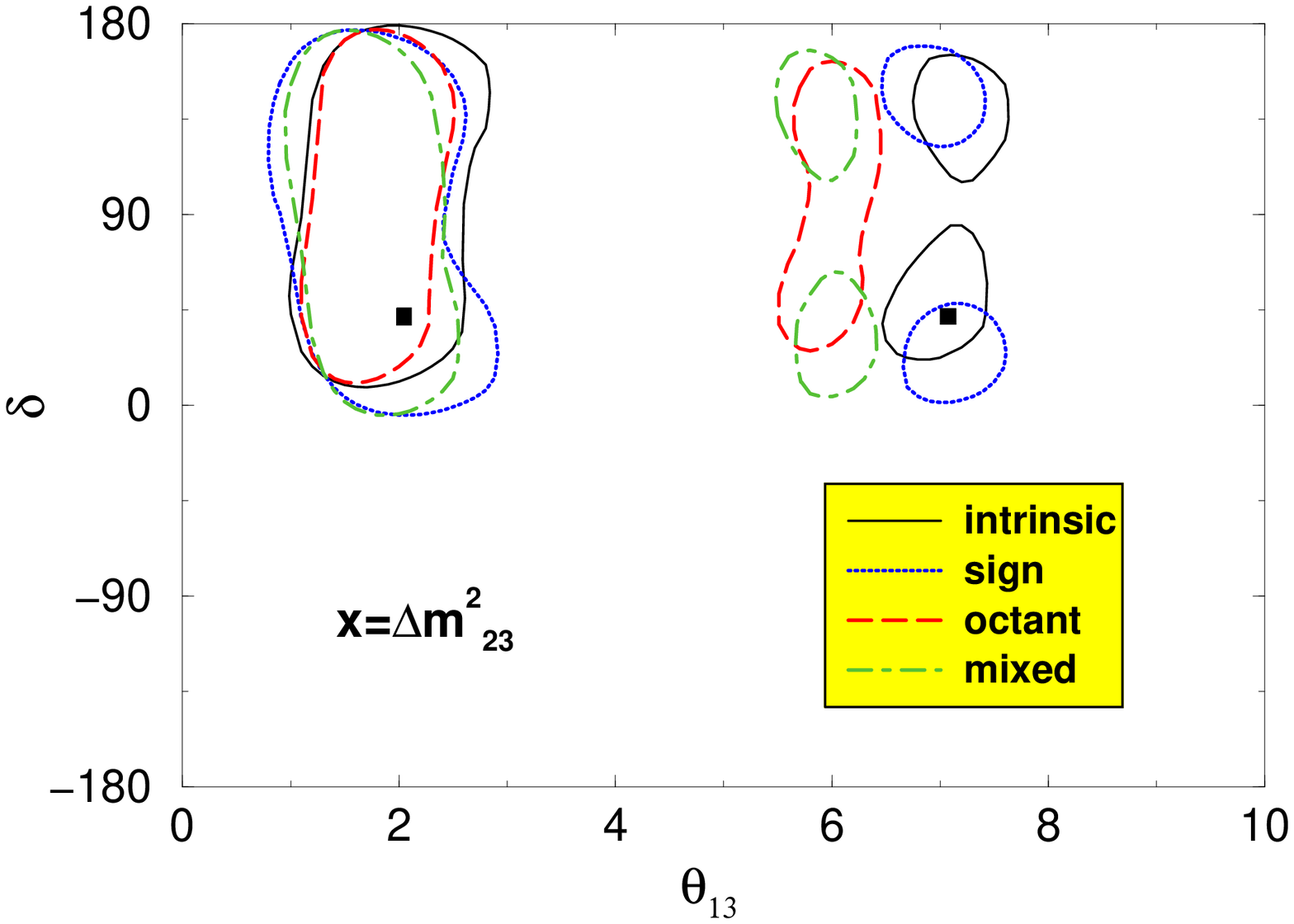} \\
\end{tabular}
\caption{\it Three-parameters 90 \% CL contours after a 2+8 years run at the Super-Beam. 
Input parameters: $\bar \theta_{13} = 2^\circ, 7^\circ$; $\bar \delta = 45^\circ$.
Left: $x = \theta_{23}$; right: $x = \Delta m^2_{23}$.}
\label{fig:atmoSB2}
\end{center}
\end{figure}

\begin{figure}[h!]
\vspace{-0.35cm}
\begin{center}
\begin{tabular}{c c}
\hspace{-2cm} 
\epsfxsize9cm\epsffile{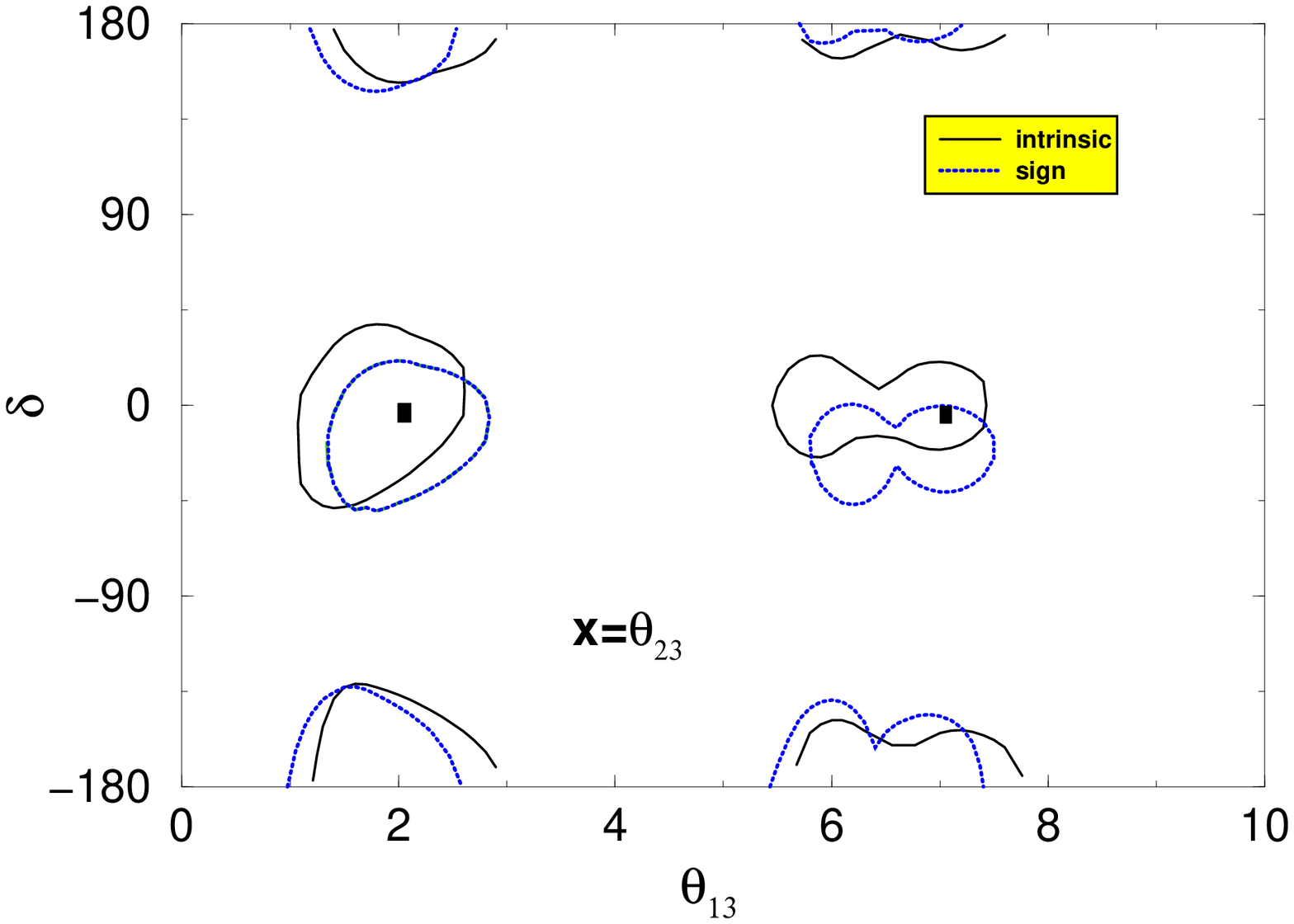} & 
\epsfxsize9cm\epsffile{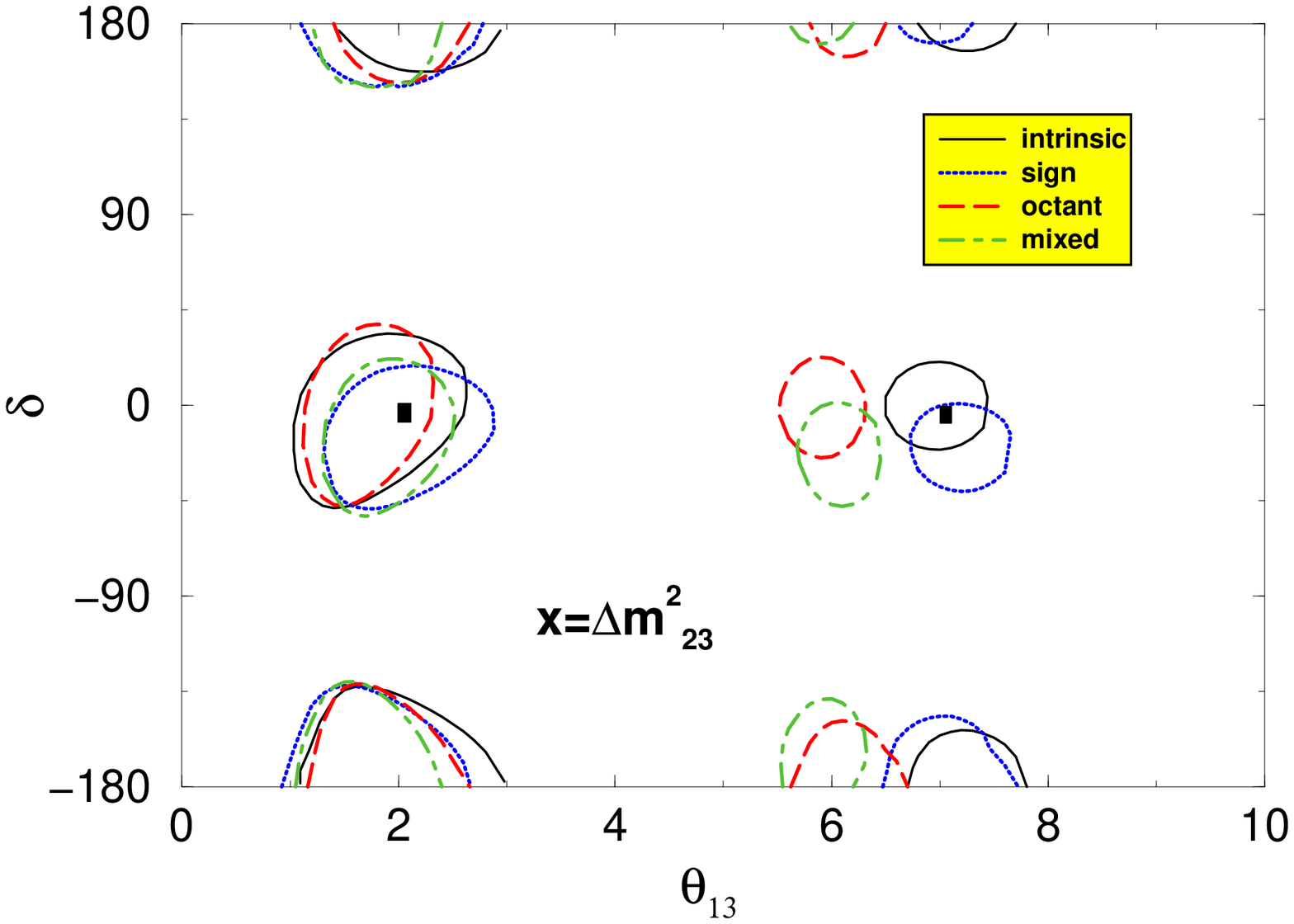} \\
\end{tabular}
\caption{\it Three-parameters 90 \% CL contours after a 2+8 years run at the Super-Beam. 
Input parameters: $\bar \theta_{13} = 2^\circ, 7^\circ$; $\bar \delta = 0^\circ$.
Left: $x = \theta_{23}$; right: $x = \Delta m^2_{23}$.}
\label{fig:atmoSB2_00}
\end{center}
\end{figure}

\begin{figure}[h!]
\vspace{-0.35cm}
\begin{center}
\begin{tabular}{c c}
\hspace{-2cm} 
\epsfxsize9cm\epsffile{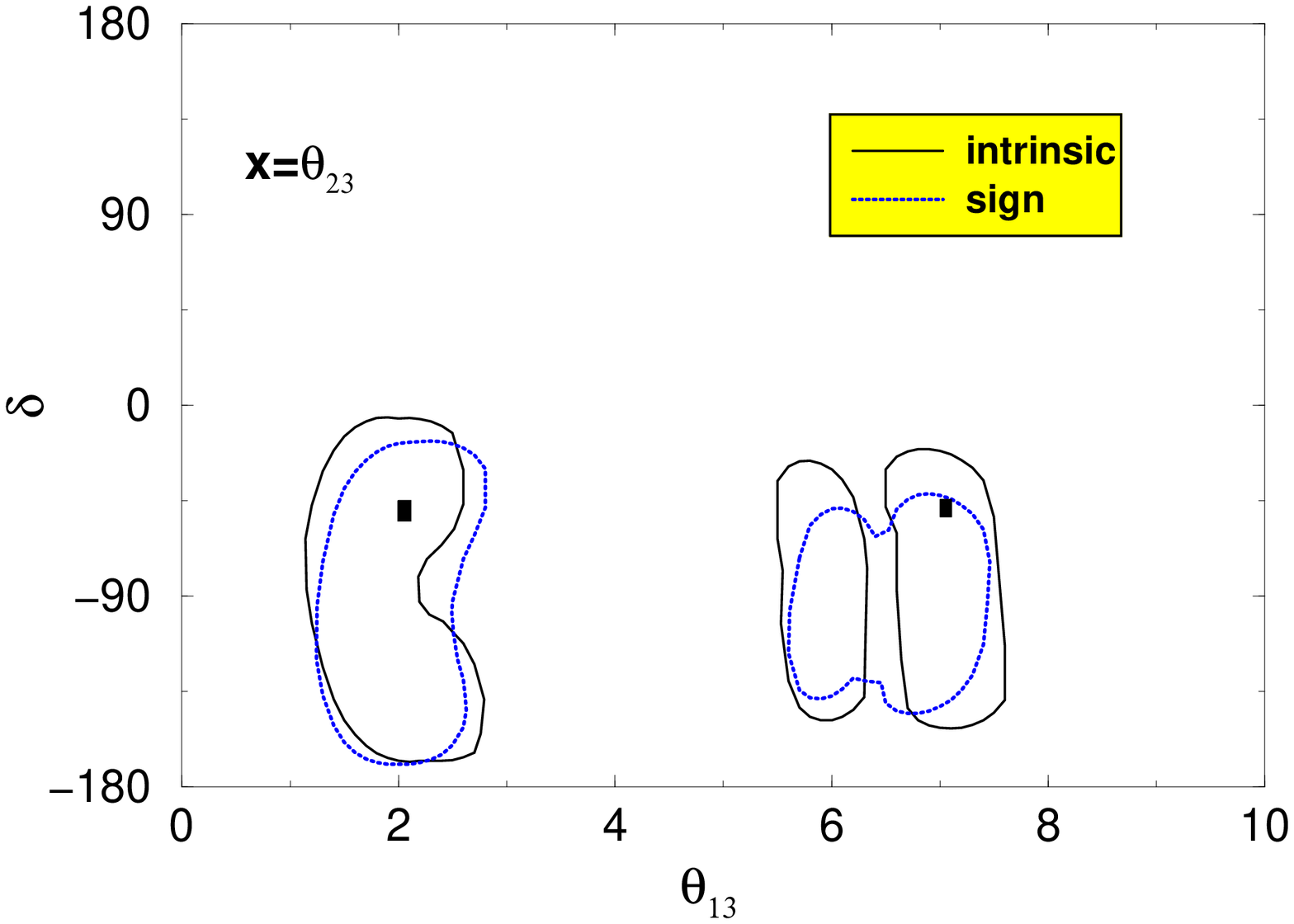} & 
\epsfxsize9cm\epsffile{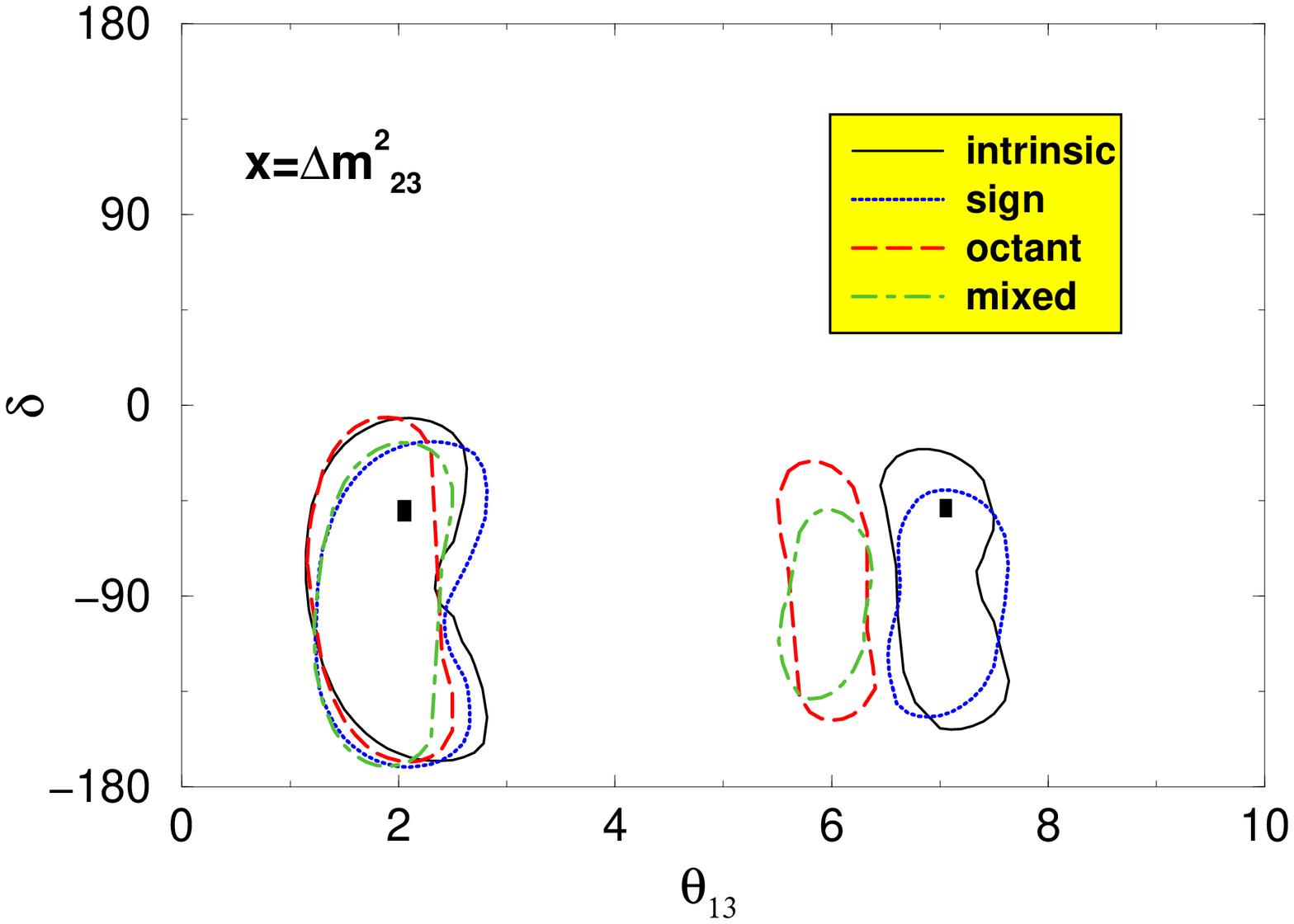} \\
\end{tabular}
\caption{\it Three-parameters 90 \% CL contours after a 2+8 years run at the Super-Beam. 
Input parameters: $\bar \theta_{13} = 2^\circ, 7^\circ$; $\bar \delta = -45^\circ$.
Left: $x = \theta_{23}$; right: $x = \Delta m^2_{23}$.}
\label{fig:atmoSB2_m5}
\end{center}
\end{figure}

\begin{figure}[h!]
\vspace{-0.35cm}
\begin{center}
\begin{tabular}{c c}
\hspace{-2cm} 
\epsfxsize9cm\epsffile{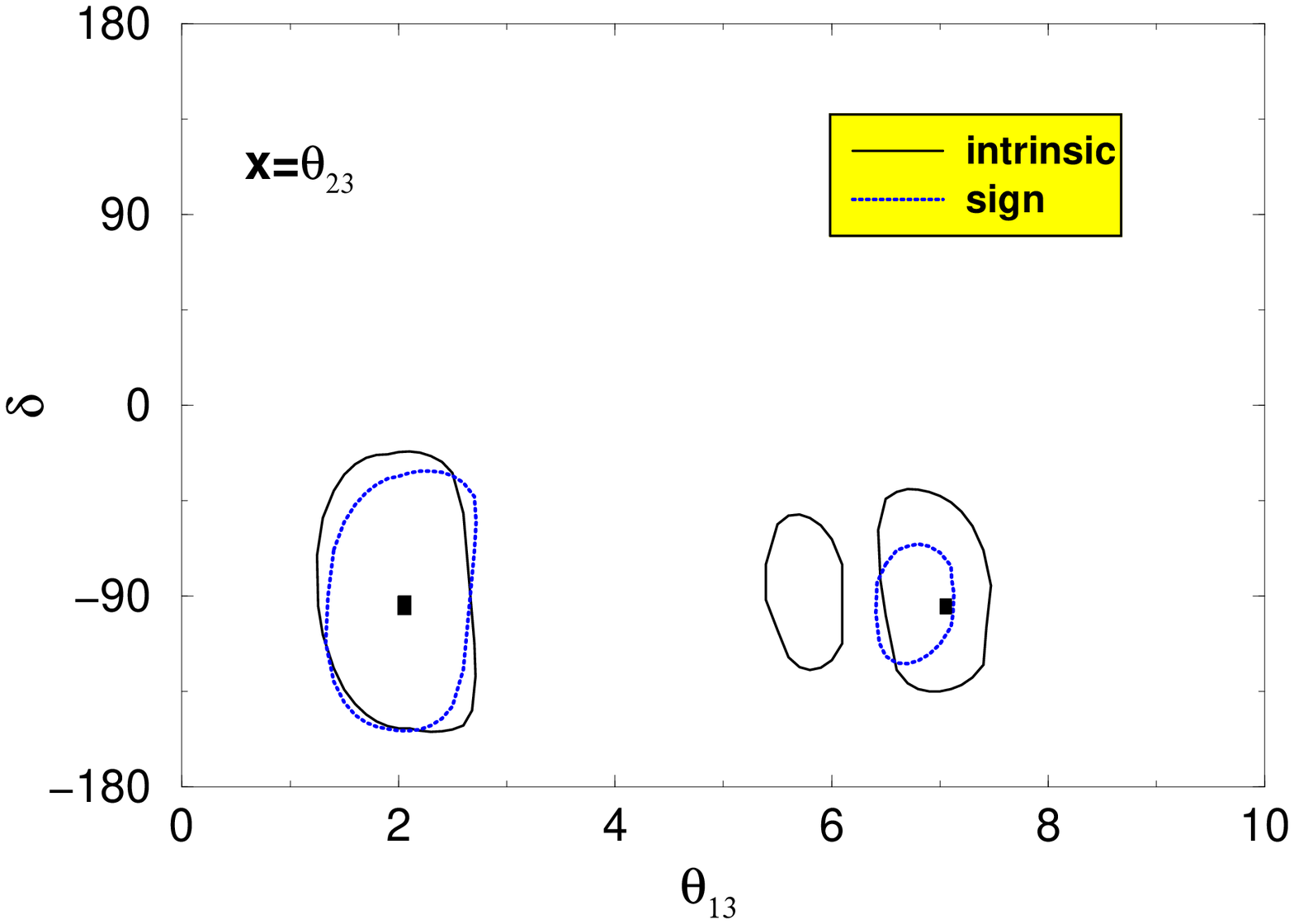} & 
\epsfxsize9cm\epsffile{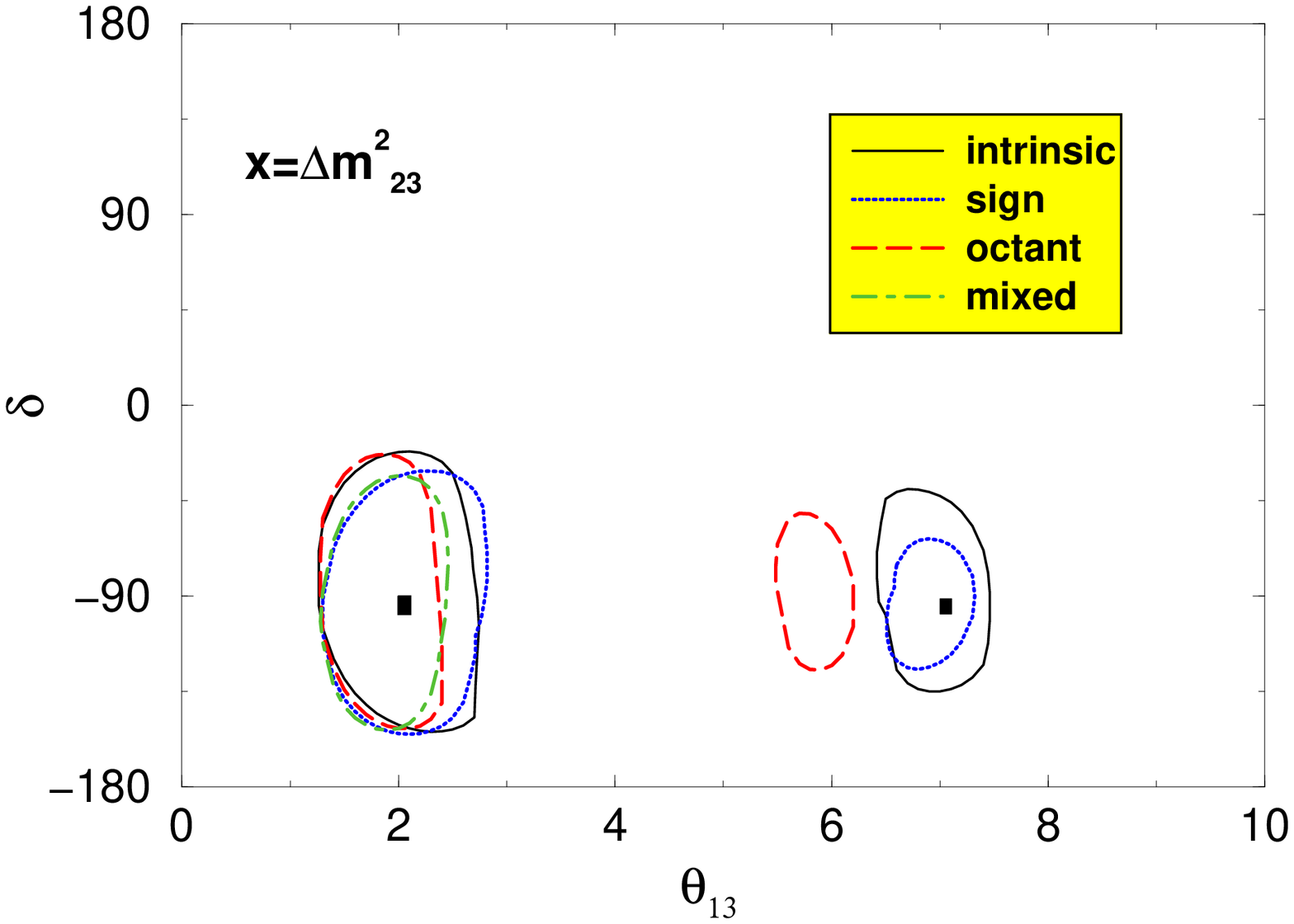} \\
\end{tabular}
\caption{\it Three-parameters 90 \% CL contours after a 2+8 years run at the Super-Beam. 
Input parameters: $\bar \theta_{13} = 2^\circ, 7^\circ$; $\bar \delta = -90^\circ$.
Left: $x = \theta_{23}$; right: $x = \Delta m^2_{23}$.}
\label{fig:atmoSB2_m0}
\end{center}
\end{figure}


\begin{figure}[h!]
\vspace{-0.35cm}
\begin{center}
\begin{tabular}{c c}
\hspace{-2cm} 
\epsfxsize9cm\epsffile{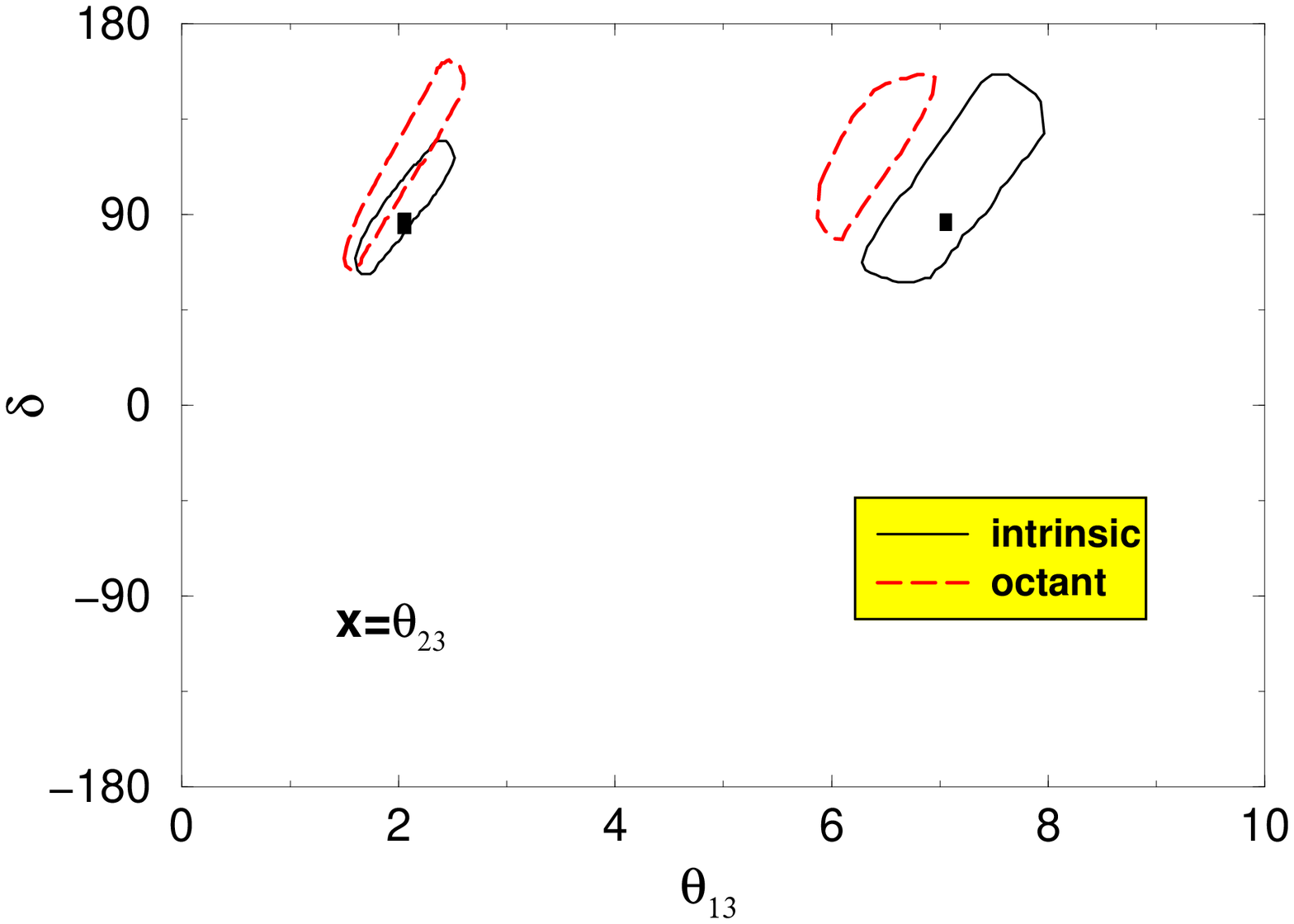} & 
\epsfxsize9cm\epsffile{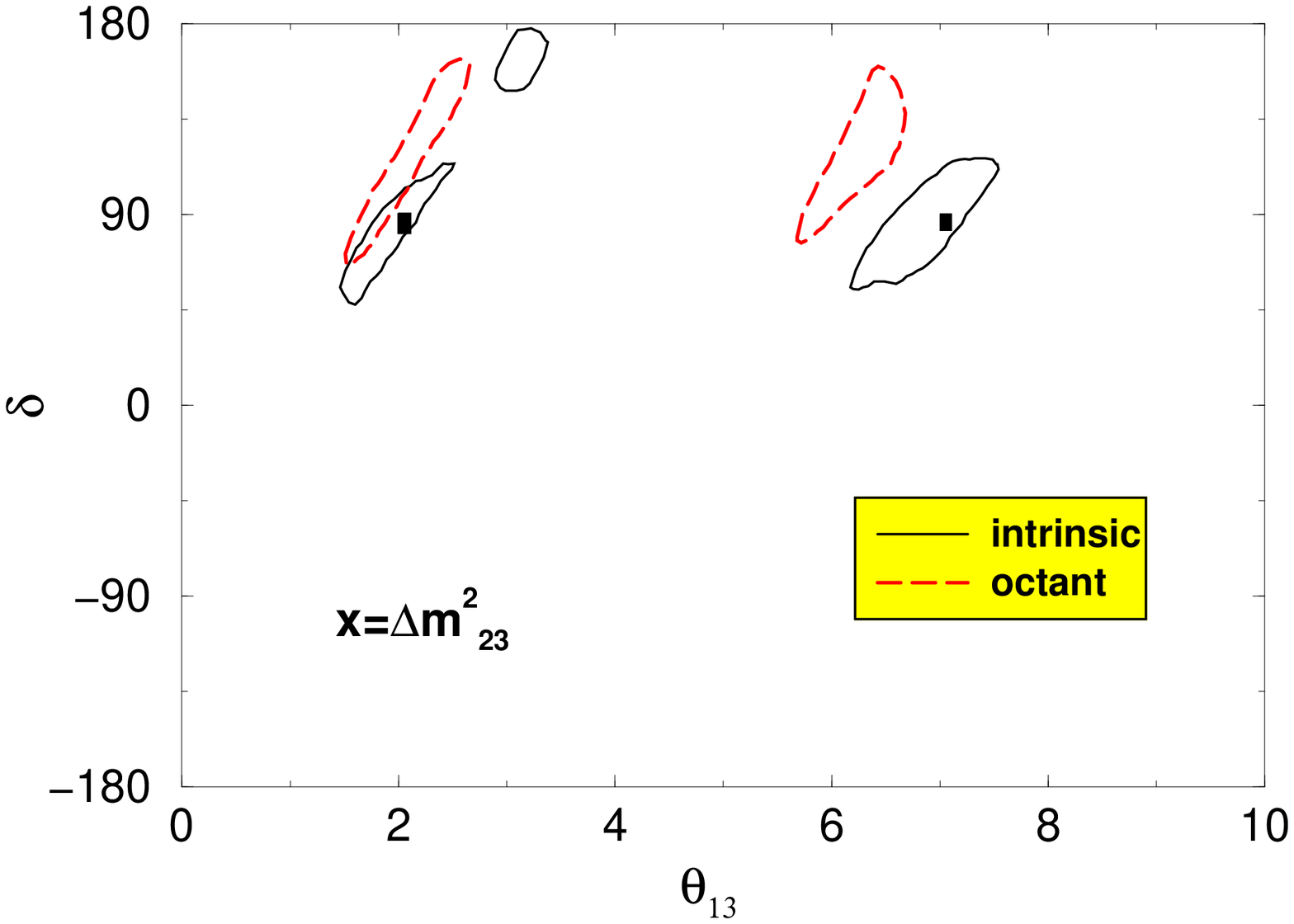} \\
\end{tabular}
\caption{\it Three-parameters 90 \% CL contours after a 5+5 years run at the Neutrino Factory. 
Input parameters: $\bar \theta_{13} = 2^\circ, 7^\circ$; $\bar \delta = 90^\circ$.
Left: $x = \theta_{23}$; right: $x = \Delta m^2_{23}$.}
\label{fig:atmoNF_90}
\end{center}
\end{figure}

\begin{figure}[h!]
\vspace{-0.35cm}
\begin{center}
\begin{tabular}{c c}
\hspace{-2cm} 
\epsfxsize9cm\epsffile{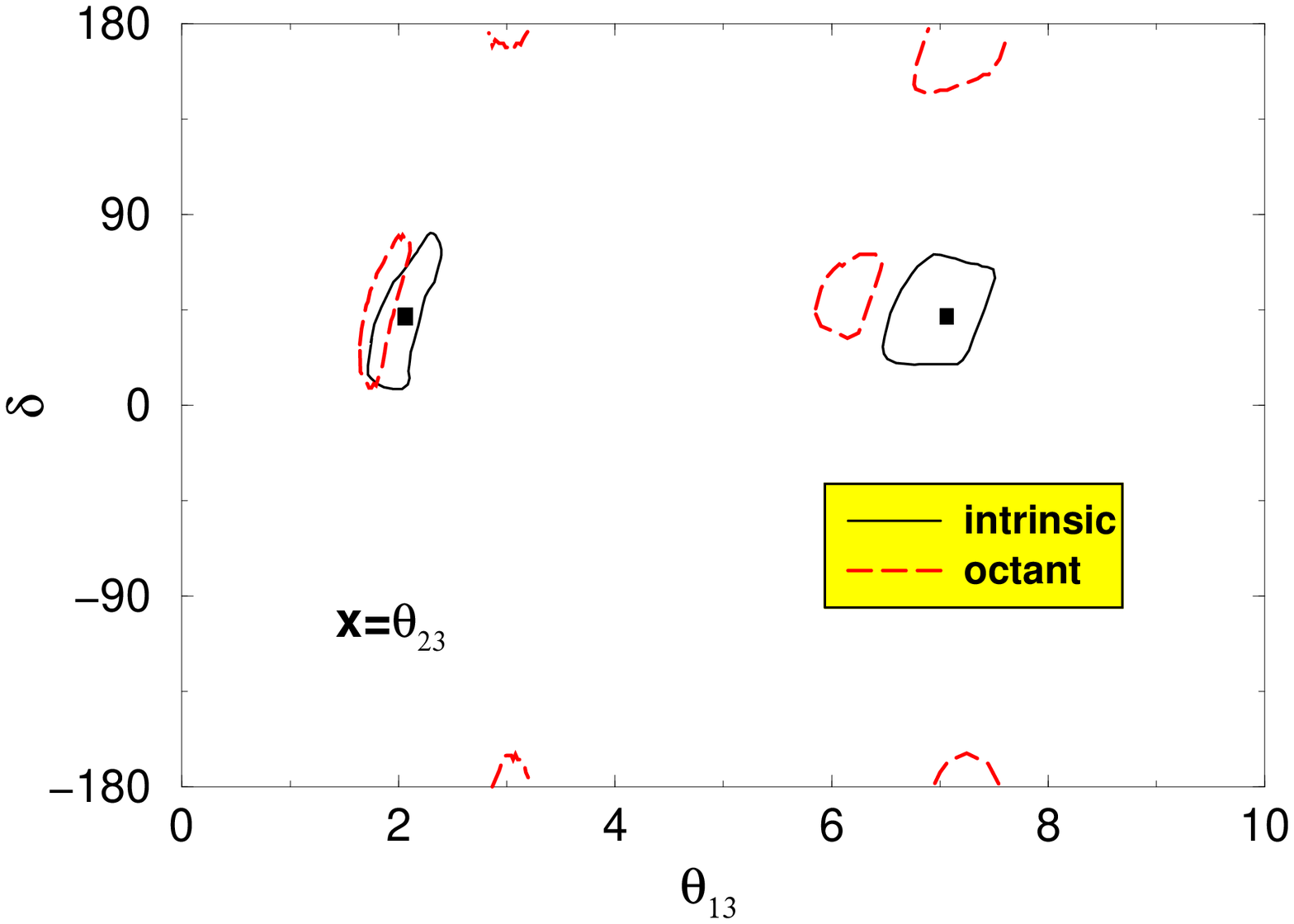} & 
\epsfxsize9cm\epsffile{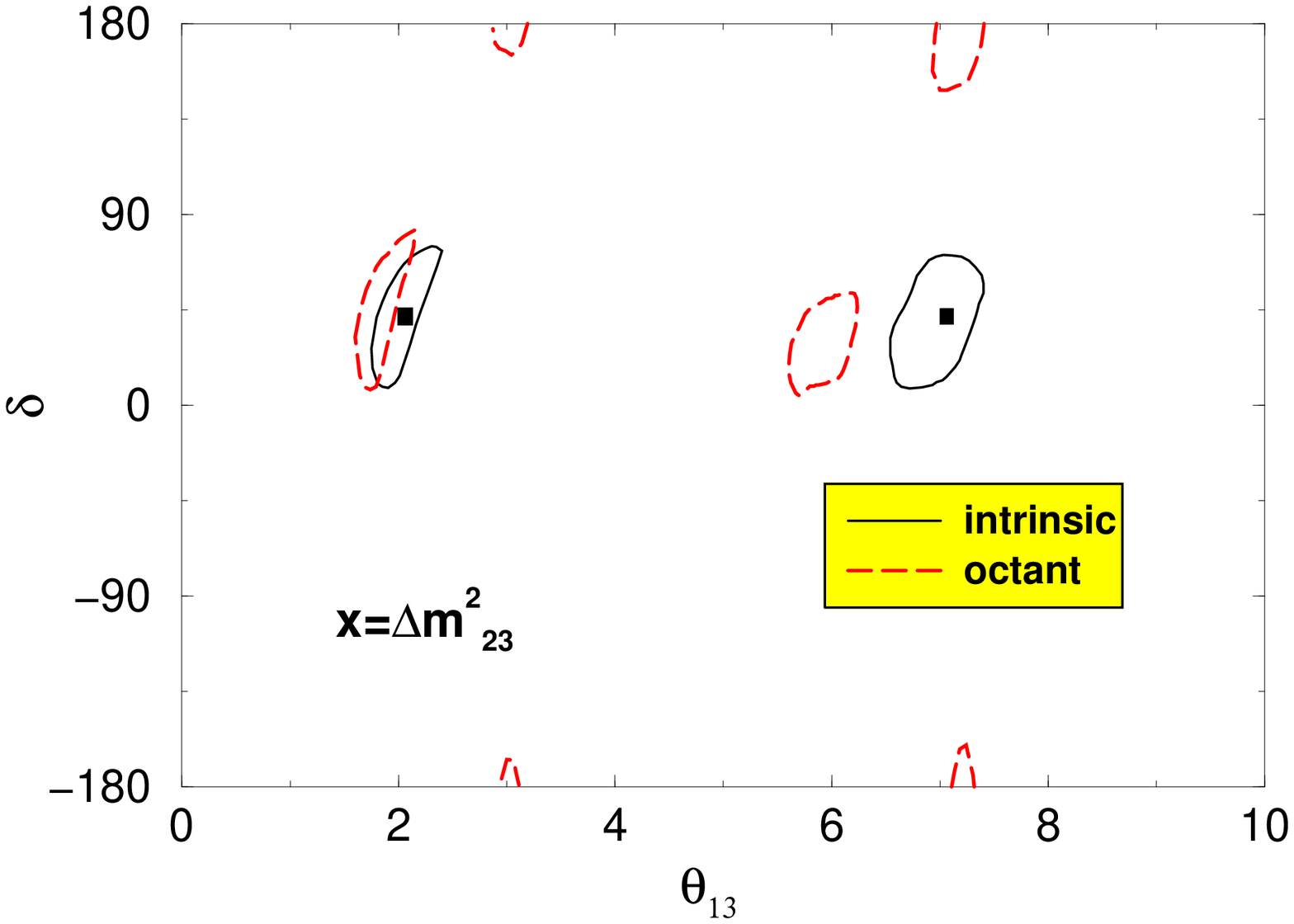} \\
\end{tabular}
\caption{\it Three-parameters 90 \% CL contours after a 5+5 years run at the Neutrino Factory. 
Input parameters: $\bar \theta_{13} = 2^\circ, 7^\circ$; $\bar \delta = 45^\circ$.
Left: $x = \theta_{23}$; right: $x = \Delta m^2_{23}$.}
\label{fig:atmoNF_45}
\end{center}
\end{figure}

\begin{figure}[h!]
\vspace{-0.35cm}
\begin{center}
\begin{tabular}{c c}
\hspace{-2cm} 
\epsfxsize9cm\epsffile{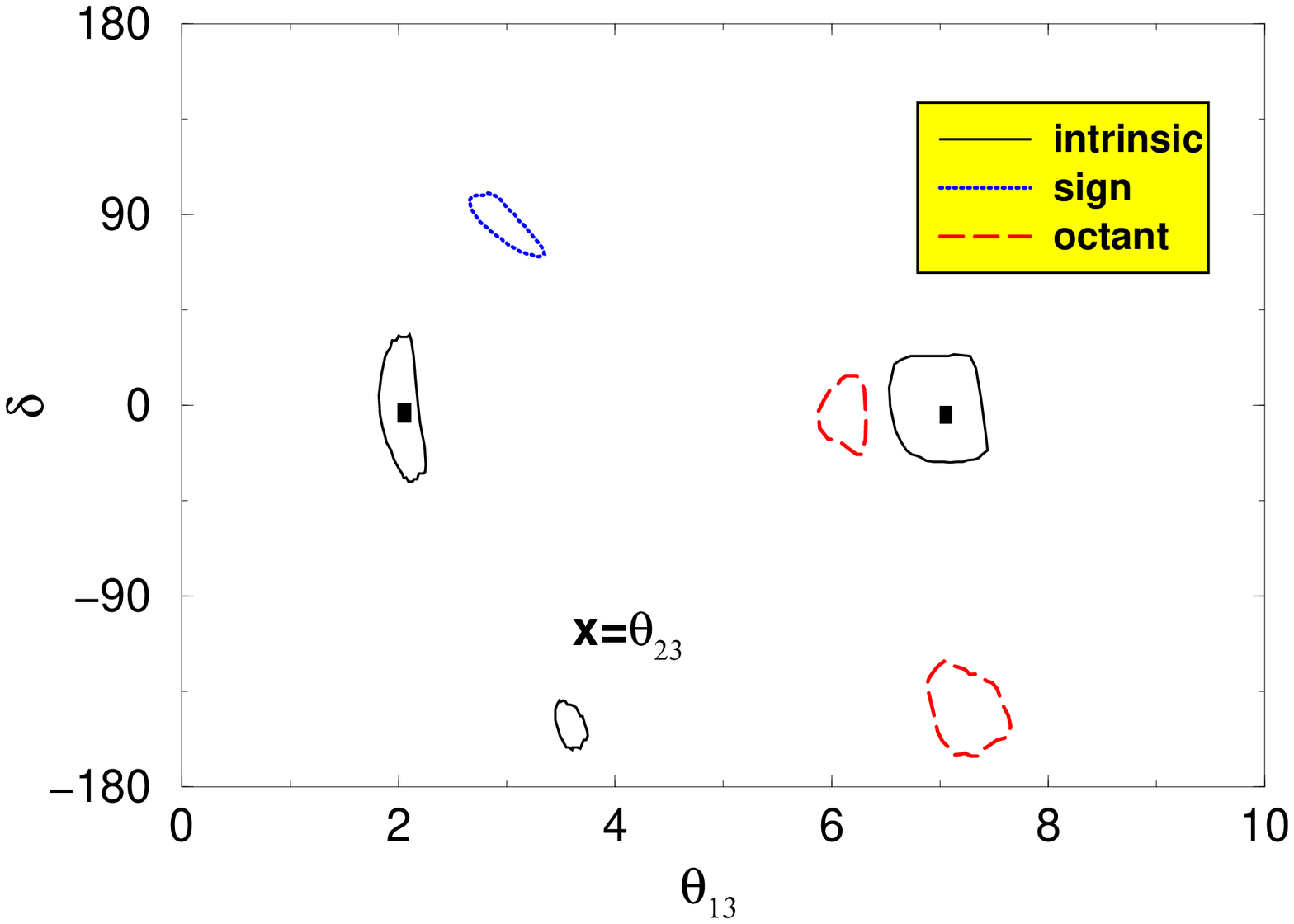} & 
\epsfxsize9cm\epsffile{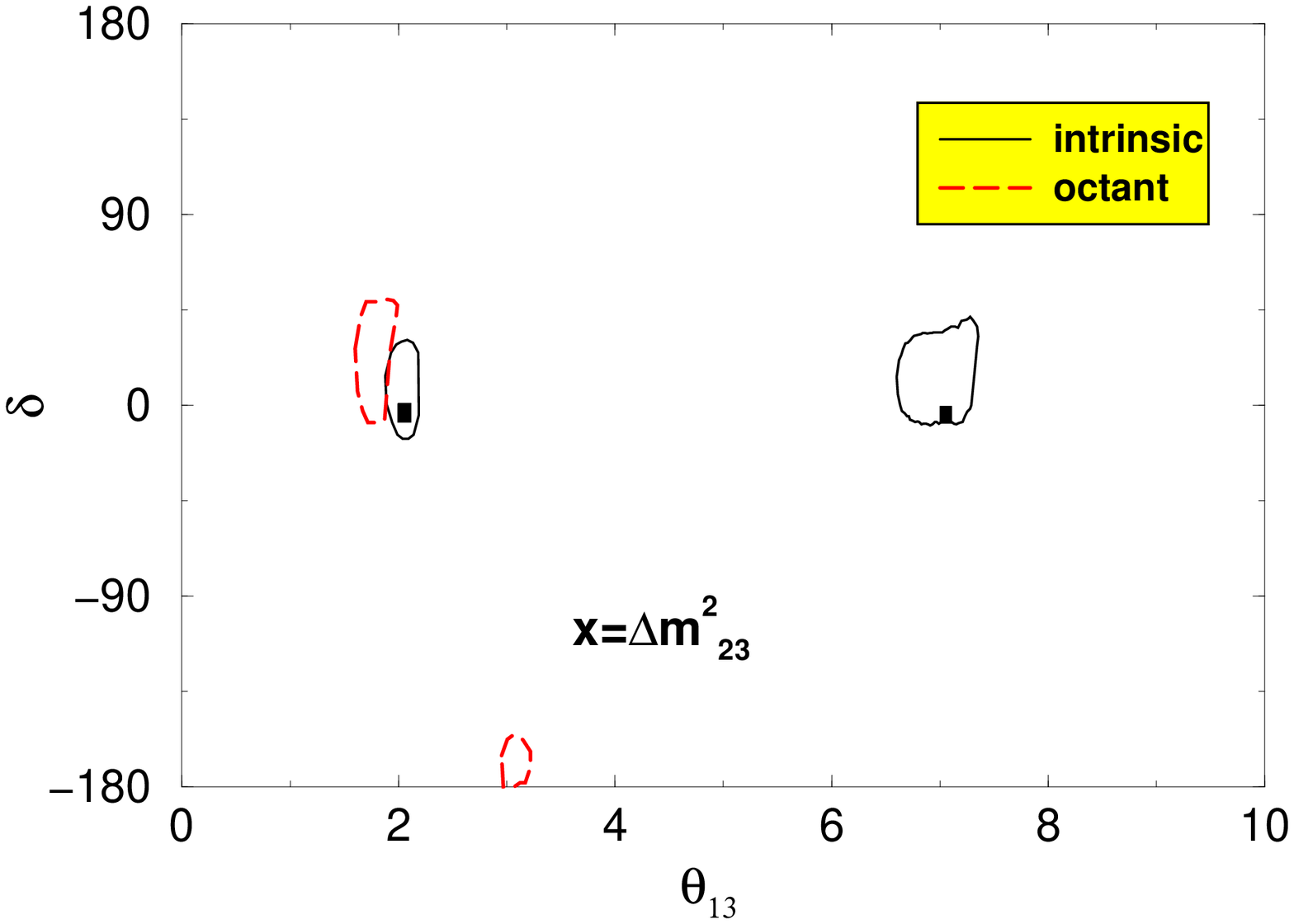} \\
\end{tabular}
\caption{\it Three-parameters 90 \% CL contours after a 5+5 years run at the Neutrino Factory. 
Input parameters: $\bar \theta_{13} = 2^\circ, 7^\circ$; $\bar \delta = 0^\circ$.
Left: $x = \theta_{23}$; right: $x = \Delta m^2_{23}$.}
\label{fig:atmoNF_00}
\end{center}
\end{figure}

\begin{figure}[h!]
\vspace{-0.35cm}
\begin{center}
\begin{tabular}{c c}
\hspace{-2cm} 
\epsfxsize9cm\epsffile{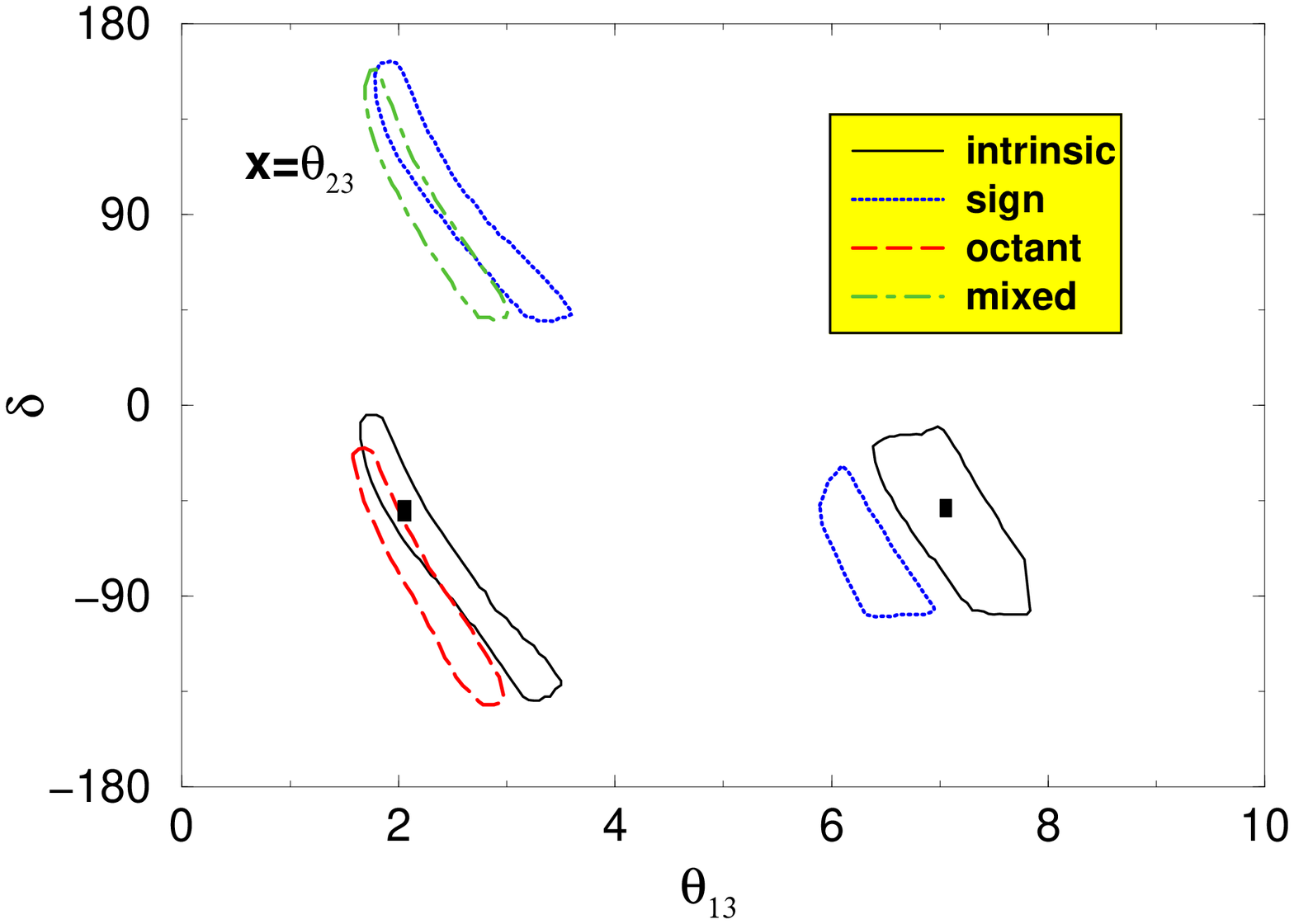} & 
\epsfxsize9cm\epsffile{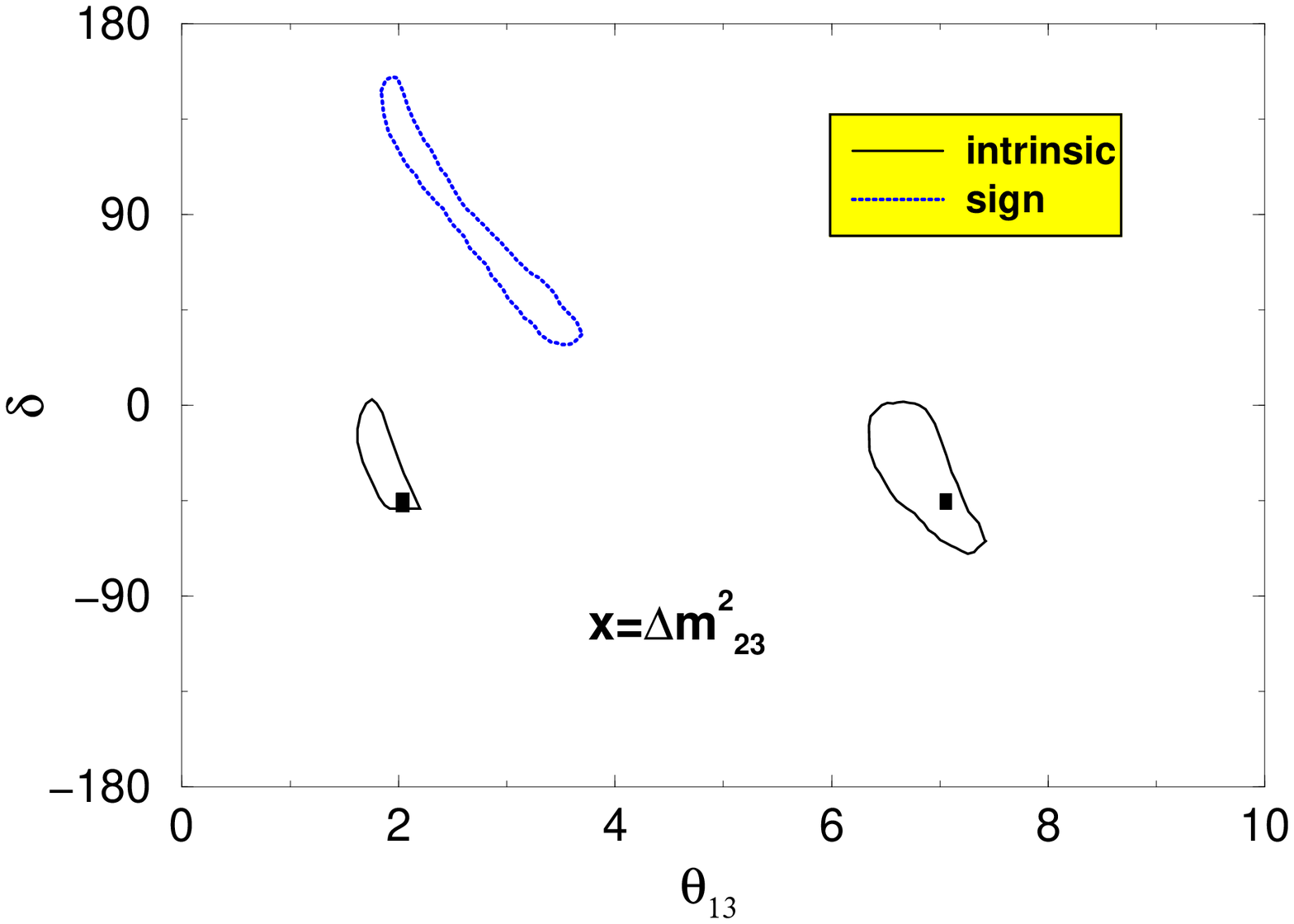} \\
\end{tabular}
\caption{\it Three-parameters 90 \% CL contours after a 5+5 years run at the Neutrino Factory. 
Input parameters: $\bar \theta_{13} = 2^\circ, 7^\circ$; $\bar \delta = -45^\circ$.
Left: $x = \theta_{23}$; right: $x = \Delta m^2_{23}$.}
\label{fig:atmoNF_m5}
\end{center}
\end{figure}

\begin{figure}[h!]
\vspace{-0.35cm}
\begin{center}
\begin{tabular}{c c}
\hspace{-2cm} 
\epsfxsize9cm\epsffile{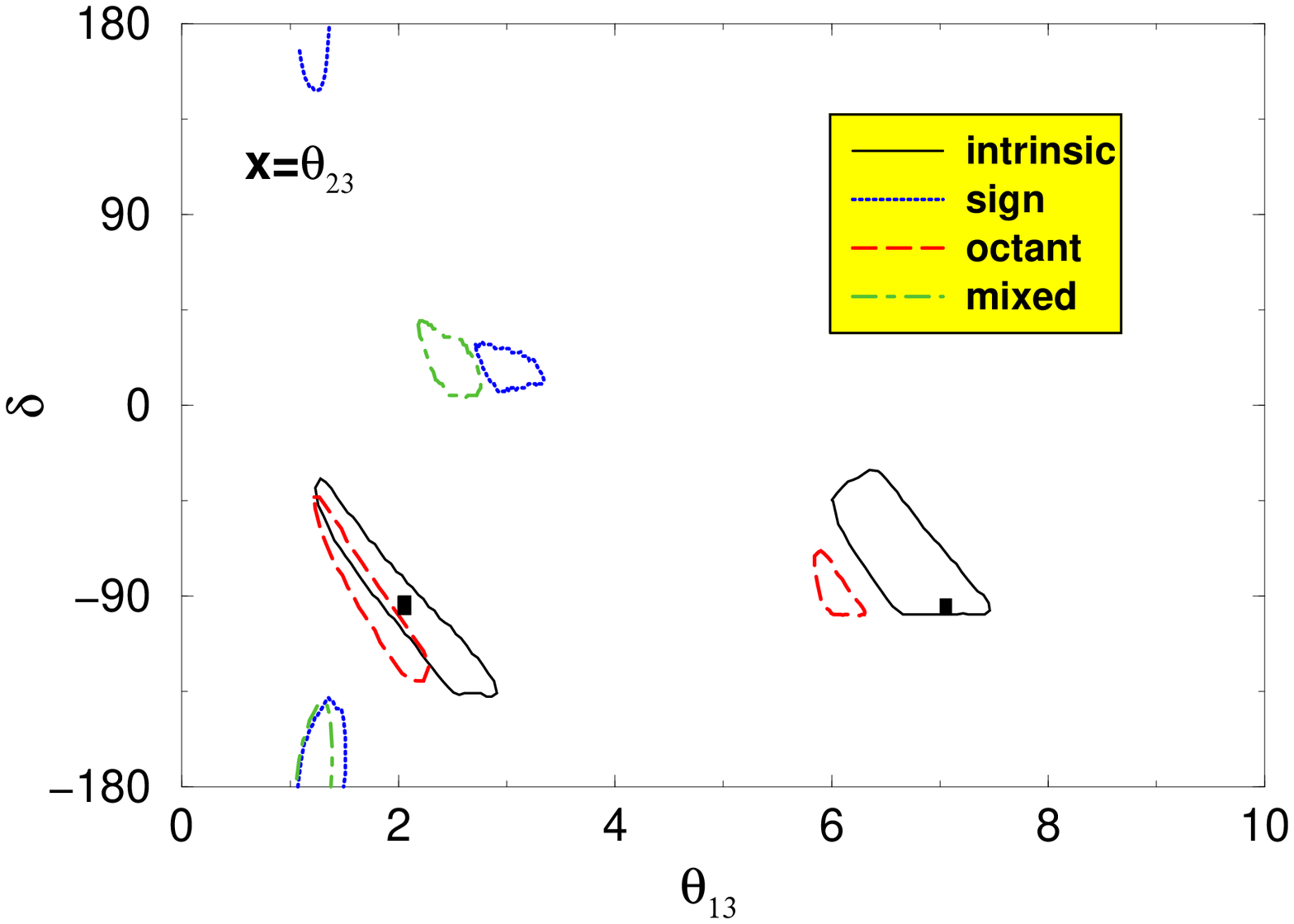} & 
\epsfxsize9cm\epsffile{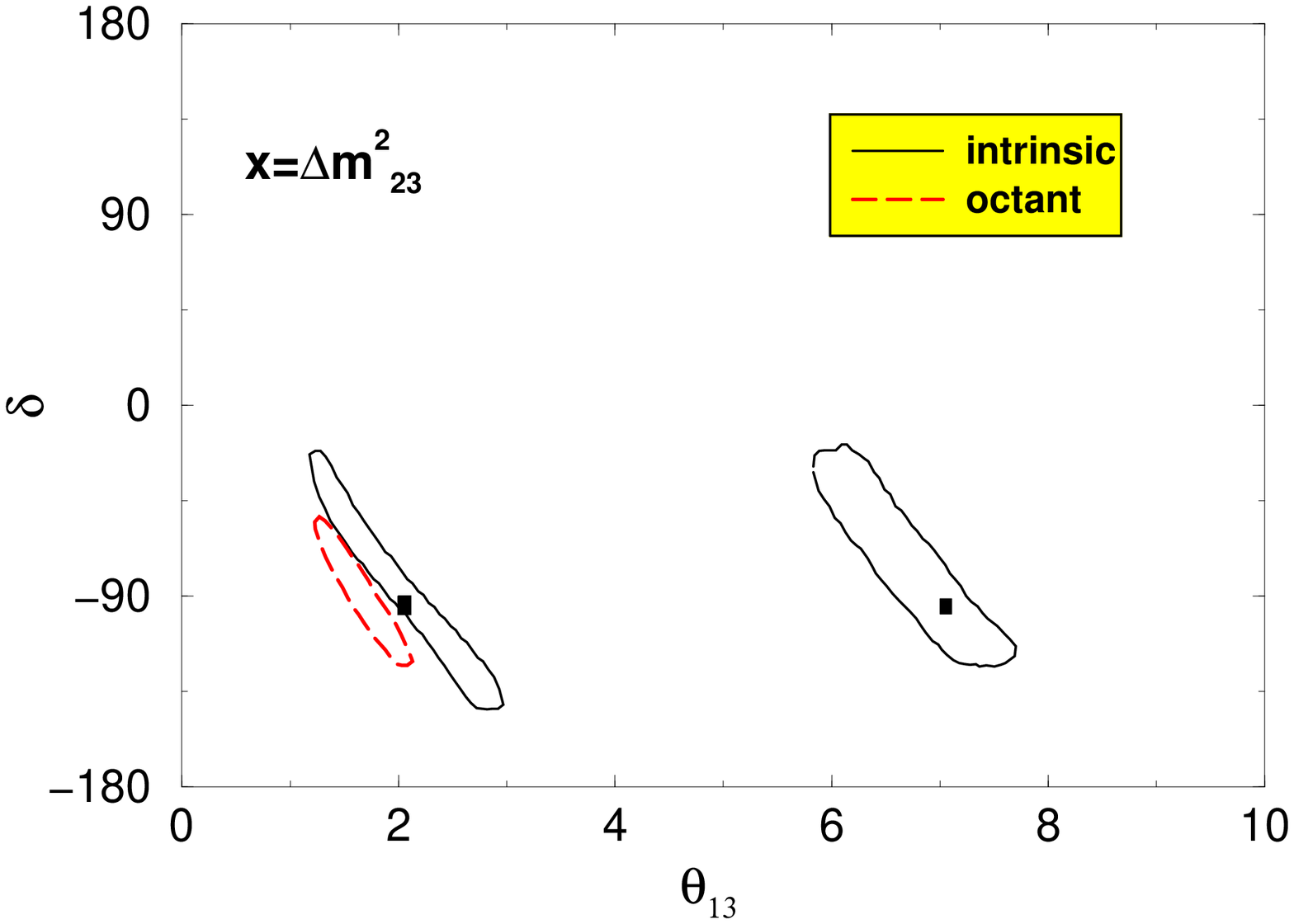} \\
\end{tabular}
\caption{\it Three-parameters 90 \% CL contours after a 5+5 years run at the Neutrino Factory. 
Input parameters: $\bar \theta_{13} = 2^\circ, 7^\circ$; $\bar \delta = -90^\circ$.
Left: $x = \theta_{23}$; right: $x = \Delta m^2_{23}$.}
\label{fig:atmoNF_m0}
\end{center}
\end{figure}

\end{document}